\documentclass[twocolumn]{aastex62}

\usepackage{upgreek}
\usepackage{enumitem}
\usepackage{comment}
\usepackage{multirow}
\usepackage{amsmath}
\usepackage{hyperref}
\usepackage{xurl}
\usepackage[normalem]{ulem}

\graphicspath{{./}{figures/}}

\shortauthors{Jin et al.}

\begin{document}


\title{Probing the disk-corona systems and broad line regions of changing-look quasars with X-ray and optical observations}

\author[0000-0002-5768-738X]{Xiangyu~Jin}
\affiliation{Department of Physics and McGill Space Institute, McGill University, 3600 University St., Montreal QC, H3A 2T8, Canada}

\author[0000-0001-8665-5523]{John~J.~Ruan}
\affiliation{Department of Physics and McGill Space Institute, McGill University, 3600 University St., Montreal QC, H3A 2T8, Canada}

\author[0000-0001-6803-2138]{Daryl~Haggard}
\affiliation{Department of Physics and McGill Space Institute, McGill University, 3600 University St., Montreal QC, H3A 2T8, Canada}
\affiliation{CIFAR Azrieli Global Scholar, Gravity \& the Extreme Universe Program, Canadian Institute for Advanced Research, 661 University
Avenue, Suite 505, Toronto, ON M5G 1M1, Canada}

\author{Marie-Joëlle~Gingras}
\affiliation{Department of Physics \& Astronomy, University of Waterloo, 200 University Avenue West, Waterloo, ON N2L 3G1, Canada}
\affiliation{Waterloo Centre for Astrophysics, University of Waterloo, 200 University Avenue West, Waterloo, ON N2L 3G1, Canada}

\author{Joseph~Hountalas}
\affiliation{Department of Physics and McGill Space Institute, McGill University, 3600 University St., Montreal QC, H3A 2T8, Canada}

\author{Chelsea~L.~MacLeod}
\affiliation{BlackSky, 1505 Westlake Ave N, Seattle, WA 98109, USA}

\author{Scott~F.~Anderson}
\affil{Department of Astronomy, University of Washington, Box 351580, Seattle, WA 98195, USA}

\author{Anh~Doan}
\affil{Department of Astronomy \& Astrophysics, The Pennsylvania State University, 525 Davey Lab, University Park, PA 16802, USA}

\author{Michael~Eracleous}
\affil{Department of Astronomy \& Astrophysics, The Pennsylvania State University, 525 Davey Lab, University Park, PA 16802, USA}
\affil{Institute for Gravitation and the Cosmos, The Pennsylvania State University, 525 Davey Lab, University Park, PA 16802, USA}

\author{Paul~J.~Green}
\affil{Harvard Smithsonian Center for Astrophysics, 60 Garden St., Cambridge, MA 02138, USA}

\author{Jessie~C.~Runnoe}
\affil{Department of Physics \& Astronomy, Vanderbilt University, 6301 Stevenson Center, Nashville, TN 37235, USA}

\correspondingauthor{Xiangyu Jin}
\email{xiangyu.jin@mail.mcgill.ca}







\begin{abstract}

``Changing-look" quasars are a new class of highly variable active galactic nuclei that have changed their spectral type over surprisingly short timescales of just a few years. The origin of this phenomenon is debated, but is likely to reflect some change in the accretion flow. To investigate the disk-corona systems in these objects, we measure optical/UV-X-ray spectral indices ($\alpha_{\rm OX}$) and Eddington ratios ($\lambda_{\rm Edd}$) of ten previously-discovered changing-look quasars at two or more epochs. 
By comparing these data with simulated results based on the behavior of X-ray binaries, we find possible similarities in spectral indices below 1\% Eddington ratio. We further investigate the Eddington ratios of changing-look quasars before and after their spectral type changes, and find that changing-look quasars cross the 1\% Eddington ratio boundary when their broad emission lines disappear/emerge. This is consistent with the disk-wind model as the origin of broad emission lines. 
\end{abstract}



\section{Introduction} \label{sec:intro}

Nearly all massive galaxies are believed to host supermassive black holes (SMBHs) at their centers, and those whose SMBHs are actively accreting are observed as active galactic nuclei (AGN). Emissions from nuclear regions of AGN are often time-variable \cite[e.\,g.,\,][]{Sesar2007AJ}, and some AGN display extreme changes in their broad Balmer emission lines and continuum fluxes in repeat optical spectroscopy (i.\,e.,\,they change from Type 1 to Type 1.9 when their broad ${\rm H \beta}$ emission lines disappear, or from Type 1.9 to Type 2 when their broad ${\rm H\alpha}$ emission lines disappear). These AGN displaying extreme spectral variability have been referred to as changing-look AGN\footnote{Though ``changing-look" AGN were first introduced to describe AGN which show variable absorption in their X-ray spectra \cite[see][]{Matt2003MNRAS}, in this work, ``changing-look" AGN are only referred to as those AGN which have changed their spectral type based on optical spectroscopy.} \citep{CS1973A&A,Tohline1976ApJ}. Apart from AGN in the local universe, the ``changing-look" phenomenon has also been observed in their more luminous and more distant counterparts $-$ quasars. The first changing-look quasar (CLQ) was identified in 2015 \citep{LaMassa2015ApJ}. Since then, dozens of CLQs have been discovered in observations from the Sloan Digital Sky Survey \cite[SDSS; see][]{MacLeod2016MNRAS,Runnoe2016MNRAS,Ruan2016ApJ,MacLeod2019ApJ}, the intermediate Palomar Transient Factory \cite[e.\,g.,\,][]{Gezari2017ApJ}, and the Zwicky Transient Facility \cite[e.\,g.,\,][]{Frederick2019ApJ}, or have been identified based on a mid-infrared color transition from a quasar-like color into a galaxy-like color or vice versa \citep{Sheng2020ApJ}. Several high-redshift ($z>2$) CLQs have also been discovered \citep{Ross2019arXiv}. Nevertheless, the physical origin of the CLQ phenomenon is still unclear. Recent observations have suggested that CLQs may be triggered by accretion state transitions \cite[e.\,g.,\,][]{ND2018MNRAS}, possibly analogous to those observed in X-ray binaries. However, the expected transition timescale ($\sim$10$^5$~years) for these state transitions in SMBHs, when directly scaled with the black hole mass \citep{Sobolewska2011MNRASa}, appears to be incompatible with
the short timescales of spectral changes in CLQs ($\sim$5--10~years). Thus, it is worthwhile to investigate whether there are indeed observable similarities between CLQs and X-ray binaries undergoing accretion state transitions.

Accretion state transitions have been detected in many Galactic X-ray binaries \cite[e.\,g.,\,][]{Maccarone2003A&A,Debnath2010A&A,Tang2011RAA,Wang2018PASJ,Kara2019Natur}. Both black hole and neutron star X-ray binaries show spectral state transitions at $\sim$ 1\% bolometric Eddington ratios ($\lambda_{\rm Edd}=L_{\rm bol}/L_{\rm Edd}$). In these different states, X-ray binaries display distinct X-ray behavior in the hardness-intensity diagram (HID). At high Eddington ratios, their X-ray spectra are dominated by thermal soft X-rays, believed to be emitted from a standard thin accretion disk \citep{SS1973A&A}. When Eddington ratios drop below 1\%, the inner parts of the accretion disk may evaporate, causing the spectra to be dominated instead by hard X-rays from either a corona \citep{Frank1987A&A}, an advection dominated accretion flow \cite[ADAF;][]{NY1995ApJ}, or a newly launched jet \citep{Markoff2001A&A}. The spectral state associated with high Eddington ratios is called the ``High/Soft" state, while the one at lower Eddington ratios is referred to as the ``Low/Hard" state. 
Similar transitions also likely occur in AGN, but due to the expected long transition scales of $10^5$~years \citep{Sobolewska2011MNRASa}, we may not expect to directly witness such a transition in AGN.
However, by observing the spectral energy distribution (SED) shapes of AGN at different Eddington ratios, and comparing these with the spectral shapes of a single black hole X-ray binary undergoing accretion state transitions, we may be able to better understand and characterize any similarities between supermassive black holes and stellar mass black holes \citep{Ruan2019arXiv}.

By assuming that supermassive black holes have analogous accretion flow structures as those in X-ray binaries, and scaling the disk and the corona emission with black hole mass, \citet{Sobolewska2011MNRASa} simulate optical/UV-X-ray spectral indices ($\alpha_{\rm OX}$) and bolometric Eddington ratios (${\lambda}_{\rm Edd}$) of AGN populations. That work predicts that AGN may display distinct spectral states similar to those observed in X-ray binaries. They predict that at ${\lambda}_{\rm Edd}\gtrsim1\%$, $\alpha_{\rm OX}$ and ${\lambda}_{\rm Edd}$ show a positive correlation, but this 
becomes an inverse correlation when ${\lambda}_{\rm Edd}$ drops below 1\%. 

Previous observations of bright quasars often show that there is indeed a positive correlation between their single-epoch $\alpha_{\rm OX}$ and ${\lambda}_{\rm Edd}$ at ${\lambda}_{\rm Edd}\gtrsim1\%$ \citep{Maoz2007MNRAS,Lusso2010A&A,Grupe2010ApJS}. However, for low-luminosity AGN (LLAGN), it is difficult to measure their black hole masses (and thus Eddington ratios), due to the difficulty of detecting broad emission lines in the spectra. 
A few studies based on LLAGN show a weak negative correlation between their $\alpha_{\rm OX}$ and $\lambda_{\rm Edd}$ \citep{Maoz2007MNRAS,Xu2011ApJ}, but additional observations are needed to probe the relation between $\alpha_{\rm OX}$ and $\lambda_{\rm Edd}$ at $\lambda_{\rm Edd}\lesssim1\%$. CLQs offer a unique opportunity to probe this relation at $\lambda_{\rm Edd}\lesssim1\%$, by combining black hole masses measured from bright state spectra of CLQs with broad Balmer emission lines present, and $\alpha_{\rm OX}$ when CLQs fade. Given that the Eddington ratios of LLAGN are usually much lower than 1\% \citep{GUCAO2009MNRAS,Xu2011ApJ}, CLQs are especially helpful to study this $\alpha_{\rm OX}-\lambda_{\rm Edd}$ relation just below 1\% $\lambda_{\rm Edd}$.
By measuring the $\alpha_{\rm OX}$ and $\lambda_{\rm Edd}$ of six ``turn-off" CLQs, \citet{Ruan2019arXiv} find a negative correlation between $\alpha_{\rm OX}$ and $\lambda_{\rm Edd}$ at $\lambda_{\rm Edd}\lesssim1\%$, as observed in X-ray binaries. This supports the notion that emission mechanisms from X-ray binaries can be directly scaled to AGN with supermassive black holes.
In this work, we report optical and X-ray observations of another ten CLQs, and test whether a similar relation is borne out by these systems at $\lambda_{\rm Edd}\lesssim1\%$. 

Dramatic changes in the optical broad emission lines (BELs) of CLQs can also probe the physical origin of the lines themselves.
Optical spectra of AGN are characterized primarily by a power-law continuum, BELs, and narrow emission lines \cite[e.\,g., ][]{VB2001AJ}. BELs are emitted from the broad line region (BLR), which is believed to be the high velocity gas gravitationally bound to the central supermassive black hole \citep{PW2000ApJ,Peterson2004ApJ}. Results from reverberation mapping \citep{Denney2009ApJ,DR2018ApJ}, and study of the quasar orientation and observed width of BEL profiles \citep{SB2017ApJ,SH2014Natur} imply that the geometry of the BLR is likely to be disk-like. Furthermore, study of BEL profiles suggests the BLR gas is structured as a smooth continuous flow \citep{Laor2006ApJ}. However, the exact origin of the BLR gas is still unclear \citep{Elvis2017ApJ}.
The disk-wind model suggests that the BLR gas comes from winds produced by the accretion disk \citep{Emmering1992ApJ,Murray1995ApJ,MC1997ApJ,EH2009ApJ,Elitzur2014MNRAS,Elitzur2016MNRAS}. When the mass accretion rate drops below a certain limit (corresponding to a luminosity of $\sim 4.7\times10^{39}\;{\rm M_{7}^{2/3}\;erg\;\!s^{-1}}$, where ${\rm M_{7}^{2/3}}$ is the black hole mass in ${\rm 10^{7}\;M_{\odot}}$), winds can no longer be sustained, such that the observed BELs ``disappear" \citep{EH2009ApJ}. 
Below the critical luminosity, AGN are expected to be ``true" Type 2 AGN (i.\,e.,\,Type 2 AGN intrinsically absent of BLRs). Although this critical luminosity is dependent on the black hole mass, the disk-wind model predicts that 
more detections of BEL disappearance
are expected below the 1\% Eddington ratio \citep{Elitzur2016MNRAS}, and a double-peaked BEL profile, the signature of a rotating disk, should emerge in the quasar spectrum when the accretion rate drops \citep{Elitzur2014MNRAS}. 
Single-epoch observations show that Eddington ratios of bright quasars with prominent BELs, are always higher than 1\% \citep{Kollmeier2006ApJ,SE2010MNRAS}. Yet it is still unknown whether Eddington ratios of individual quasars varying around 1\% will 
display the appearance/disappearance of broad emission lines.
By studying the Eddington ratios of the single changing-look AGN UGC~3223, \citet{Wang2020AJ} find this object crosses the 1\% Eddington ratio when its AGN type changes. 
However, it is still necessary to investigate whether this 1\% Eddington ratio is associated with appearance/disappearance of BELs with multi-epoch observations of CLQs, and with a larger dataset. In this work, we also study the changes in the Eddington ratio distributions from the brightest to the faintest optical spectra of CLQs, 
investigate the possible connection with the 1\% Eddington ratio, and study the BEL profiles of those CLQs.

We present optical and X-ray observations of ten CLQs identified via repeat SDSS photometry and follow-up optical spectroscopy \citep{MacLeod2016MNRAS,MacLeod2019ApJ}. 
We obtain new optical data from the Multiple Mirror Telescope, the Magellan Telescope, the Apache Point Observatory Astrophysical Research Consortium 3.5m, and the Hobby-Eberly Telescope. 
We also obtain new X-ray data from the {\it Chandra X-ray Observatory}, along with archival data from the {\it XMM-Newton} and the {\it ROSAT}. This paper is organised as follows: we present our optical data reduction procedures in \S\ref{sec:opticaldata}, X-ray measurements in \S\ref{sec:Xraydata}, results and discussion in \S\ref{sec:resultndiscuss}, and a brief conclusion in \S\ref{sec:conclusion}. Throughout this paper, we adopt the cosmological parameters $\Omega_{m}=0.286$, $H_0=69.6\;\!{\rm km\;\!s^{-1}\;\!Mpc^{-1}}$, $\Omega_{\Lambda}=0.714$ \citep{Bennett2014ApJ}.

\begin{figure*}[t!]
    \centering
    \includegraphics[width=0.45\textwidth]{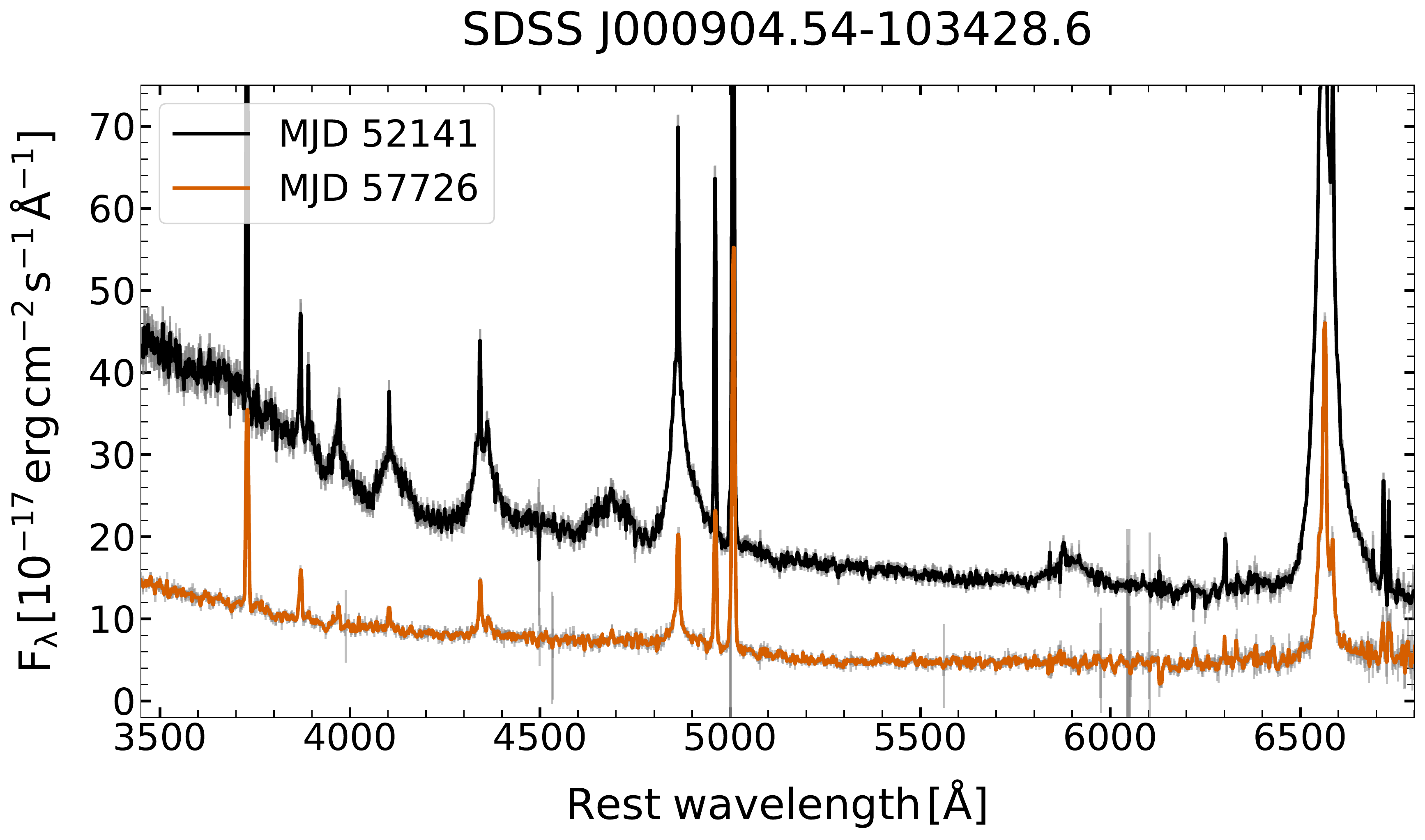}
    \includegraphics[width=0.45\textwidth]{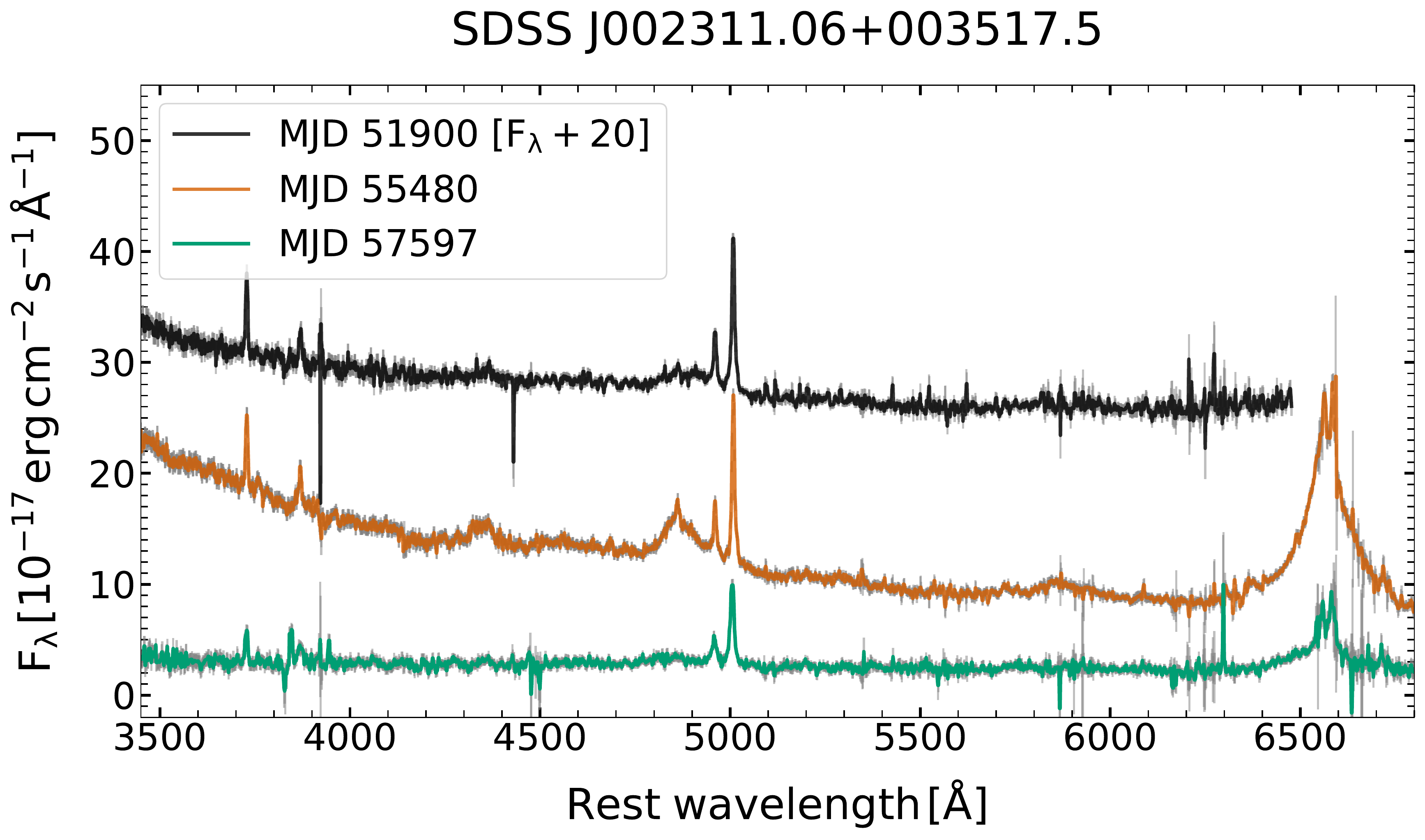}
    \caption{Original optical spectra of the changing-look quasars SDSS J000904.54-103428.6 and SDSS J002311.06+003517.5. See Table \ref{tab:Optical} for more details about the instruments and the transition directions. Noise is shown in grey. We apply a median filter to smooth the spectra for better visualization.}
\end{figure*}\label{fig:optical_ori_spec_1}
\renewcommand{\thefigure}{\arabic{figure} (Cont.)}
\addtocounter{figure}{-1}

\begin{deluxetable*}{cccccc}[!htb]\centering 
\tabletypesize{\scriptsize}
\tablecaption{Optical Data of Changing-look Quasars}
\tablewidth{2\columnwidth}
\tablehead{
\colhead{Target Name} & \colhead{Redshift ($z$)} & \colhead{Instrument} & \colhead{Observation Date} & \colhead{Note} & \colhead{Transition}\\
\colhead{(1)} & \colhead{(2)} & \colhead{(3)} & \colhead{(4)} & \colhead{(5)} & \colhead{(6)}}
\startdata
SDSS J000904.54-103428.6 & 0.2406 & SDSS & 52141 & Bright & \multirow{2}{*}{Off} \\
 & & MMT & 57726 & Faint &  \\
SDSS J002311.06+003517.5 & 0.4221 & SDSS & 51900 &  & \multirow{3}{*}{On/Off}  \\
 & & SDSS & 55480 & Bright &   \\
 & & Magellan & 57597 & Faint & \\
SDSS J022556.08+003026.7 & 0.5039 & SDSS & 52200 &  & \multirow{4}{*}{On/Off/On}  \\
 & & SDSS & 52944 & Bright &  \\
 & & SDSS & 55208 & Faint &  \\
 & & ARC 3.5m & 58814 &  &  \\
SDSS J132457.29+480241.2 & 0.2716 & SDSS & 52759 & Bright & \multirow{2}{*}{Off} \\
 & & HET & 58127 & Faint & \\
 SDSS J160111.25+474509.6 & 0.2970 & SDSS & 52354 & Bright & \multirow{2}{*}{Off}\\
 & & MMT & 57895 & Faint &  \\
SDSS J164920.79+630431.3 & 0.3221 & SDSS & 51699 & Bright  & \multirow{2}{*}{Off} \\
 & & ARC 3.5m & 58276 & Faint &  \\
 SDSS J214613.30+000930.8 & 0.6220 & SDSS & 52968 &  & \multirow{3}{*}{On/Off} \\
 & & SDSS & 55478 & Bright & \\
 & & ARC 3.5m & 57663 & Faint & \\
SDSS J220537.71-071114.5 & 0.2950 & SDSS & 52468 & Bright & \multirow{2}{*}{Off} \\
 & & MMT & 57989 & Faint & \\
 SDSS J225240.37+010958.7 & 0.5335 & SDSS & 52178 &  & \multirow{4}{*}{On/Off}\\
 & & SDSS & 55500 & Bright & \\
 & & Magellan & 57598 &  &  \\
 & & ARC 3.5m & 58814 & Faint  &  \\
SDSS J233317.38-002303.5 & 0.5130 & SDSS & 52199 &  & \multirow{3}{*}{On/Off} \\
& & SDSS & 55447 & Bright & \\
 & & ARC 3.5m & 58429 & Faint & \\
\enddata
\tablecomments{(1) Name of changing-look quasars in SDSS, in order of increasing R.A.; (2) Spectroscopic redshift $z$, note that all 1$\sigma$ uncertainties in spectroscopic redshifts are smaller than the last digit; (3) Instrument of observations; (4) Date of optical observations, in modified Julian date (MJD); (5) ``Bright" denotes the brightest optical spectrum of the object, while ``Faint" denotes the faintest optical spectrum of the object; (6) Transition direction(s) of changing-look quasars; ``On" means that the changing-look quasar brightens (i.\,e., a ``Turn-on" changing-look quasar), while ``Off" means the changing-look quasars fades (i.\,e., a ``Turn-off" changing-look quasar).}
\end{deluxetable*}\label{tab:Optical}

\begin{figure*}[hb!]
    \centering
    \includegraphics[width=0.45\textwidth]{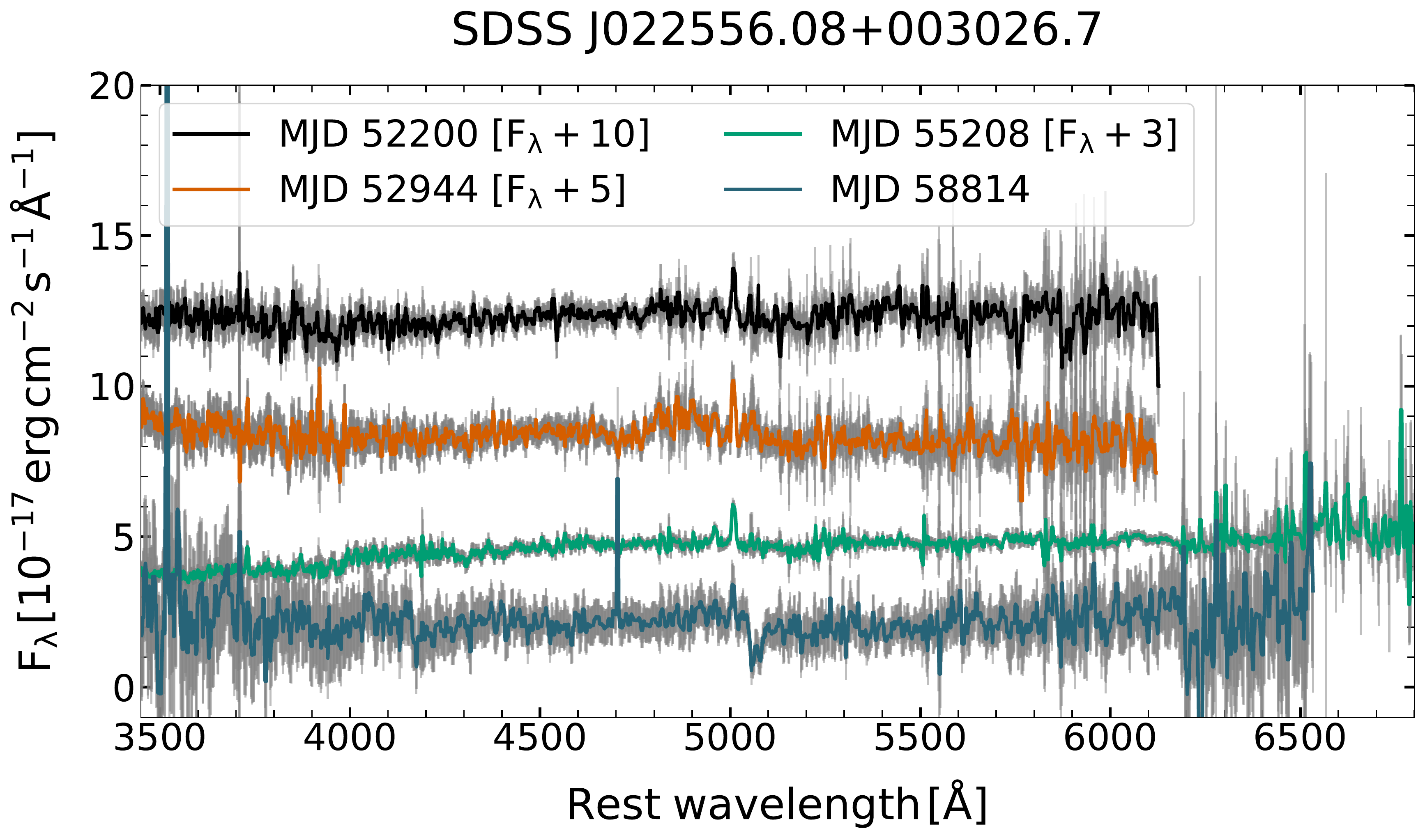}
    \includegraphics[width=0.45\textwidth]{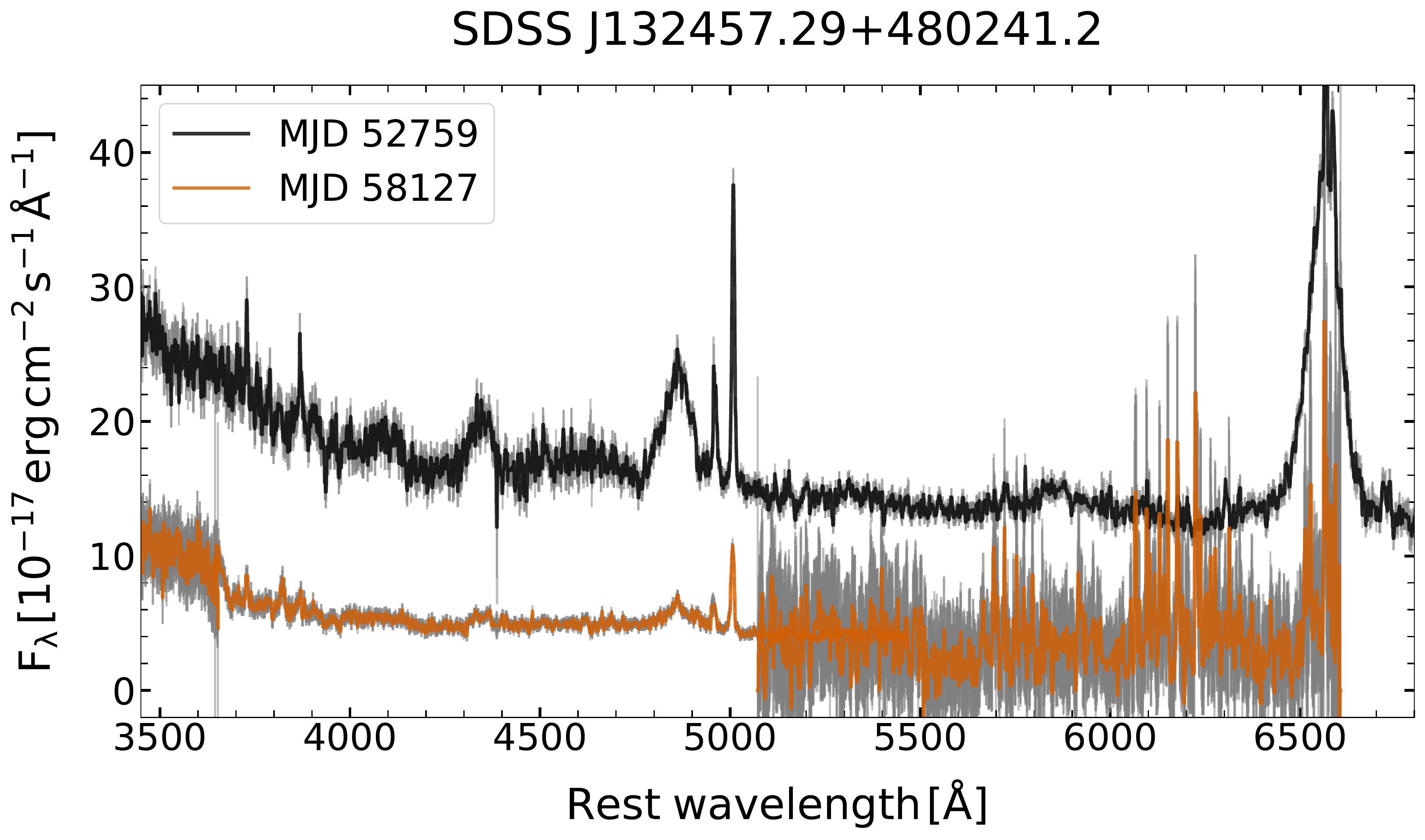}\\
    \vspace{0.3cm}
    \includegraphics[width=0.45\textwidth]{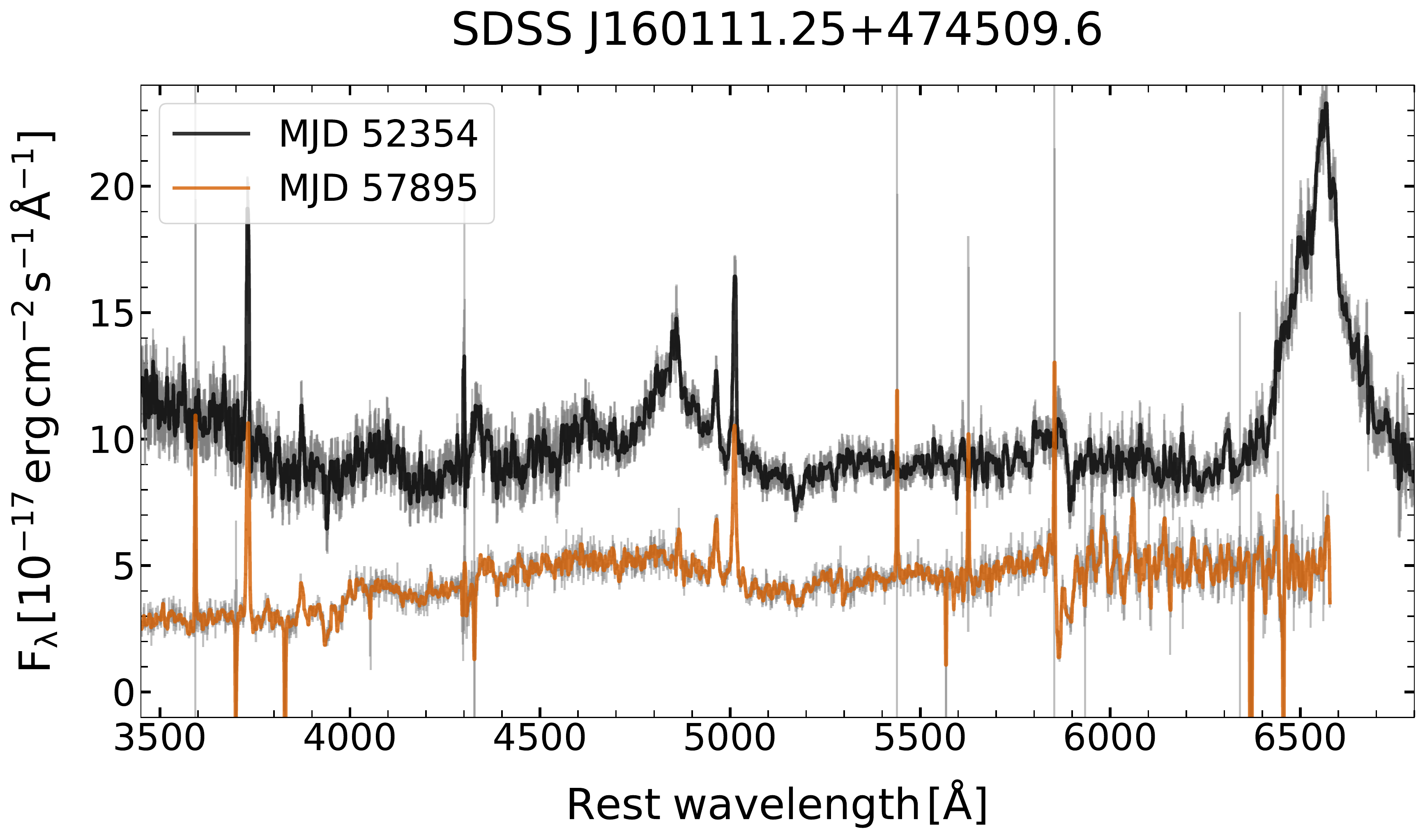}
    \includegraphics[width=0.45\textwidth]{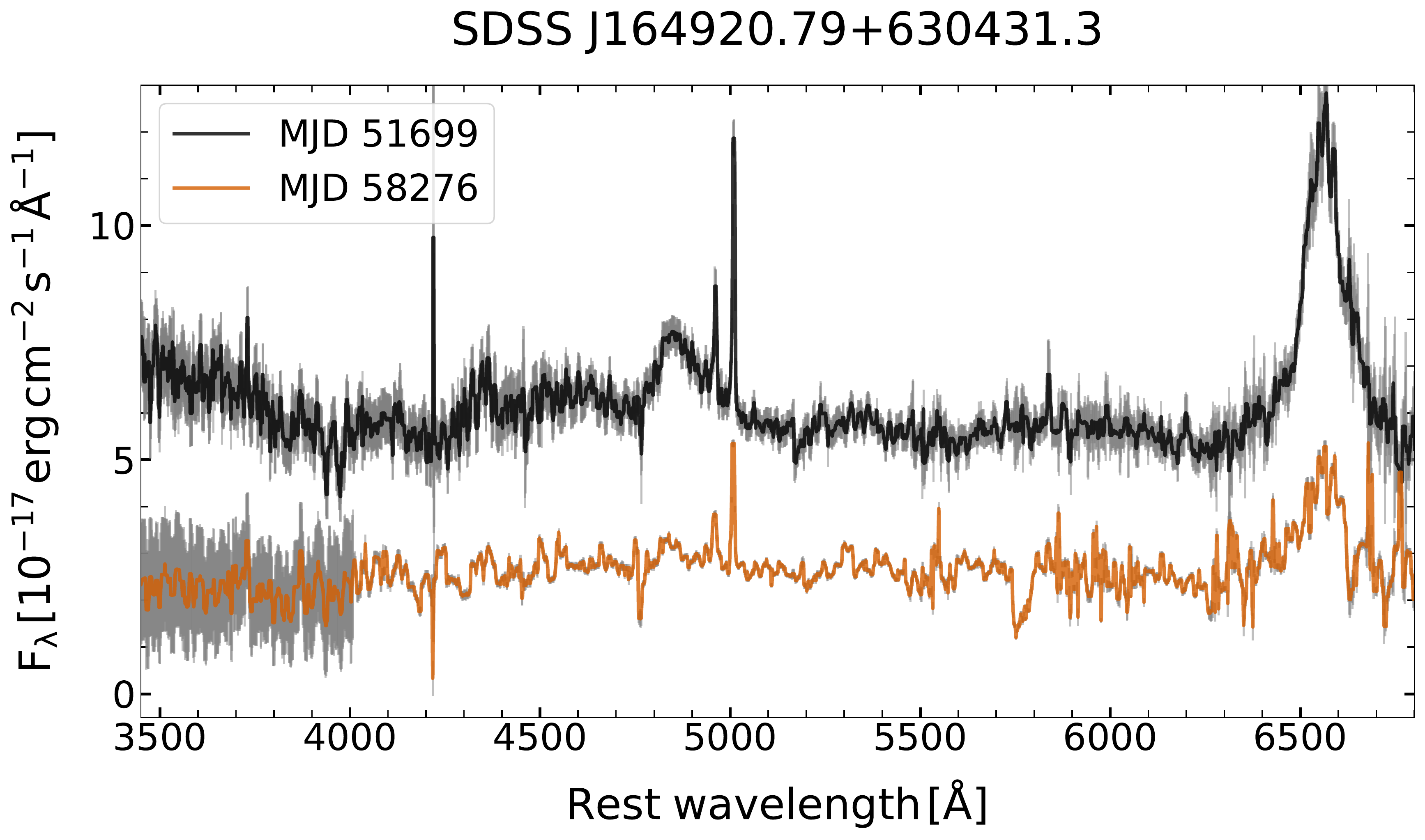}\\
    \vspace{0.3cm}
    \includegraphics[width=0.45\textwidth]{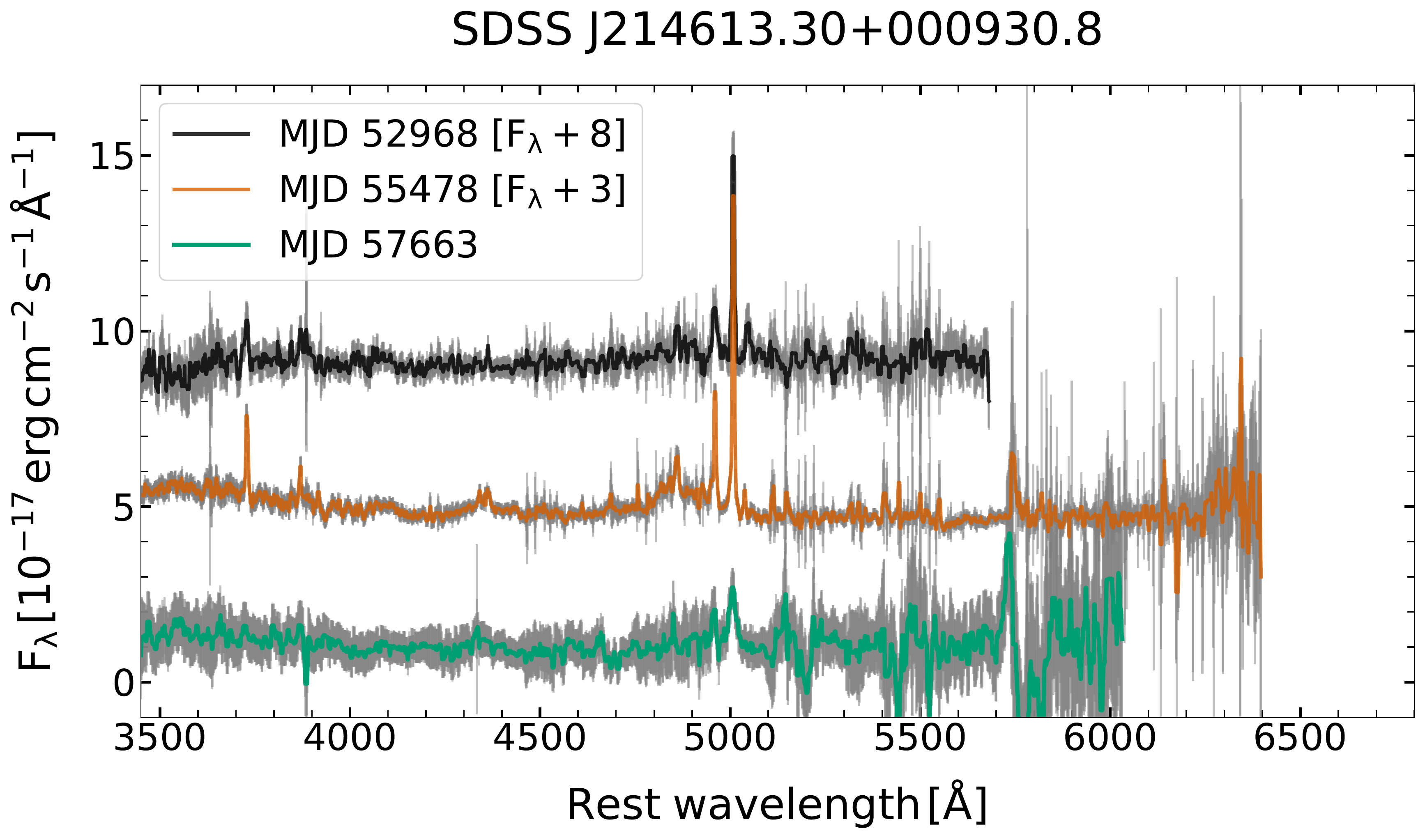}
    \includegraphics[width=0.45\textwidth]{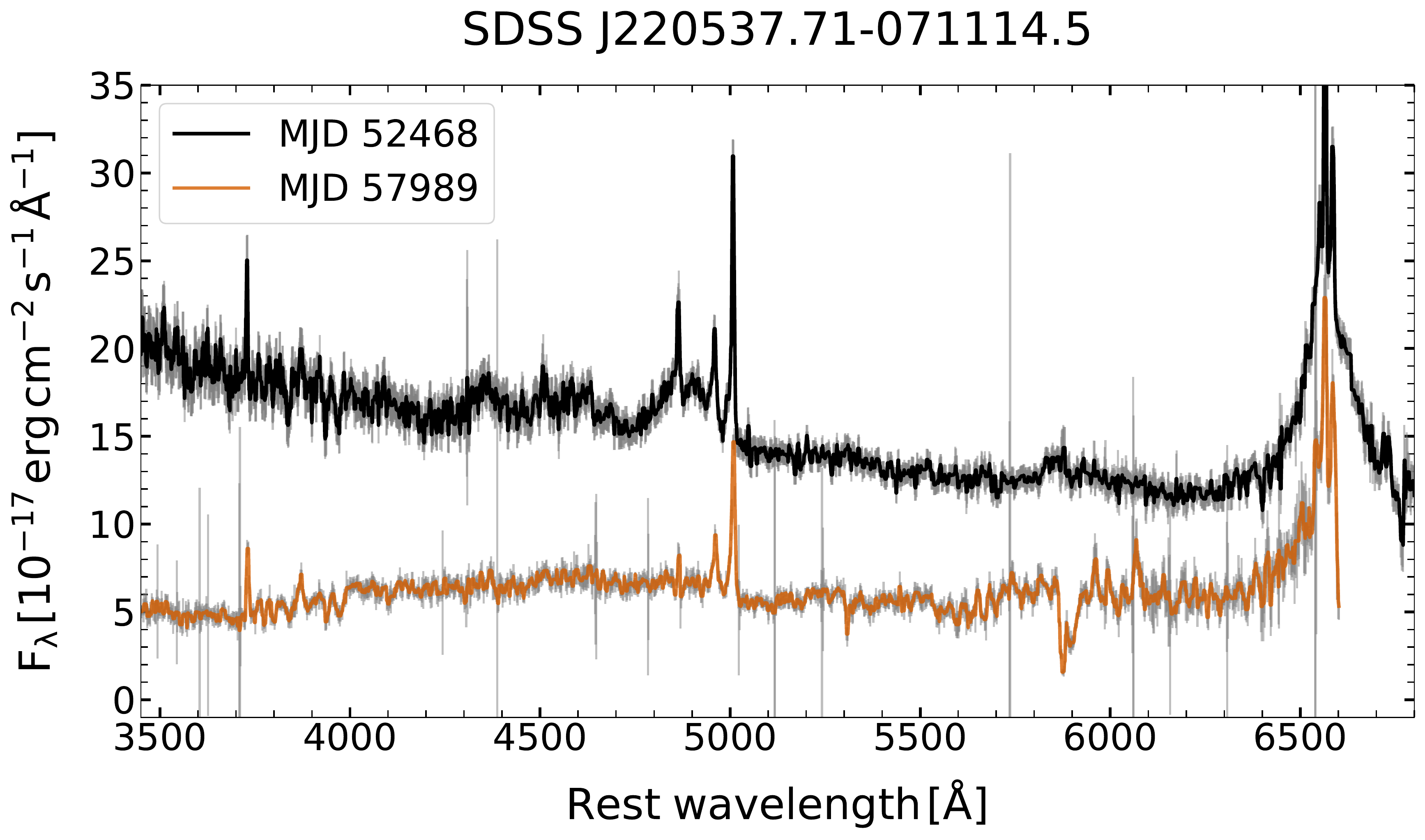}\\
    \vspace{0.3cm}
    \includegraphics[width=0.45\textwidth]{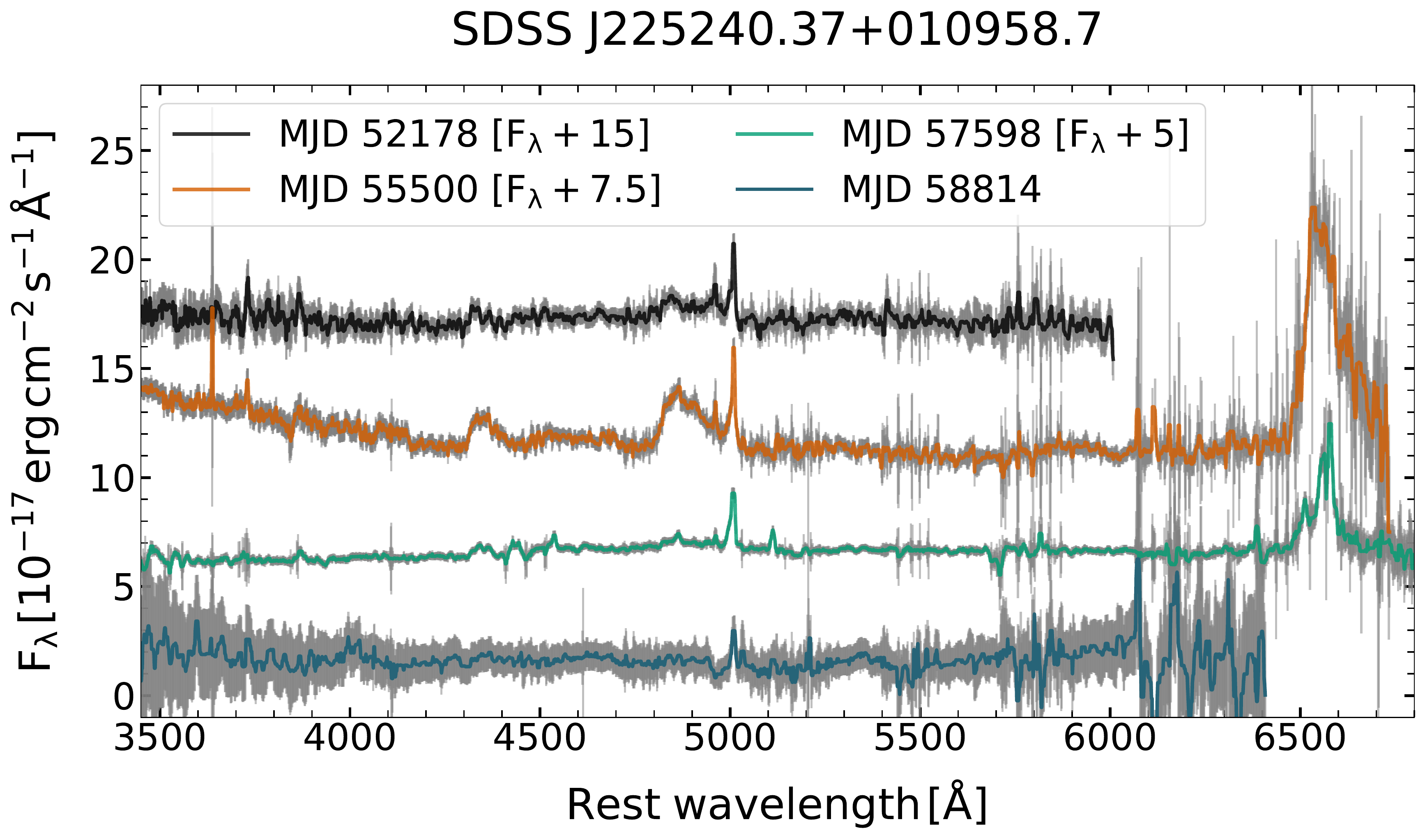}
    \includegraphics[width=0.45\textwidth]{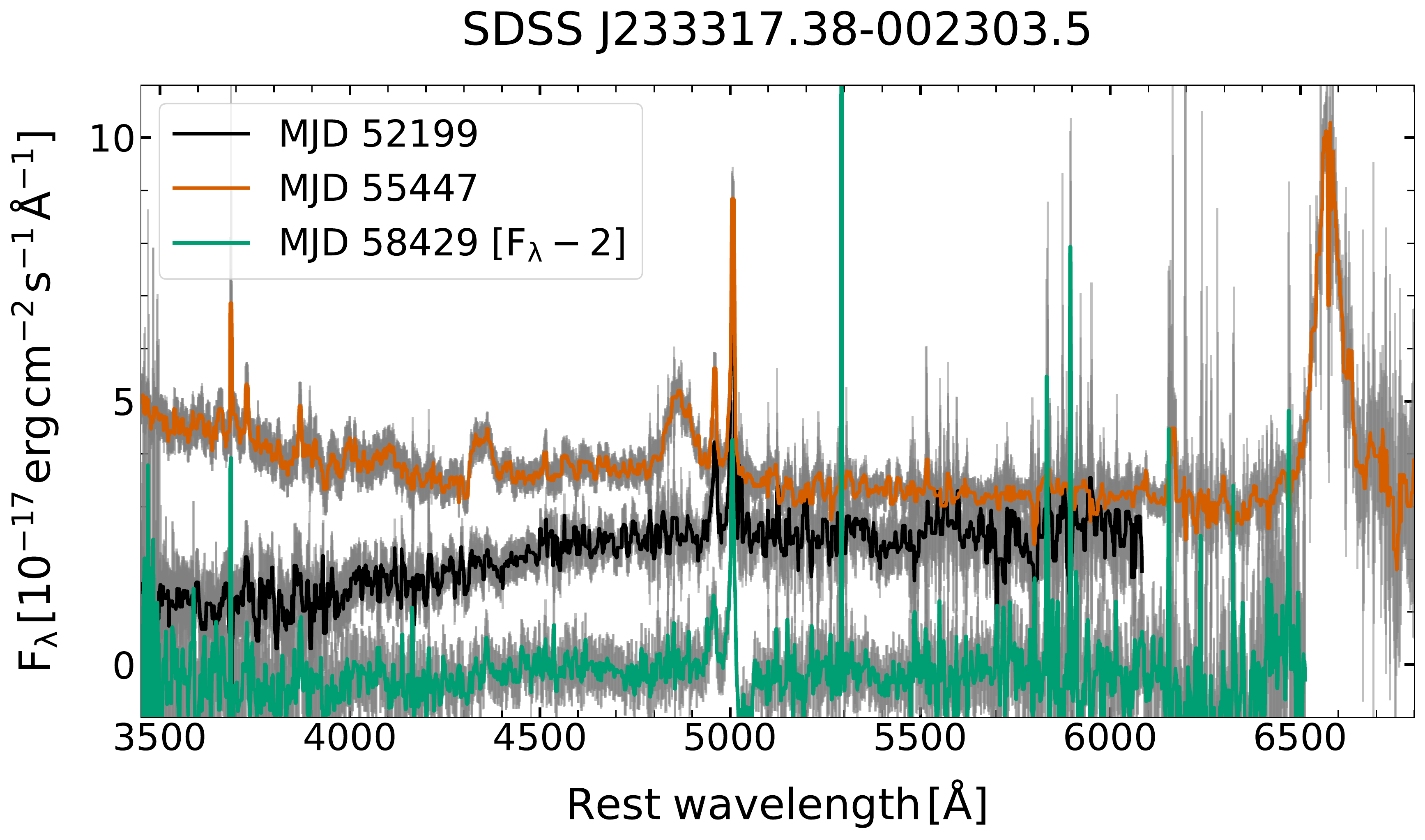}l
    \caption{Optical spectral of eight changing-look quasars.}
\end{figure*} \label{fig:optical_ori_spec_2}

\renewcommand{\thefigure}{\arabic{figure}}

\section{Optical Data Reduction}\label{sec:opticaldata}

\subsection{Targets and Data}

The ten CLQs studied here were selected from the SDSS \citep{York2000AJ,Abazajian2009ApJS}
based on their large amplitude photometric changes, and then confirmed by follow-up optical spectroscopy \citep{MacLeod2016MNRAS,MacLeod2019ApJ}. We also obtain new optical spectroscopy from several different telescopes, which we briefly summarize below. 

\subsubsection{MMT and Magellan}
We observe three CLQs with the Blue Channel Spectrograph on the 6.5m Multiple Mirror Telescope (MMT) during 2016 to 2018 \citep{Angel1979SAOSR}, with the 300$\;l\;\!{\rm mm}^{-1}$ grating, covering the wavelength range 3,300--8,500~${\rm \AA}$ at a spectral resolution\footnote{\url{http://www.mmto.org/instrument-suite/blue-red-channel-spectrographs/blue-channel-details}} of 6.47~${\rm \AA}$. We use {\tt pydis}\footnote{\url{https://jradavenport.github.io/2015/04/01/spectra.html}} to reduce the MMT data, and calibrate the flux of the observed spectra using a standard star observed on the same night. 

We also acquire two spectra from the 6.5m Magellan Clay telescope, with the Low Dispersion Survey Spectrograph 3 (LDSS3)-C spectrograph. We use the VPH-All grism, which covers the wavelength range 4,250--10,000~${\rm \AA}$, with a spectral resolving power\footnote{\url{http://www.lco.cl/Members/gblanc/ldss-3/ldss-3-user-manual-tmp}} $R\sim860$. We use both {\tt pydis} and IRAF, and follow the data reduction procedures in \citet{MacLeod2019ApJ} to reduce the data, and perform wavelength and flux calibration. 

\subsubsection{ARC 3.5m}
For the five Astrophysical Research Consortium (ARC) 3.5-meter telescope observations, we use the Dual Imaging Spectrograph (DIS) with the B400/R300 grating for each spectrum, along with spectra of HeNeAr lamps to perform wavelength calibration. Those spectra cover
3,400--9,200~{\rm \AA} at a spectral resolving power of $R$~$\sim1,000$, with a $1.\!\!^{\prime\prime}$5 slit. Spectra of standard stars from the same night are used for flux calibration. 
We use IRAF to perform bias and flat field corrections, aperture extraction, and wavelength and flux calibration \citep{Tody1986SPIE,Tody1993ASPC}. After we reduce the blue/red spectra, we normalize the flux of the blue spectra to match the corresponding flux of the red spectra via the overlapping observed-frame wavelength region (5,300--5,500~${\rm \AA}$). 

\subsubsection{Hobby-Eberly Telescope}
We obtain a new spectrum of SDSS J132457.29+480241.2
from the Hobby-Eberly Telescope \cite[HET;][]{Ramsey1998SPIE}, with the Low Resolution Spectrograph 2 \cite[LRS2;][]{Chonis2016SPIE}.
The LRS2 is a fiber-fed, integral-field spectrograph made up
of two units, the LRS2-B and LRS2-R, each with dual channels, that
operate as independent instruments and observe separately. Thus, we obtain spectra that cover four separate but overlapping bands with
comparable spectral resolution: the UV band, covering the range
3,700--4,700~\AA\ at a spectral resolution of 2.2~\AA, the orange
band, covering the range 4,600--7,000~\AA\ at a spectral resolution of
5.1~\AA, the red band, covering the range 6,500--8,470~\AA\ at a
spectral resolution of 4.2~\AA, and the far-red band, covering the
range 8,230--10,500~\AA\ at a spectral resolution of 4.9~\AA. The
throughput of the far-red arm is the lowest, yielding spectra with low
signal-to-noise ratio that are not useful for our purposes. The
spectra are extracted from a circular aperture matching the seeing, with a diameter of $\sim
1.\!\!^{\prime\prime}5$--$2.\!\!^{\prime\prime}2$, depending on the
weather conditions. The reductions consist of flat-field division,
extraction of spectra, and subtraction of scattered light from adjacent
fibers, and wavelength and flux calibration. These steps are
carried out with the {\tt Panacea}\footnote{\tt
\url{https://github.com/grzeimann/Panacea}} software package written by
G. Zeimann. At the end of these calibrations the effects of continuous
telluric absorption are rectified using standard tables and the
discrete O$_2$ and H$_2$O telluric absorption bands are corrected with
the help of templates constructed from bright standard stars observed
on the same night as the targets. In the end, we re-normalize the flux
scale of the spectra from different arms so that the flux density in
overlapping regions matches.
\subsection{Spectral Decomposition}\label{sec:decompose}
We summarize the optical data used in this work in Table \ref{tab:Optical}, and the brightest and the faintest optical spectra are noted with ``Bright" and ``Faint", respectively. We show all the optical spectra used in this study in Figure \ref{fig:optical_ori_spec_1}. 

To study the properties of the quasars while avoiding starlight contamination from their host galaxies, we use a Markov Chain Monte-Carlo (MCMC) method to decompose every optical spectrum into a host galaxy spectrum and a quasar spectrum. We assume that these two spectra are linear combinations of galaxy/quasar eigenspectra. The galaxy and quasar eigenspectra are derived from principal component analysis based on 170,000 SDSS galaxy and 16,707 quasar spectra, respectively \citep{Yip2004AJ2,Yip2004AJ}. We test selections of different numbers of galaxy and quasar eigenspectra used in the spectral decomposition, and find that 7 galaxy eigenspectra and 18 quasar eigenspectra are sufficient to represent all the necessary features in our spectra. The second quasar eigenspectrum is not used in our decomposition, since it primarily represents a host-galaxy component instead of a quasar spectrum \citep{Yip2004AJ}. We fit for the normalization factors of the first 7 galaxy eigenspectra \citep{Yip2004AJ2}, and 18 quasar eigenspectra \citep{Yip2004AJ}. 

We perform Galactic extinction correction for all spectra before the decomposition \citep{Cardelli1989ApJ,Schlegel1998ApJ}. Due to the limited wavelength range of the eigenspectra, we only decompose spectra within the rest-frame wavelength range 3,450$-$6,800${\rm \;\!\AA}$. 

To perform our fit, we use {\tt emcee v2.2.1} to run the MCMC \citep{FM2013PASP}, with a flat prior, and a logarithmic likelihood function: 
\begin{equation}
{\rm ln}\;\! p=-\frac{1}{2}\;\!\sum_{n}\;\![\frac{(y_n-{\scriptstyle \sum\limits_{i}}\;m_i x_{i,n})^2}{s_n^2}+{\rm ln}\;\!(2\pi s_n^2)],
\end{equation} \label{eq:likelihood_func}
where $p$ is the likelihood, $m_i$ is the fitted amplitude parameter for the $i$\;\!th eigenspectrum, $x_{i,n}$ is the flux of $i$th eigenspectrum at the $n$\;\!th pixel in the spectrum, $y_n$ is the flux of the observed spectrum at the $n$\;\!th pixel, and $s_n^2$ is the variance of the observed spectrum at the $n$\;\!th pixel.
 
For the selected optical spectra of the same object (see Table \ref{tab:Optical}), we decompose the two spectra together by assuming that their host galaxy spectra are the same shape. If more than two optical spectra are used in our analysis, for other optical spectra of that object, we subtract the best-fit host galaxy spectrum to derive their quasar spectrum. We present the decomposed spectra of SDSS J000904.54-103428.6 in Figure \ref{fig:optical_data_decompose_1}. We present decomposed spectra for our remaining targets in Appendix \ref{Appendix:optical}.

For every successful decomposition (i.\,e., the Markov Chain has converged), we subtract the derived host galaxy spectrum from the extinction-corrected, observed spectrum yielding the quasar spectrum used in the following analysis. 

\begin{figure*}[pt]
    \centering
    \includegraphics[width=0.9\textwidth]{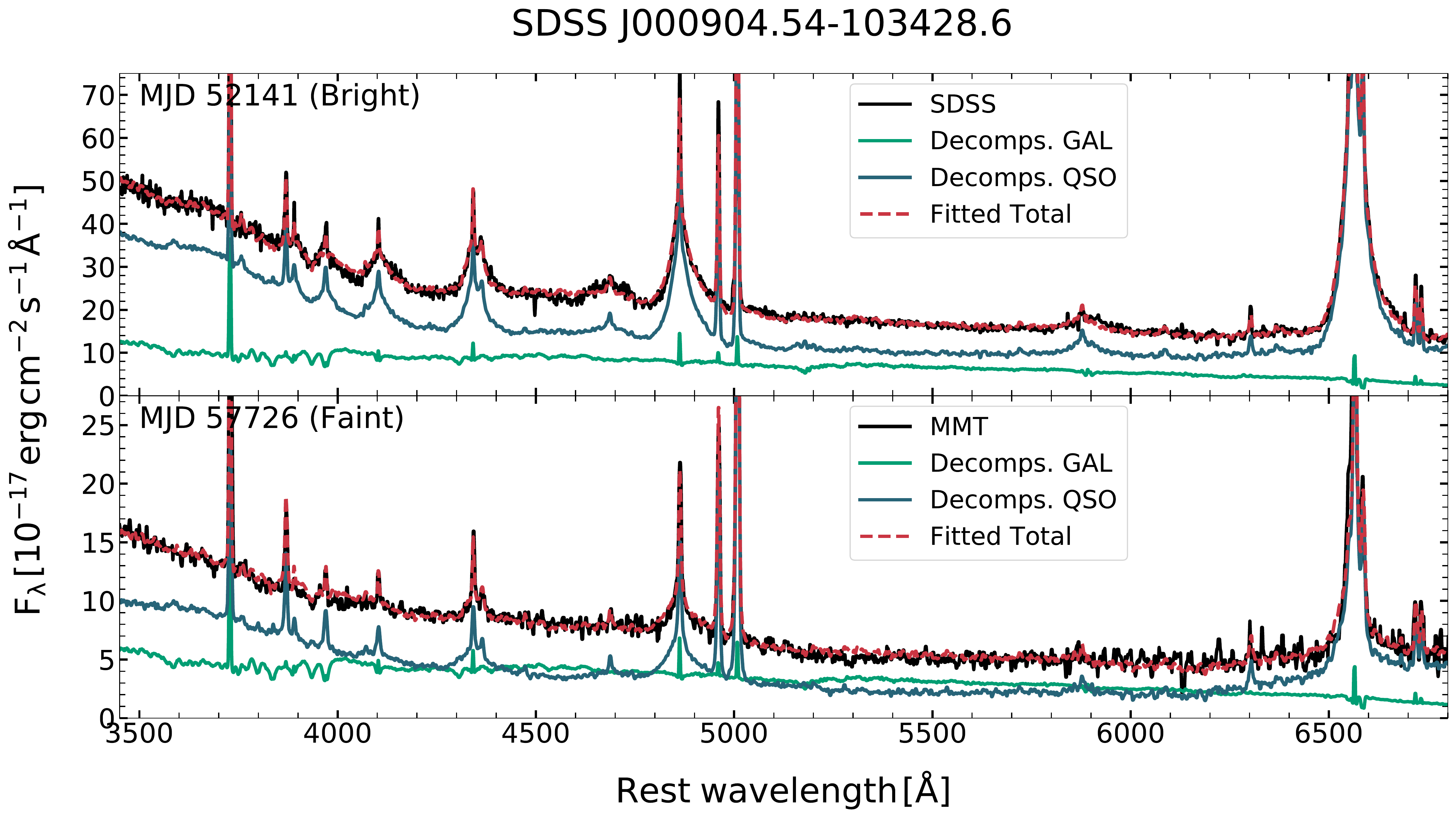}
    \caption{The decomposed spectra of SDSS J000904.54-103428.6. In each panel, we show the extinction-corrected observed spectrum in black. Noise is shown in grey. The decomposed host galaxy and the decomposed quasar spectrum are shown in green and blue, respectively. The red dashed line denotes the sum of the host galaxy and the quasar spectrum.}
\end{figure*} \label{fig:optical_data_decompose_1}

\begin{figure*}[!ptb]
    \centering
    \includegraphics[width=0.9\textwidth]{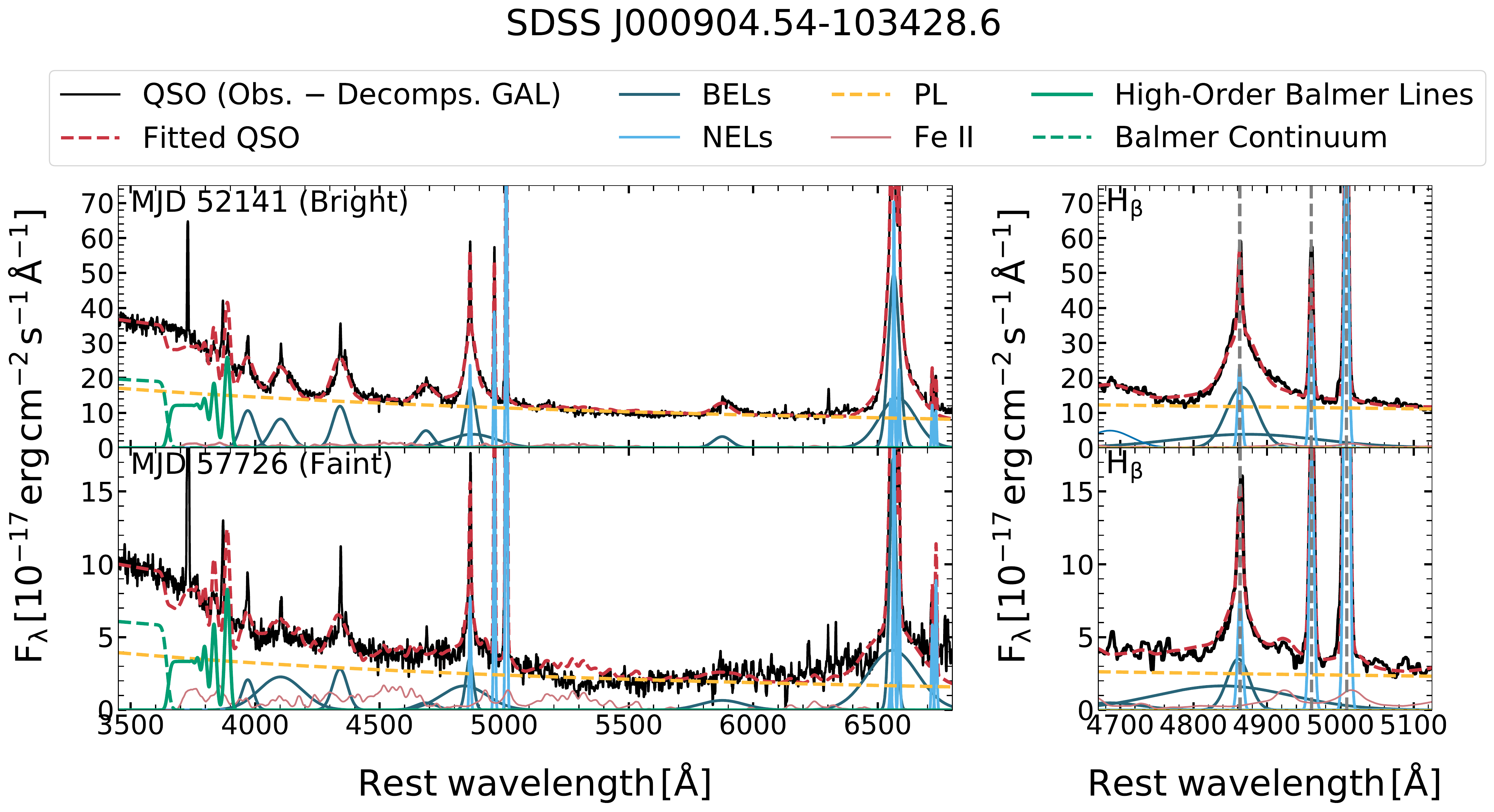}
    \caption{The spectral fitting of SDSS J000904.54-103428.6. The quasar spectrum is in black. The best-fit broad and narrow emission lines are shown in dark and light blue, respectively. The best-fit power-law continuum and blended iron lines are denoted by a yellow dashed line and a pink line. High-order Balmer lines and the Balmer continuum are in solid green and dashed green lines. The red dashed line represents the sum of the fitted models. {\it Left Panels} - Spectral fitting in 3,450$-$6,800${\rm \AA}$; {\it Right Panels} - Spectral fitting around ${\rm H\beta}$. Wavelength of fitted narrow ${\rm H\beta}$, [\ion{O}{3}]~$\lambda\lambda$4959,5007 emission lines are denoted by vertical dashed lines.}
\end{figure*} \label{fig:optical_data_qso_fit_1}

\subsection{Quasar Spectrum Fitting}\label{sec:qsofitting}
A quasar optical/UV spectrum can be represented by a power-law continuum and various emission lines \citep{VB2001AJ}. To investigate the properties of quasars, we need to measure the characteristic power-law index and the normalisation of the continuum, and full widths at half-maximum (FWHMs) of the broad emission lines. For the extracted quasar spectra, we use the MCMC ({\tt emcee} v2.2.1 python package) method again \citep{FM2013PASP}, applying a flat prior, and a likelihood function:
\begin{equation}
{\rm ln}\;\! p=-\frac{1}{2}\;\!\sum_{n}\;\![\frac{(y_n-{\scriptstyle \sum\limits_{i}}\;f_{i,n})^2}{s_n^2}+{\rm ln}\;\!(2\pi s_n^2)], 
\end{equation} \label{eq:likelihood_func}
where $p$ is the likelihood, $f_{i,n}$ is the flux of $i$th component at the $n$\;\!th pixel in the spectrum, $y_n$ is the flux of the quasar spectrum at the $n$\;\!th pixel, and $s_n^2$ is the variance of the observed spectrum at the $n$\;\!th pixel.


We fit the quasar spectra with the following parameters:
\begin{enumerate}
    \item Gaussian broad Balmer lines: Heights, widths of ${\rm H}{\alpha}$-${\rm H}{\varepsilon}$, and line centers of ${\rm H}{\alpha}$-${\rm H}{\beta}$. For most spectra in our analysis, their broad ${\rm H}{\alpha}$ and ${\rm H}{\beta}$ can be fitted with one Gaussian function. However, there are some spectra showing more complex features in their broad ${\rm H}{\alpha}$ and ${\rm H}{\beta}$. The broad ${\rm H}{\alpha}$ and ${\rm H}{\beta}$ lines of those quasars are fitted with two Gaussian functions, with heights, widths, and line centers as free parameters.
    \item Gaussian narrow emission lines: Heights and widths of two Balmer lines ${\rm H}{\alpha}$ and ${\rm H}{\beta}$ and forbidden lines [\ion{O}{3}]~$\lambda\lambda$4959, 5007, [\ion{N}{2}]~$\lambda\lambda$6548, 6584 and [\ion{S}{2}]~$\lambda\lambda$6717, 6731. We allow there to have a small wavelength offset ($-0.001<{\delta\;\! \lambda}/\lambda<0.001$) for all narrow emission lines due to uncertainties in spectroscopic redshifts or small wavelength shifts caused by imprecise wavelength calibration. We assume all the narrow lines have the same width ($1\;\!{\rm \AA}<\sigma<11\;\!{\rm \AA}$). 
    \item A Balmer continuum \citep{Grandi1982ApJ,Wills1985ApJ}: The shape of the Balmer continuum depends sensitively on the optical depth but not on the electron temperature $T_e$. Therefore, 
    we assume $T_e=15,000~{\rm K}$, and then calculate the Balmer continuum of optical depth from $0.1-2.0$ with $0.1$ spacing. The strength of the Balmer continuum is a parameter in our fit. 
    \item High-order Balmer lines: We adopt a high-order Balmer line model which includes energy levels from n=8 to n=50 in \citealt{SH1995MNRAS} (Case B, electron temperature $T_e=15,000~{\rm K}$, and electron density $n_e=10^{11}~{\rm cm^{-3}}$), and calculate a series of templates based on ${\rm H{\beta}}$ velocity from 1,000~${\rm km\;s^{-1}}$ to 11,000~${\rm km\;s^{-1}}$. We fit the normalisation of the high-order Balmer line template.
    \item Blended iron lines \citep{BG1992ApJS}: The contribution from \ion{Fe}{2} is also important for most quasars. We adopt the iron line template from \citet{BG1992ApJS} and fit its normalization. The width of the iron line template is associated with the FWHM of H$\beta$. During our analysis, we run a few test MCMCs to derive the FWHM of H$\beta$, and use the matched iron line template in the final MCMC (see below).
    \item A power-law continuum: a power-law index and a normalization factor are fitted for the power-law continuum.
    \item Some quasars have non-negligible broad helium emission lines, such as \ion{He}{2}~$\lambda$4686 or/and \ion{He}{1}~$\lambda$5876. In this case, we also fit \ion{He}{2}~$\lambda$4686 or/and \ion{He}{1}~$\lambda$5876 as broad Gaussian emission lines, and fit their strengths and widths with fixed line centers.
\end{enumerate}

\begin{deluxetable*}{ccccccccc}\centering 
\tabletypesize{\scriptsize}
\tablecaption{Best-fit Results of Optical Changing-look Quasar Spectra}
\tablewidth{2\columnwidth}
\colnumbers
\tablehead{
\colhead{Target Name} & \colhead{Obs. Date} & \colhead{PL Index} & \colhead{PL 
Norm.} & \colhead{$\lambda\;\!L{\rm {\scriptstyle2500\AA}}$} &\colhead{$\lambda\;\!L{\rm {\scriptstyle5100\AA}}$} & \colhead{FWHM (H${\beta}$)} &\colhead{${\rm M_ {BH}}$} &\colhead{Ref. ${\rm M_ {BH}}$} \\
\colhead{(SDSS)} & \colhead{(MJD)} & \colhead{} & \colhead{$(10^{-15}\;{\rm erg\;\!s^{-1}\;\!cm^{-2}\;\!\AA^{-1}})$} &\colhead{$(10^{44}\;{\rm erg\;\!s^{-1}})$}& \colhead{$(10^{44}\;{\rm erg\;\!s^{-1}})$} &\colhead{($1000\;{\rm km\;\!s^{-1}}$)} & \colhead{($10^8\;{\rm M_{\odot}}$)} & \colhead{($10^8\;{\rm M_{\odot}}$)}}
\startdata
J000904.54-103428.6&52141&$-1.08^{+0.22}_{-0.20}$&$1200^{+5200}_{-1000}$&$1.07^{+0.19}_{-0.18}$&$1.01^{+0.02}_{-0.02}$&$3.5^{+0.1}_{-0.1}$&$1.1^{+0.1}_{-0.1}$& $1.0^{+0.1}_{-0.1}$\\
 &57726&$-1.34^{+0.27}_{-0.18}$&$2100^{+7900}_{-1900}$&$0.27^{+0.04}_{-0.06}$&$0.21^{+0.01}_{-0.01}$& & & \\
J002311.06+003517.5&51900&$-0.90^{+0.02}_{-0.03}$&$140^{+30}_{-30}$&$1.98^{+0.04}_{-0.03}$&$2.13^{+0.01}_{-0.01}$&$13.6^{+0.4}_{-0.5}$&$25.0^{+2.5}_{-2.5}$& $17.0^{+2.5}_{-2.2}$\\
 &55480&$-1.02^{+0.16}_{-0.13}$&$600^{+1400}_{-500}$&$3.45^{+0.40}_{-0.42}$&$3.39^{+0.05}_{-0.06}$& & & \\
 &57597&$-0.96^{+0.05}_{-0.03}$&$80^{+20}_{-30}$&$0.75^{+0.02}_{-0.02}$&$0.78^{+0.01}_{-0.01}$& & & \\
 J022556.08+003026.7&52200&$-1.05^{+0.03}_{-0.01}$&$100^{+10}_{-20}$&$0.68^{+0.01}_{-0.01}$&$0.66^{+0.01}_{-0.01}$& & & \\
 &52944&$-0.99^{+0.01}_{-0.01}$&$110^{+10}_{-10}$&$1.17^{+0.01}_{-0.01}$&$1.18^{+0.01}_{-0.01}$&$8.1^{+2.9}_{-3.4}$&$6.5^{+4.7}_{-5.5}$& $2.2^{+5.5}_{-1.6}$ \\
 &55208&$-1.10^{+0.10}_{-0.10}$&$50^{+60}_{-30}$&$0.21^{+0.01}_{-0.01}$&$0.19^{+0.01}_{-0.01}$& & & \\
  &58814&$-1.08^{+0.02}_{-0.02}$&$110^{+20}_{-10}$&$0.57^{+0.02}_{-0.02}$&$0.53^{+0.01}_{-0.01}$& & & \\
J132457.29+480241.2&52759&$-0.68^{+0.02}_{-0.04}$&$50^{+20}_{-10}$&$1.36^{+0.04}_{-0.02}$&$1.71^{+0.01}_{-0.01}$&$5.4^{+0.1}_{-0.1}$&$3.5^{+0.3}_{-0.3}$& $3.2^{+0.5}_{-0.4}$\\
 &58127&$-0.87^{+0.05}_{-0.05}$&$60^{+30}_{-20}$&$0.39^{+0.01}_{-0.01}$&$0.43^{+0.01}_{-0.01}$& & & \\
J160111.25+474509.6&52354&$-0.96^{+0.05}_{-0.04}$&$210^{+90}_{-70}$&$0.84^{+0.02}_{-0.03}$&$0.87^{+0.01}_{-0.01}$&$7.5^{+0.3}_{-0.2}$&$4.8^{+0.4}_{-0.4}$& $3.9^{+3.0}_{-1.7}$ \\
 &57895&$-1.16^{+0.08}_{-0.07}$&$210^{+160}_{-100}$&$0.17^{+0.01}_{-0.01}$&$0.15^{+0.01}_{-0.01}$ & & & \\
J164920.79+630431.3&51699&$-0.90^{+0.02}_{-0.02}$&$90^{+10}_{-20}$&$0.69^{+0.01}_{-0.01}$&$0.74^{+0.01}_{-0.01}$&$8.8^{+0.5}_{-0.8}$&$6.0^{+0.7}_{-1.2}$& $6.0^{+0.7}_{-0.7}$\\
 &58276&$-1.16^{+0.03}_{-0.08}$&$140^{+140}_{-30}$&$0.14^{+0.01}_{-0.01}$&$0.13^{+0.01}_{-0.01}$& & & \\
J214613.30+000930.8&52968&$-1.14^{+0.01}_{-0.01}$&$110^{+10}_{-10}$&$0.57^{+0.02}_{-0.02}$&$0.52^{+0.01}_{-0.02}$& & & $8.7^{+5.1}_{-3.2}$\\
 &55478&$-1.06^{+0.01}_{-0.02}$&$110^{+20}_{-10}$&$1.17^{+0.01}_{-0.01}$&$1.12^{+0.01}_{-0.01}$&$11.8^{+0.7}_{-0.9}$&$13.5^{+1.7}_{-2.1}$& \\
 &57663&$-1.18^{+0.01}_{-0.01}$&$110^{+10}_{-10}$&$0.44^{+0.03}_{-0.03}$&$0.39^{+0.02}_{-0.03}$& & & \\
J220537.71-071114.5&52468&$-1.07^{+0.13}_{-0.16}$&$800^{+2500}_{-500}$&$1.35^{+0.20}_{-0.14}$&$1.28^{+0.03}_{-0.02}$&$12.6^{+0.7}_{-0.7}$&$16.5^{+2.2}_{-2.1}$& $9.8^{+2.0}_{-1.6}$ \\
 &57989&$-1.23^{+0.10}_{-0.12}$&$400^{+600}_{-200}$&$0.17^{+0.02}_{-0.01}$&$0.14^{+0.01}_{-0.01}$& & & \\
J225240.37+010958.7&52178&$-1.07^{+0.02}_{-0.02}$&$110^{+10}_{-20}$&$0.71^{+0.02}_{-0.02}$&$0.67^{+0.02}_{-0.02}$& & & $7.6^{+3.6}_{-2.5}$\\
 &55500&$-1.01^{+0.04}_{-0.05}$&$160^{+90}_{-50}$&$1.77^{+0.07}_{-0.05}$&$1.76^{+0.01}_{-0.01}$&$6.0^{+0.3}_{-0.3}$&$4.4^{+0.5}_{-0.5}$& \\
 &57598&$-1.10^{+0.10}_{-0.02}$&$90^{+20}_{-50}$&$0.47^{+0.01}_{-0.03}$&$0.44^{+0.01}_{-0.01}$& & & \\
 &58814&$-1.17^{+0.02}_{-0.02}$&$110^{+20}_{-10}$&$0.33^{+0.02}_{-0.02}$&$0.29^{+0.02}_{-0.02}$& & & \\
J233317.38-002303.5 &52199&$-1.08^{+0.07}_{-0.04}$&$80^{+30}_{-30}$&$0.44^{+0.01}_{-0.01}$&$0.42^{+0.01}_{-0.01}$& & & \\
& 52525 & - & - & - & - & - & - & $141.3^{+115.8}_{-63.6}$\\
 &55447 &$-1.25^{+0.17}_{-0.19}$&$1000^{+4200}_{-800}$&$1.45^{+0.24}_{-0.19}$&$1.21^{+0.02}_{-0.02}$&$5.8^{+0.5}_{-0.3}$&$3.4^{+0.6}_{-0.5}$& \\
 &58429&$-1.24^{+0.06}_{-0.10}$&$160^{+210}_{-60}$&$0.27^{+0.03}_{-0.03}$&$0.23^{+0.02}_{-0.02}$ & & & \\
\enddata
\tablecomments{(1) Name of CLQs in SDSS; (2) Date of optical observations in MJD; (3)-(4) The best-fit power-law index and normalization of the power-law continuum; (5) The monochromatic luminosity of the power-law continuum at 2500\AA; (6) The monochromatic luminosity of the power-law continuum at 5100\AA; (7) The full width at half maximum (FWHM) of broad H$\beta$; (8) The measured black hole mass; (9) The referenced black hole mass is adopted from the fiducial black hole mass in \citet{Shen2011ApJS}. All the errors represent 1$\sigma$ confidence intervals.}
\end{deluxetable*}\label{tab:fitteddata}

Since we do not correct telluric absorption for ARC 3.5m spectra and MMT spectra, we mask some wavelength ranges in those spectra when performing the quasar spectral fitting, to avoid biases from telluric absorption. The masked wavelength ranges are shown as pink shaded regions in the figures of Appendix \ref{Appendix:optical}.

We first run 20 separate test MCMCs for different optical depth Balmer continua from $0.1-2.0$ by assuming the high-order Balmer line template and the iron line template associated with ${\rm H{\beta}}$ have a FWHM of 4,500~${\rm km\;s^{-1}}$. We find the lowest $\chi^2$ among those 20 tests, and use the corresponding optical depth Balmer continuum in the final MCMC, and calculate the fitted ${\rm H{\beta}}$ FWHM. Then we find the high-order Balmer line and the iron line templates that most closely match that FWHM, and apply those high-order Balmer line and iron line templates in our final MCMC fit. 

Since the high-order Balmer line model only includes energy levels to n=50, there can be some discontinuity at $\sim$~3,600$\;\!\rm \AA$ between the Balmer continuum and high-order Balmer lines in our fitted spectra. A high-order Balmer line model with more energy levels can solve this issue \citep{Kova2014AdSpR,Kova2015ApJS}. However, for the purposes of our work, a high-order Balmer line model with energy levels to n=50 is adequate. 

We obtain the best-fit power-law continuum luminosity at 2,500~${\rm \AA}$ and 5,100~${\rm \AA}$ ($\lambda\;\!L{\rm{\scriptstyle 2500\AA}}$ and $\lambda\;\!L{\rm {\scriptstyle 5100\AA}}$), and their $1\sigma$ errors from the sample drawn from the MCMC. Based on the best-fit parameters and their $1\sigma$ errors, we use Monte-Carlo resampling to generate 5000 profiles to calculate the uncertainties in the FWHM of broad H$\beta$ which is fitted with two Gaussian functions. For broad H$\beta$ which can be fitted with a single Gaussian function, we directly use the best-fit width of broad H$\beta$ to calculate its FWHM. 

For each object, we choose a spectrum with a prominent broad H$\beta$ emission line to measure its black hole mass, since broad H$\beta$ is the best calibrated broad emission line from reverberation mapping \citep{Peterson2004ApJ}. 
We then measure the black hole mass based on the calculated FWHM of broad H$\beta$ and $\lambda\;\!L{\rm {\scriptstyle 5100\AA}}$, using the following equation \citep{Greene2010ApJ}: 
\begin{equation}
    \begin{split}
    {\rm M_{BH,H\beta}=(9.1\pm0.5)\times10^{6}\;\!\left(\frac{FWHM~(H\beta)}{1000\;km\;s^{-1}}\right)^{2}}\\
    \times\left(\frac{\lambda\;\! L {\rm {\scriptstyle5100\AA}}}{10^{44}\;{\rm erg\;s^{-1}}}\right)^{0.519\pm0.07}\;{\rm M_{\odot}}.
   \end{split}
\end{equation}
The errors of the black hole mass are calculated using propagation of errors in FWHM of H$\beta$ and $\lambda\;\!L{\rm {\scriptstyle 5100\AA}}$.

We present our best-fit power-law indices, best-fit power-law normalizations, $\lambda\;\!\!L{\rm {\scriptstyle2500\AA}}$, $\lambda \;\!\!L{\rm {\scriptstyle5100\AA}}$, FWHM of ${\rm H{\beta}}$, and black hole mass ${\rm M_{BH}}$ in Table \ref{tab:fitteddata}. We also provide the fiducial black hole mass measurement from \citet{Shen2011ApJS} for comparison; the latter incorporates black hole mass prescriptions and measurements from previous works \citep{McLure2004MNRAS,VP2006ApJ,VO2009ApJ,Shen2011ApJS}. 
Our measured results are consistent with their measurements within 1$\sigma$ errors or within 0.4~dex, the intrinsic scatter in mass scaling relationship of broad emission lines \citep{VP2006ApJ}, except for SDSS J233317.38-002303.5. For this object, we adopt a different SDSS spectrum than \citet{Shen2011ApJS} to measure its black hole mass, in which the broad H$\beta$ emission is more prominent, and thus we adopt our measured black hole mass as the black hole mass of SDSS J233317.38-002303.5 in the following analysis.

In Figure \ref{fig:optical_data_qso_fit_1}, we show our spectral fitting of SDSS J000904.54-103428.6. Spectral fittings of other objects are included in Appendix \ref{Appendix:optical}.

\section{X-ray data}\label{sec:Xraydata}
For our X-ray analysis of CLQs, we search the \textit{XMM-Newton} data archive, and obtain new \textit{Chandra} Cycle 19 observations. The new \textit{Chandra} and archival \textit{XMM-Newton} observations are close in time to our new optical spectra, and thus it is likely that they reflect the X-ray properties of the same states as the optical spectra. Further discussion about the uncertainties on $\alpha_{\rm OX}$ caused by the time difference between optical and X-ray observations can be found in \S\ref{sec:aoxuncer}. 

We also used X-ray data from the \textit{ROSAT} data archive to constrain the X-ray flux of changing-look quasars close in time to their first optical observations from SDSS.
The \textit{ROSAT} observations were obtained approximately 10 years before the first SDSS observations; we use them only to loosely constrain the X-ray flux associated with the first SDSS observation of each CLQ. Of course, these CLQs could have changed their X-ray fluxes within these 10 years. Therefore, we only tentatively connect \text{ROSAT} X-ray fluxes with the early optical data, and present our results in Appendix \ref{Appendix:ROSAT}.

\begin{figure*}[!htb]\centering
\includegraphics[width=0.45\textwidth]{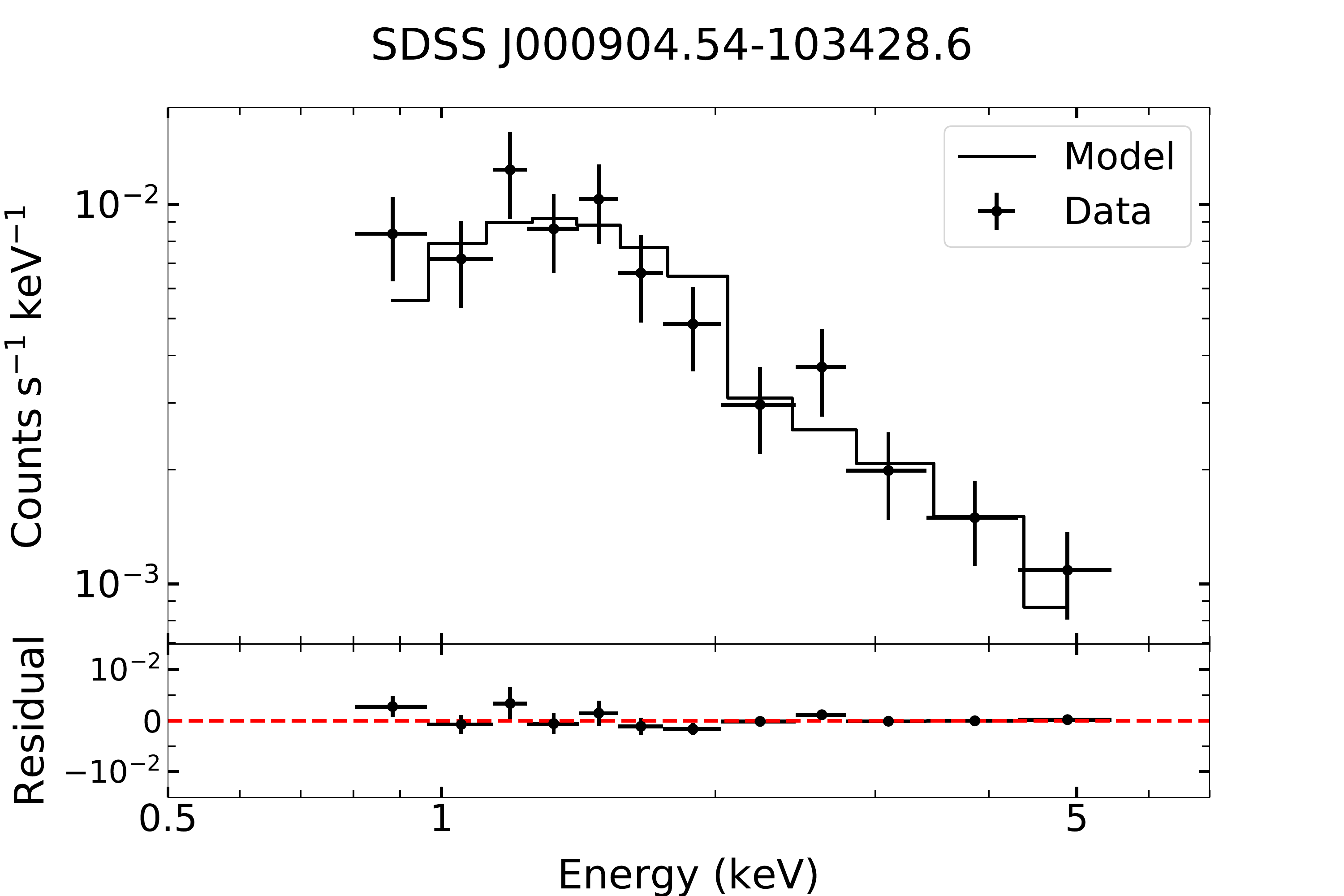}
\includegraphics[width=0.45\textwidth]{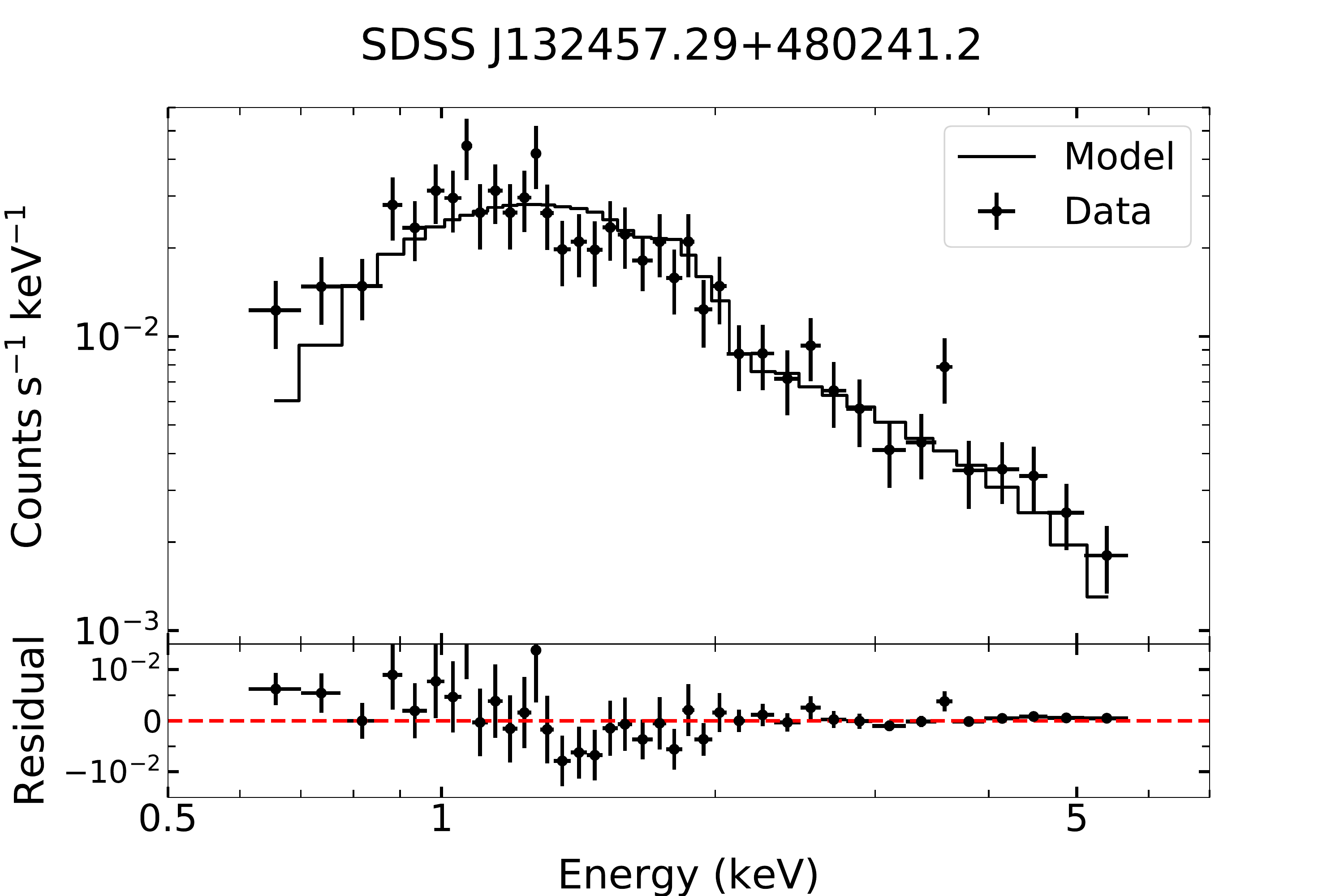}\\
\vspace{0.4cm}
\includegraphics[width=0.45\textwidth]{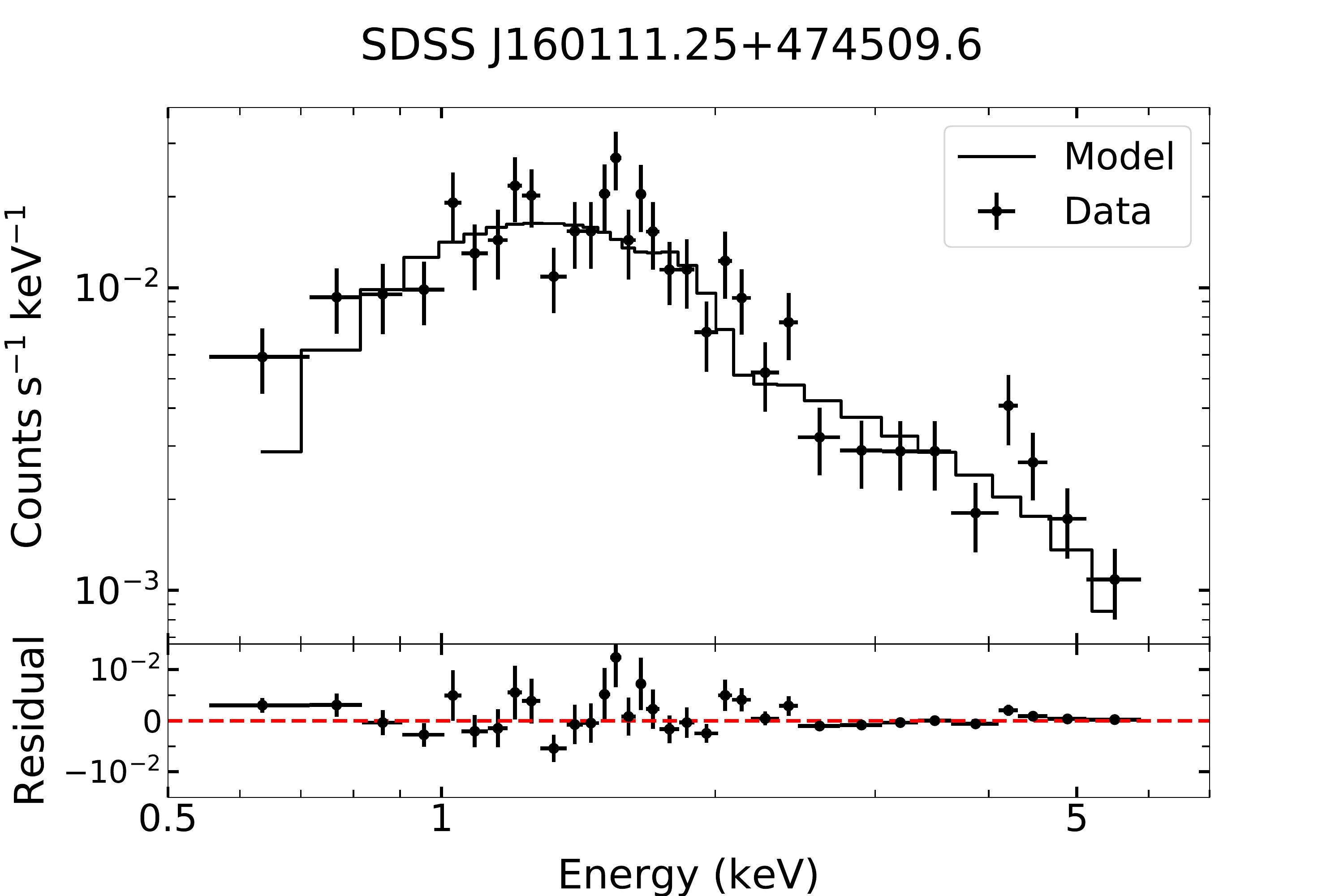}
\includegraphics[width=0.45\textwidth]{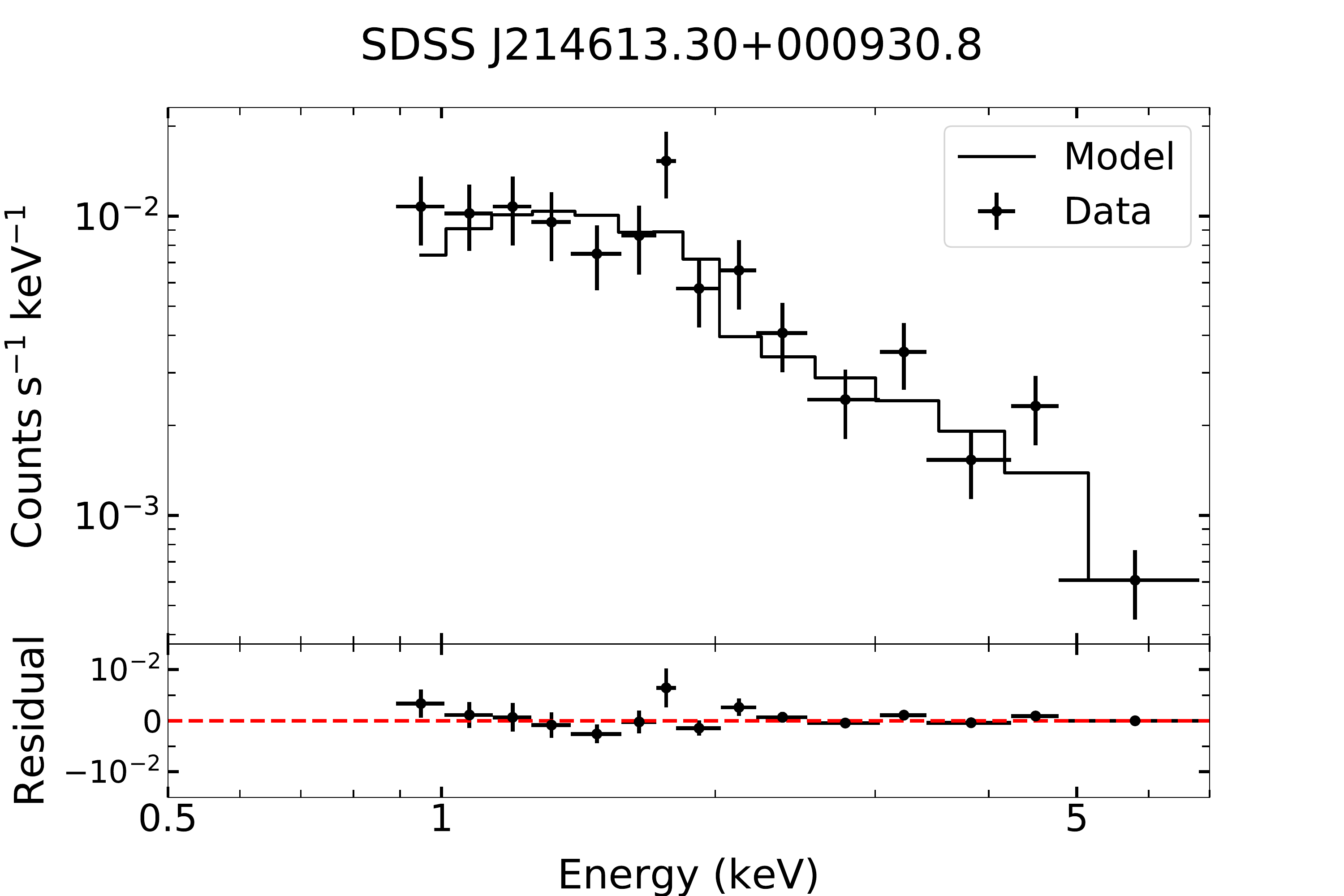}\\
\caption{{\it Chandra} spectra of four changing-look quasars. Every upper panel shows the best-fit model and the observed X-ray spectrum, and each lower panel displays the residuals.}
\end{figure*}\label{fig:chandraxspec}

\begin{figure*}[!htb]\centering
\includegraphics[width=0.45\textwidth]{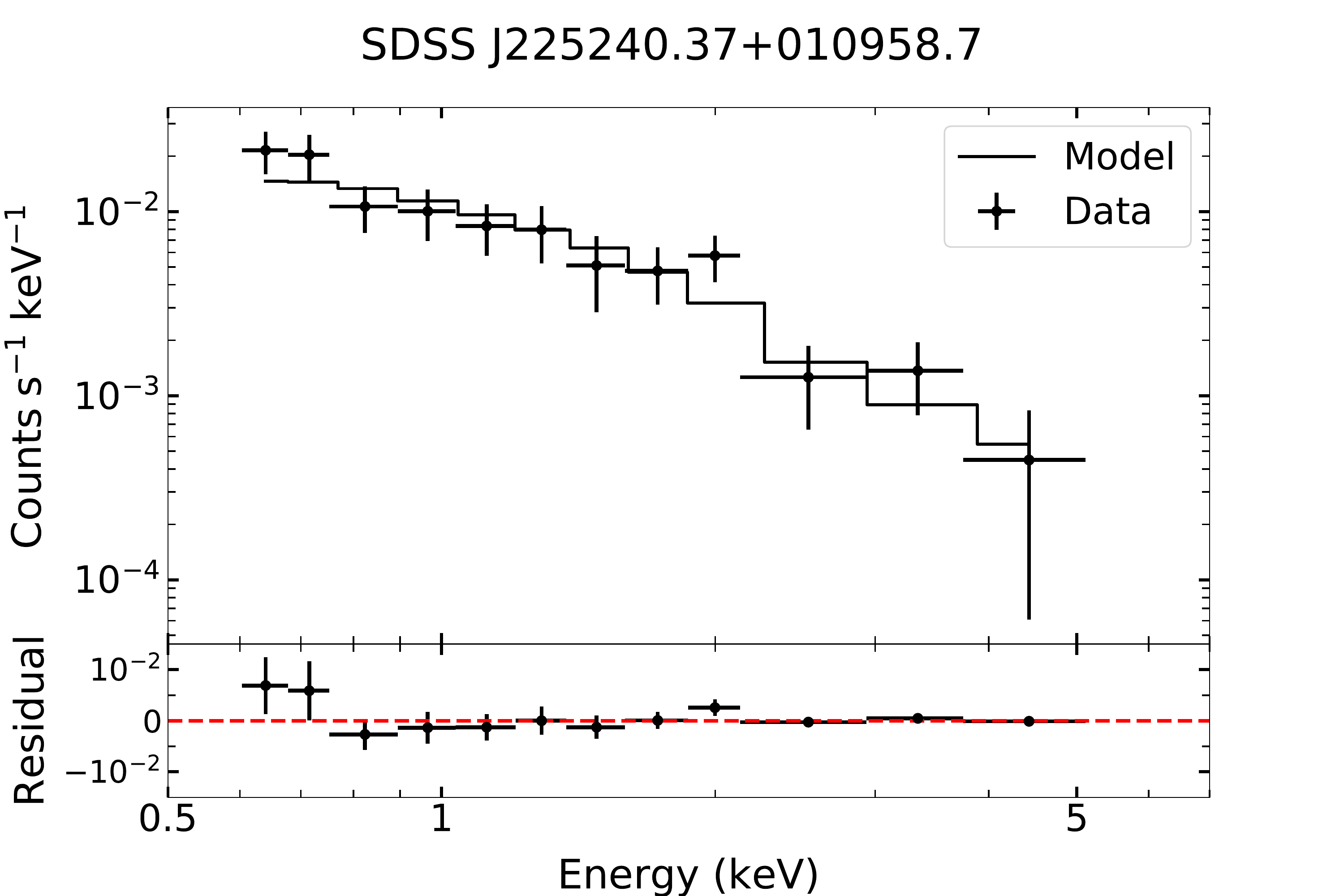}
\includegraphics[width=0.45\textwidth]{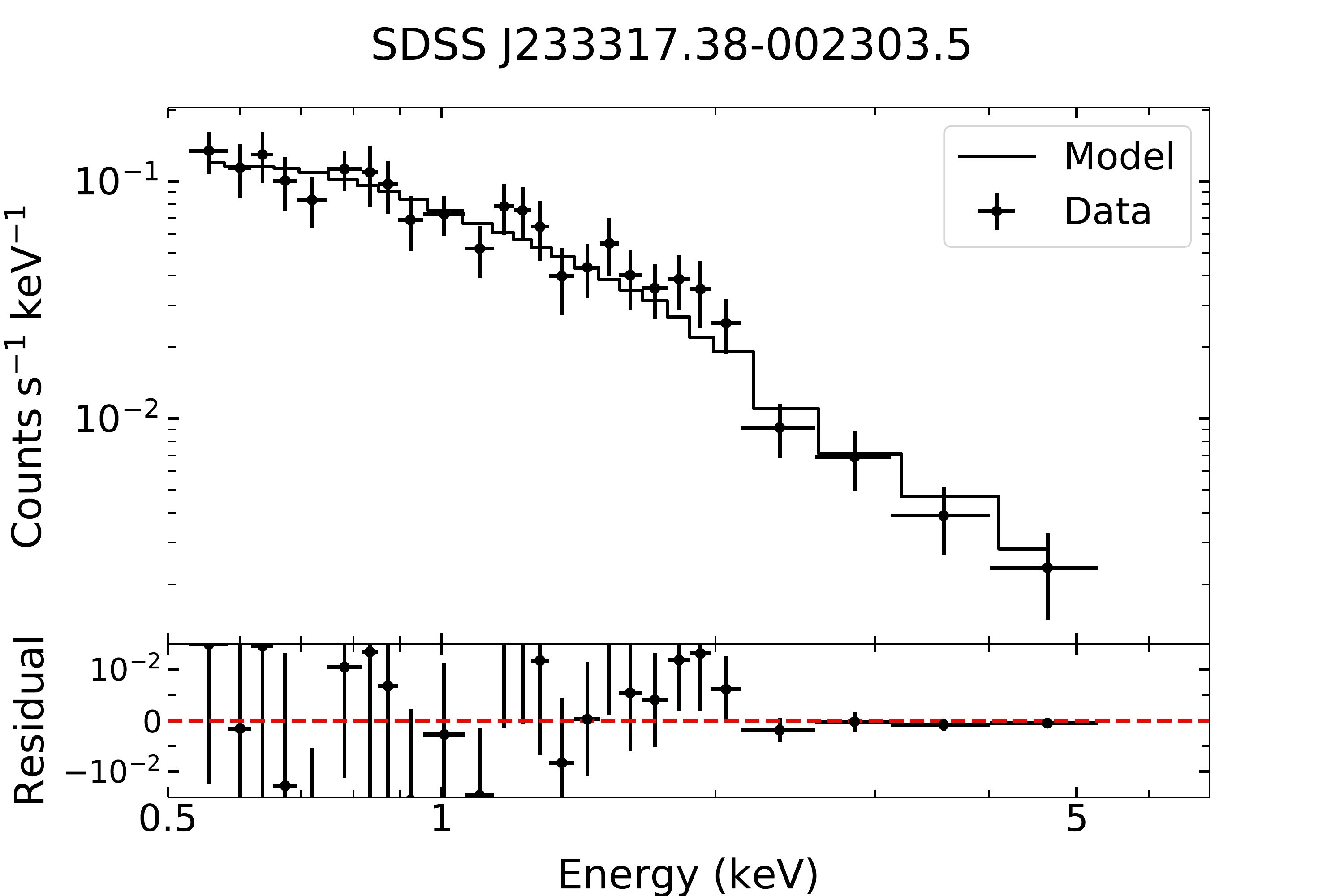}
\caption{The {\it XMM-Newton} spectra of SDSS J225240.37+010958.7 and SDSS J233317.38-002303.5. The observed X-ray spectrum and the best-fit model are in every upper panel, and every lower panel displays the residuals.}
\end{figure*}\label{fig:xmmxspec}

\begin{deluxetable*}{cccccCc}\centering 
\tabletypesize{\scriptsize}
\tablecolumns{7}
\tablecaption{X-ray Properties of Changing-look Quasars}
\tablewidth{0pt}
\tablehead{
\colhead{Target Name} &\colhead{Instrument} & \colhead{Observation Date} & \colhead{Exp.} & \colhead{Count Rate} &\colhead{Photon Index} & \colhead{$\nu L{\rm \scriptstyle 2\;\!keV}$}\\
\colhead{(SDSS)} & \colhead{} & \colhead{(MJD)} & \colhead{(ks)} &   \colhead{$(10^{-2}\;{\rm cts\;s^{-1}})$} &  \colhead{} & \colhead{$(10^{42}\;{\rm erg\;s^{-1}})$}
}
\colnumbers
\startdata
J000904.54-103428.6 
& Chandra & 58363 & 11.9 & $1.7\pm0.1$ & 1.7\pm0.2^a & $15_{-1}^{+2}$ \\
J002311.06+003517.5 
&  Chandra & 58254 & 4.0 & $3.7\pm0.3$ & 1.8 & $140\pm12$ \\
J022556.08+003026.7 
& Chandra & 58182 & 7.0 & $0.6\pm0.1$ & 1.8 & $39\pm6$\\
J132457.29+480241.2 
& Chandra & 58385 & 13.9 & $4.8\pm0.2$ & $1.9\pm0.1^{a}$ & $59_{-2}^{+3}$ \\
J160111.25+474509.6 
& Chandra & 58359 & 17.8 & $3.0\pm0.1$ & 1.8\pm0.1^{a} & $44\pm2$\\
J164920.79+630431.3 
& Chandra & 58213 & 22.8 & $0.47\pm0.05$ & 1.8 & $8.8\pm0.9$\\
J214613.30+000930.8 
& Chandra & 58230 & 11.9 & $2.0\pm0.1$ & 1.6^{+0.2^{a}}_{-0.1} & $192_{-13}^{+17}$\\
J220537.71-071114.5 
& Chandra & 58023 & 18.8 & $0.65\pm0.06$ & 1.8 & $9.5\pm0.9$\\
J225240.37+010958.7 
& XMM-Newton & 57559 & 9.3 & $1.5\pm 0.2$ & {1.9_{-0.2}^{+0.3^{a}}} & $69_{-9}^{+14}$ \\
& Chandra & 58027 &  8.0 & $1.1\pm0.1$ & 1.8 & $77\pm8$\\
J233317.38-002303.5 
 & XMM-Newton & 56072 & 3.7 & $11.4\pm0.6$ & 2.1\pm0.1^{a} & $ 258^{+23}_{-14}$ \\
 & Chandra & 58256 & 7.0 & $1.3\pm0.1$ & 1.8 & $87\pm9$
\enddata
\tablenotetext{\it a}{The photon index is derived from an absorbed power-law ({\tt wabs*powerlaw}) fitting.}
\tablecomments{(1) Name of CLQs in SDSS; (2) Instrument of X-ray observations; (3) Date of X-ray observations in MJD; (4) Exposure time in kilo-seconds (ks). \textit{XMM-Newton} observations only include the PN instrument exposure time; (5) 0.5--7~keV count rate in units of $10^{-2}\;{\rm cts\;s^{-1}}$. 
(6) Best-fit photon index. $1.8$ is for the sources which do not have enough counts to perform X-ray spectral fitting; (7) Unabsorbed rest-frame 2~keV luminosity, in units of $10^{42}\;{\rm erg\;s^{-1}}$, calculated by \textit{Chandra} WebPIMMs by assuming a fixed galactic absorption and a photon index in the column (6). All the errors are 1$\sigma$, and all the upper limits are 3$\sigma$.}
\end{deluxetable*}\label{tab:xray}

\subsection{Chandra Observations}
We obtained new {\it Chandra} Cycle 19 ACIS-S observations (ObsID 20459$-$20468) to measure the X-ray properties of quasars associated with the new optical spectra (PI: Ruan; Program NO: 19700565). Each observation has an exposure time of at least 4~ks, up to $\sim$20~ks, and is in ``VFAINT" mode to ensure the best sensitivity, with the targeted quasar located at the aim point of ACIS-S3 chip. We use \textit{Chandra} Interactive Analysis of Observations (CIAO) v4.10 (CALDB~v4.7.9) for data reduction \citep{Fruscione2006SPIE}. 
We reprocess level-1 data with the calibration files, following the standard data reduction procedure. 
We check each light curve, and find there was no significant particle flare during the observation. 
We produce a 0.3--8~keV counts map of each observation and use {\tt wavdetect} to perform source detection. The targeted quasar is detected in every dataset. Due to possible physical offsets between the optical and X-ray center or limited pointing accuracy, there can be some small offsets ($\lesssim2\arcsec$) between the coordinates detected by SDSS and {\it Chandra}. For this reason, we use X-ray source coordinates detected by {\tt wavdetect} as the source centers when extracting spectra with {\tt specextract}. 
The source counts are extracted within 5{\arcsec} of each source's center, and the corresponding background counts are extracted from a 40{\arcsec}--50{\arcsec} annulus. 
To avoid point source contamination in the background region, we use a 20{\arcsec}--30{\arcsec} annulus for SDSS J132457.29+480241.2 and SDSS J160111.25+474509.6 for the extraction of background counts.  

For four sources which have sufficient 0.5--7.0~keV counts \cite[$\gtrsim150$;][]{Evans2010ApJS}, we group the X-ray counts into 15 per bin and use Xspec v12.10.1 to perform spectral fits \citep{Arnaud1996ASPC}, assuming an absorbed power-law with a fixed Galactic absorption \cite[{\tt wabs*powerlaw};][]{MM1983ApJ}. We present the best-fit results in Table \ref{tab:xray}, and we present the X-ray spectra in Figure \ref{fig:chandraxspec}, with the models superposed. We then calculate the 0.5--7.0~keV energy flux, based on the best-fit results. For the other six CLQs, their X-ray counts are not sufficient to perform X-ray spectral fitting. We instead measure their 0.5--7.0~keV count rate, and then convert the 0.5--7.0~keV count rate into the 0.5--7.0~keV energy flux by assuming a photon index of 1.8, which is the median X-ray power-law photon index from BAT AGN Spectroscopic Survey \cite[BASS;][]{Trakhtenbrot2017MNRAS}, and a fixed Galactic absorption. We calculate their rest-frame 2~keV flux density using the \textit{Chandra} WebPIMMs\footnote{\url{https://cxc.harvard.edu/toolkit/pimms.jsp}} tool. We list our measured rest-frame 2~keV luminosity ($\nu L{\rm \scriptstyle 2\;\!keV}$) in Table \ref{tab:xray}.

\subsection{Archival XMM-Newton Data}
 We search for additional archival X-ray observations
and find that SDSS J225240.37+010958.7 and SDSS J233317.38-002303.5 were previously observed with {\it XMM-Newton}. Due to the existence of nearly contemporaneous optical observations with the {\it XMM-Newton} observations, we include these two \textit{XMM-Newton} observations in our analysis. 
 To ensure the best spectral resolution, we only use the data from the PN instrument. We use the Science Analysis System (SAS) v17.0.0 to reduce the {\it XMM-Newton} PN observations with newly calibrated files, following the SAS Threads, and reprocess the observation data files (ODFs) with {\tt epproc}. We then use {\tt evselect} to filter the event lists for flaring particle background. We adopt a circular source region of 32{\arcsec} radius, and use the same CCD chip to measure the background from a region of the same size without any discrete sources. We then use {\tt evselect} to extract the corresponding spectrum. A redistribution matrix and an ancillary file are generated based on the extracted spectra by {\tt rmfgen} and {\tt arfgen}. We group the X-ray counts into 15 counts per bin and use Xspec v12.10.1 to perform X-ray spectral fitting. We fit an absorbed power-law ({\tt wabs*powerlaw}) to the extracted spectra, accounting for a fixed Galactic absorption. We present our fitted results in Table \ref{tab:xray} and the fitted spectrum in \ref{fig:xmmxspec}. The 0.5--7~keV energy flux is calculated based on the best-fit results, and then we calculate the rest-frame 2~keV flux density using the \textit{Chandra} WebPIMMs tool, and present rest-frame 2~keV luminosity ($\nu L{\rm \scriptstyle 2\;\!keV}$) in Table \ref{tab:xray}.

\section{Results and Discussion}\label{sec:resultndiscuss}
\subsection{$\alpha_{\rm OX}-\lambda_{\rm Edd}$}\label{subsec:aox_ledd}

We use the \textit{Chandra} and \textit{XMM-Newton} X-ray data together with optical measurements that are close in time
to analyze the optical/X-ray spectral shapes. 

We also leverage the earliest optical measurements with X-ray flux constraints derived from \textit{ROSAT} PSPC all-sky survey. We emphasize that due to the large difference between the time of the earliest SDSS observations and \textit{ROSAT} observations, these measurements should be regarded as crude constraints on $\alpha_{\rm OX}$. We present these results using \textit{ROSAT} observations in Appendix \ref{Appendix:ROSAT}.

We use the following definition to calculate the optical/X-ray spectral index $\alpha_{\rm OX}$ \citep{Tananbaum1979ApJ}: \begin{equation}\label{eq:aox}
\alpha_{\rm OX}=-\frac{{\rm log}\;\!(\lambda L{\rm \scriptstyle 2500\AA})-{\rm log}\;\!(\nu L{\rm \scriptstyle 2\;\!keV})}{{\rm log}\;\!\nu{\rm \scriptstyle 2500\AA}-{\rm log}\;\!\nu {\rm \scriptstyle 2\;\!keV}}+1, 
\end{equation}
where the monochromatic 2,500~${\rm \AA}$ luminosity ($\lambda L{\rm \scriptstyle 2500\AA}$) is calculated by extrapolating the best-fit power-law continuum to 2,500~${\rm \AA}$ in the rest-frame (see Table \ref{tab:fitteddata}), and the rest-frame 2~keV luminosity ($\nu L{\rm \scriptstyle 2\;\!keV}$) is calculated as in \S\ref{sec:Xraydata} (see Table \ref{tab:xray}).

We then use the 2--10~keV luminosity and $\alpha_{\rm OX}$ to obtain the bolometric luminosity, with the following bolometric correction \citep{Lusso2010A&A}:
\begin{equation}\label{eq:lbolxo}
{\rm log}\;\!L {\rm \scriptstyle bol}={\rm log}\;\!L {\rm \scriptstyle [2-10]\;\!keV}+1.561-1.853\;\!\alpha_{\rm OX}+1.226\;\!\alpha_{\rm OX}^2, 
\end{equation}
where $L {\rm \scriptstyle [2-10]\;\!keV}$ is the rest-frame 2-10 keV luminosity. $L {\rm \scriptstyle [2-10]\;\!keV}$ is calculated based on the best-fit X-ray spectrum if there are enough X-ray counts in that observation, otherwise it is calculated by assuming an X-ray photon index of 1.8. We then calculate the Eddington ratio ($\lambda_{\rm Edd}$) as the ratio of the bolometric luminosity to the Eddington luminosity, i.\,e.,\,$\lambda_{\rm Edd}=L_{\rm bol}/L_{\rm Edd}$, where $L_{\rm Edd}=1.3\times10^{38}\;\!({\rm M/M_{\odot}})~{\rm erg\;s^{-1}}$. 
We present the measured $\alpha_{\rm OX}$ values and Eddington ratios in Table \ref{tag:aoxledd_1}. 

\begin{deluxetable*}{cccCCc}\centering 
\tabletypesize{\scriptsize}
\tablecolumns{6}
\tablecaption{Measured $\alpha_{\rm OX}$ and $\lambda_{\rm Edd}$ of Changing-look Quasars}
\tablewidth{0pt}
\tablehead{
\colhead{Target Name (SDSS)} & \colhead{Optical MJD} & \colhead{X-ray MJD} & \colhead{$\alpha_{\rm OX}$} & \colhead{${\rm log}\;\!(\lambda_{\rm Edd})$} & \colhead{Label}}
\colnumbers
\startdata
J000904.54-103428.6 & 57726 & 58363 & 1.10^{+0.03}_{-0.04} & -1.65^{+0.13}_{-0.13} & A \\
 J002311.06+003517.5 & 57597 & 58254 & 0.90^{+0.01}_{-0.01} & -2.21^{+0.07}_{-0.07} & B \\
 J022556.08+003026.7 & 58814 & 58182  & 1.06^{+0.04}_{-0.04}  & -2.08^{+0.39}_{-0.34} & C  \\
 J132457.29+480241.2 & 58127 & 58385 & 0.93^{+0.01}_{-0.01} & -1.75^{+0.06}_{-0.05} & D \\
 J160111.25+474509.6 & 57895 & 58359 & 0.84^{+0.01}_{-0.01} & -2.00^{+0.06}_{-0.06} & E \\
 J164920.79+630431.3 & 58276 & 58213 & 1.08^{+0.02}_{-0.02} & -2.68^{+0.11}_{-0.09} & F \\
J214613.30+000930.8 & 57663 & 58230 & 0.76^{+0.01}_{-0.02} & -1.74^{+0.09}_{-0.09} & G \\
J220537.71-071114.5 & 57989 & 58023 & 1.10^{+0.02}_{-0.02} & -3.07^{+0.09}_{-0.10} & H \\
J225240.37+010958.7 & 57598 & 57559 & 0.94^{+0.02}_{-0.04} & -1.79^{+0.20}_{-0.14} & I1  \\
& 58814 & 58027 & 0.86^{+0.02}_{-0.02} & -1.72^{+0.09}_{-0.09} & I2 \\
J233317.38-002303.5 & 55447 & 56072 & 0.90^{+0.03}_{-0.03} &  -1.16^{+0.11}_{-0.12} & J1 \\
& 58429 & 58256 & 0.81^{+0.02}_{-0.02} & -1.57^{+0.11}_{-0.11} & J2 \\
\enddata
\tablecomments{(1) Name of CLQs in SDSS; 
(2) MJD of the optical data, see Table \ref{tab:Optical} for details; (3) MJD of the X-ray data, see Table \ref{tab:xray} for details; (4) Optical/UV-X-ray spectral indices $\alpha_{\rm OX}$ ($\lambda L{\rm \scriptstyle 2500\AA}$ can be found in Table \ref{tab:fitteddata}, and $\nu L{\rm \scriptstyle 2\;\!keV}$ can be found in Table \ref{tab:xray}); 
(5) Logarithmic Eddington ratio ${\rm log}\;\!(\lambda_{\rm Edd})$, using a bolometric luminosity calculated with optical data and X-ray data (i.e.,\,Equation \ref{eq:lbolxo}); (6) Label of green square data points presented in Figure \ref{fig:aoxledd}. All error bars denote $1\sigma$ errors.}
\vspace{-4mm}
\end{deluxetable*}\label{tag:aoxledd_1}

\begin{figure}[!b]\centering
\includegraphics[width=0.45\textwidth]{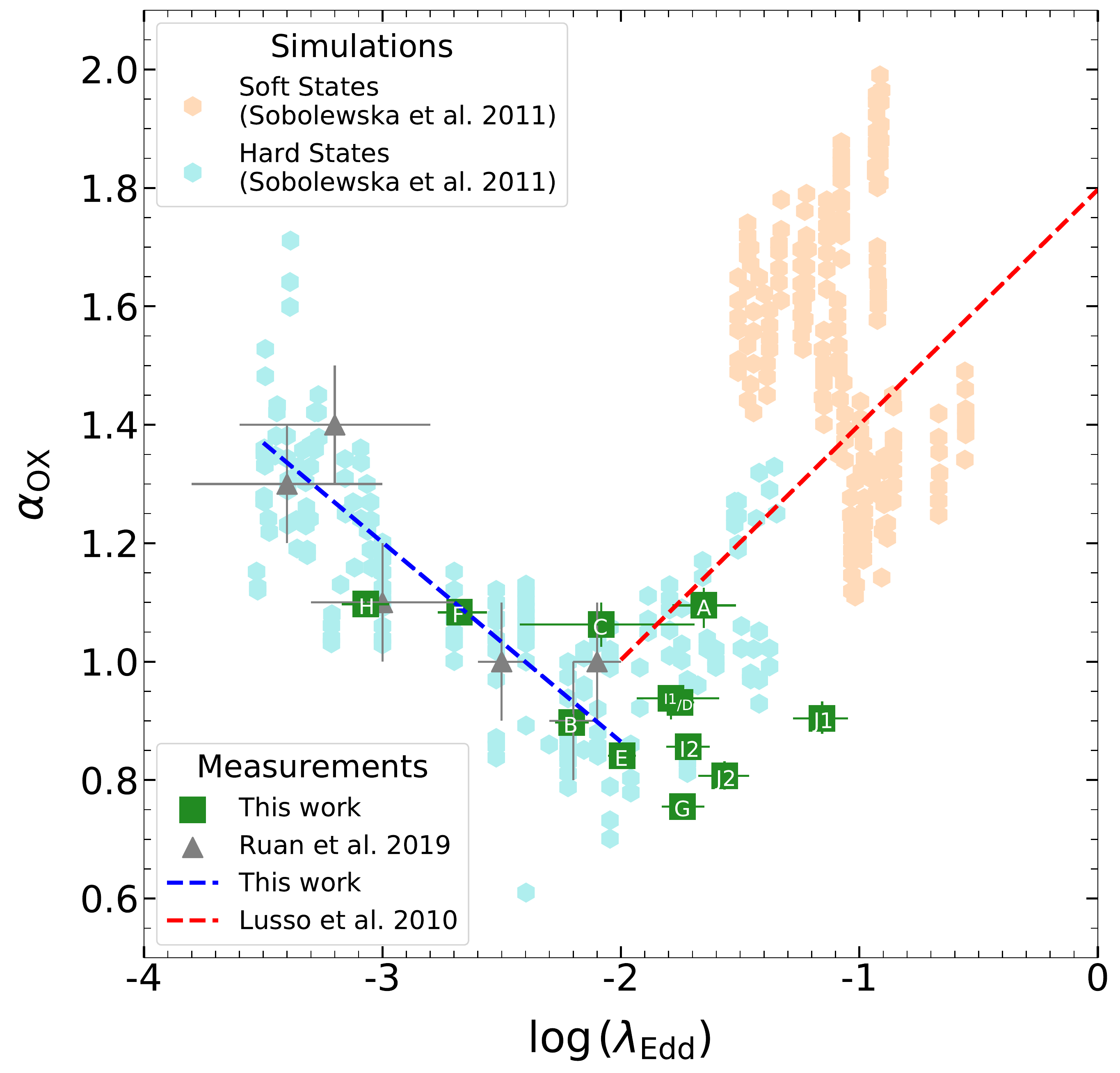}
\caption{Measured $\alpha_{\rm OX}$ and $\lambda_{\rm Edd}$ of changing-look quasars. Green squares are measurements of ten changing-look quasars in this work (see Table \ref{tag:aoxledd_1} for more details). Labels of individual changing-look quasars can be found in the column (6) of Table \ref{tag:aoxledd_1}. Latest measurements of $\alpha_{\rm OX}$ and $\lambda_{\rm Edd}$ from \citet{Ruan2019arXiv} are denoted by grey triangles. The blue dashed line displays the fitted $\alpha_{\rm OX}-\lambda_{\rm Edd}$ relation (i.\,e.,\,Equation \ref{eq:aox-ledd}) at $\lambda_{\rm Edd}\leqslant1\%$ from this work and \citet{Ruan2019arXiv}. The $\alpha_{\rm OX}-\lambda_{\rm Edd}$ relation of bright quasars derived from the XMM-COSMOS survey is shown in the red dashed line \citep{Lusso2010A&A}. Peach and turquoise hexagons are simulated results from \citet{Sobolewska2011MNRASa}, based on \textit{RXTE} observations of the black hole X-ray binary GRO J1655-40. }
\end{figure}\label{fig:aoxledd}

Figure \ref{fig:aoxledd} presents our measurements of $\alpha_{\rm OX}$ and $\lambda_{\rm Edd}$. We include measurements of 6 ``turn-off" CLQs from \citet{Ruan2019arXiv} as well. 
We compare these measurements with results of simulations from \citet{Sobolewska2011MNRASa}.  
\citet{Sobolewska2011MNRASa} use a multi-color blackbody model \cite[{\tt DISKBB};][]{Mitsuda1984PASJ} for the emission from the accretion disk and a Comptonization model \cite[{\tt EQPAIR};][]{Coppi1999ASPC} for the coronal emission to fit the X-ray spectrum of the black hole X-ray binary GRO J1655-40 during its outburst in 2005. This analysis assumes the emission from the accretion disk and the corona can be directly scaled by black hole masses at the same Eddington ratio (for a Shakura-Sunyaev disk, the scaling relation between the accretion disk temperature and the black hole mass is ${T_{\rm disk}\propto M_{\rm BH}^{-1/4}}$), and it scales the emission from this stellar mass black hole to the supermassive black hole case with a mass distribution from the zCOSMOS survey \citep{Merloni2010ApJ}. It predicts a positive correlation between $\alpha_{\rm OX}$ and $\lambda_{\rm Edd}$ at Eddington ratios $\gtrsim1\%$, a critical Eddington ratio 1\%, following the critical Eddington ratio of X-ray binaries \citep{Maccarone2003A&A}, and an anti-correlation between $\alpha_{\rm OX}$ and $\lambda_{\rm Edd}$ at low Eddington ratios $\lesssim1\%$.

Our measurements are generally consistent with the simulations of \citet{Sobolewska2011MNRASa}. 
At $\lambda_{\rm Edd}\gtrsim1\%$, there is also an observed positive correlation between $\alpha_{\rm OX}$ and $\lambda_{\rm Edd}$ \citep{Maoz2007MNRAS,Lusso2010A&A,Grupe2010ApJS}. This relation given by the XMM-COSMOS survey \citep{Lusso2010A&A} is : $\alpha_{\rm OX}=(0.397\pm0.043)\;{\rm log}\;\!\lambda_{\rm Edd}+(1.797\pm0.047)$. The null hypothesis is rejected at $\sim9\sigma$ confidential level. 
At $\lambda_{\rm Edd}\lesssim1\%$, $\alpha_{\rm OX}$ and $\lambda_{\rm Edd}$ show an anti-correlation, and the Pearson correlation coefficient is $-0.85$. For eleven measurements at $\lambda_{\rm Edd}\lesssim1\%$ (in this work and \citealt{Ruan2019arXiv}), we use linear regression analysis, considering measurements errors in both $\alpha_{\rm OX}$ and ${\rm log}\;\!\lambda_{\rm Edd}$ (i.\,e.,\,the bivariate correlated errors and intrinsic scatter [BCES] method, see \citealt{AB1996ApJ,Nemmen2012Sci}), to fit the relation between $\alpha_{\rm OX}$ and ${\rm log}\;\!\lambda_{\rm Edd}$. Our fitted result (bisector) is: 
\begin{equation} \label{eq:aox-ledd}
\alpha_{\rm OX}=(-0.34\pm0.07)\;{\rm log}\;\!\lambda_{\rm Edd}+(0.19\pm0.18).
\end{equation} 
This anti-correlation\footnote{We also try fitting $\alpha_{\rm OX}$ and ${\rm log}\;\!\lambda_{\rm Edd}$ relation by using only five data points at $\lambda_{\rm Edd}\lesssim1\%$ in this work, the fitted result (bisector) from BCES is: $\alpha_{\rm OX}=(-0.29\pm0.08)\;{\rm log}\;\!\lambda_{\rm Edd}+(0.30\pm0.17)$.} again suggests similarities in the spectral indices below 1\% $\lambda_{\rm Edd}$ between CLQs and X-ray binaries undergoing accretion state transitions. 
Though the transition timescale of CLQs is not consistent with the expected accretion state transition timescale for AGN ($\sim$10$^{5}$ years), \citet{ND2018MNRAS} suggest that in AGN, radiation pressure and magnetic pressure can play more important roles than in X-ray binaries, which can dramatically reduce the variability timescale. 

\subsection{Sources of uncertainty in $\alpha_{\rm OX}$}\label{sec:aoxuncer}
In Table \ref{tag:aoxledd_1}, the uncertainties in $\alpha_{\rm OX}$ are from propagation of uncertainties in measured 
$\lambda L{\rm \scriptstyle 2500\AA}$ (reported in Table \ref{tab:fitteddata}) and $\nu L{\rm \scriptstyle 2\;\!keV}$ (reported in Table \ref{tab:xray}). 
However, there are additional sources of uncertainties in $\alpha_{\rm OX}$. We briefly discuss them here:
\begin{enumerate}
    \item Selection of photon indices: For sources that do not have sufficient X-ray counts for spectral fitting, we assume a photon index 1.8 to calculate the rest-frame 2~keV luminosity $\nu L{\rm \scriptstyle 2\;\!keV}$. The assumption of different photon indices will result in different $\alpha_{\rm OX}$, and thus we investigate the possible uncertainty in $\alpha_{\rm OX}$. We adopt the photon index distribution from the BASS, which has a mean value 1.8, and a standard deviation 0.27 \citep{Trakhtenbrot2017MNRAS}. We thus use different photon indices (1.53, 1.8, and 2.07) to calculate $\nu L{\rm \scriptstyle 2\;\!keV}$, and we find they are consistent with 6\%, which will change $\alpha_{\rm OX}$ by only 0.01. 
    \item Time difference between optical and X-ray observations: 
    The observation dates of X-ray (\textit{Chandra} or \textit{XMM-Newton}) and optical data are close but not exactly the same, which can also cause uncertainty in $\alpha_{\rm OX}$. The maximal time lag between optical and X-ray observation dates in the rest-frame is~$\sim$500~days ($\sim$1.4~years). Since there are no existing studies of CLQ optical and X-ray variability over a yearly timescale, to investigate the uncertainty in $\lambda L{\rm \scriptstyle 2500\AA}$, we instead use the structure function (SF) for quasar variability in 2,000--3,000~${\rm \AA}$ \citep{MacLeod2012ApJ}, and we find the corresponding average SF in mag is 0.2 for a rest-frame time lag of 500 days. For the uncertainty in $\nu L{\rm \scriptstyle 2\;\!keV}$, we use an X-ray structure function for quasar variability in 0.2--2~keV band \citep{Middei2017A&A}, and we find the average SF for a rest-frame time lag of 1.4 years is~$\sim$0.23 in a base-10 log scale.
    Combining these two uncertainties and using the propagation of errors, we estimate that a rest-frame time lag of 500 days between the optical and X-ray observations can change $\alpha_{\rm OX}$ up to 0.09. 
\end{enumerate}

Based on the estimation of these two uncertainties, we fit the $\alpha_{\rm OX}-{\rm log}\;\!\lambda_{\rm Edd}$ relation by including an extra uncertainty of 0.09 in $\alpha_{\rm OX}$, apart from the given uncertainties in $\alpha_{\rm OX}$ in Table \ref{tag:aoxledd_1}. The fitted result from BCES (bisector) is: $\alpha_{\rm OX}=(-0.31\pm0.08)\;{\rm log}\;\!\lambda_{\rm Edd}+(0.27\pm0.20).$ The slope is consistent with the slope in Equation \ref{eq:aox-ledd} within 1$\sigma$ errors. However, the fitted linear relation of only five data points below $1\%~\lambda_{\rm Edd}$ in this work becomes: $\alpha_{\rm OX}=(-0.16\pm0.07)\;{\rm log}\;\!\lambda_{\rm Edd}+(0.61\pm0.19)$. Therefore, more measurements below $1\%~\lambda_{\rm Edd}$ are still required to derive a more robust $\alpha_{\rm OX}-{\rm log}\;\!\lambda_{\rm Edd}$ relation in this regime.

\subsection{Changes in the BLR associated with changes in $\lambda_{\rm Edd}$}\label{subsec:ledd_changes}
To study the changes in Eddington ratios of CLQs, associated with changes in their broad emission lines, we select the brightest and faintest optical spectra, and measure their Eddington ratios. 

For every object, we define the brightest optical spectrum as the ``bright" state, while the faintest optical spectrum as the ``faint" state, as shown in Table \ref{tab:Optical}. For nine bright states and for one faint state, contemporaneous X-ray observations are not directly available. To avoid systematic offsets, we calculate all the bolometric luminosity with only optical data, by adopting a linear bolometric correction \citep{Runnoe2012MNRAS}: 
\begin{equation}\label{eq:lbolo}
L_{\rm bol}=(8.1\pm0.4)\;\!\lambda\;\!L{\rm {\scriptstyle5100\AA}}.
\end{equation}
The Eddington ratio is then calculated as the ratio of the bolometric luminosity to the Eddington luminosity. 

We present all the measured Eddington ratios in Table \ref{tag:ledd}. To study distributions of Eddington ratios in the bright and faint states, 
we use extreme deconvolution to derive the underlying distributions that is convolved with the statistical uncertainties to produce the observed Eddington ratio values \citep{Bovy2011AnApS}. To obtain a larger data sample of CLQs, we include the measurements of 6 ``turn-off" CLQs in \citet{Ruan2019arXiv} to derive a more robust result. They adopt a different bolometric correction than our work, so we re-calculate Eddington ratios of these 6 CLQs in both bright and faint states, based on their reported $\lambda\;\!L{\rm {\scriptstyle5100\AA}}$. 
We assume the distribution in Eddington ratios of the bright and faint states can be described by Gaussian mixtures in a logarithmic scale of Eddington ratios. We then use the XDGMM \citep{Holoien2017AJ} package to calculate the Bayesian information criterion \cite[BIC;][]{Schwarz1978AnSta} for the Gaussian mixtures with different number of Gaussian functions from 1 to 10. 
We find that fitting with a single Gaussian function has the lowest BIC for both bright and faint states, and thus we use a single Gaussian function to fit the Eddington ratio distributions of both states and re-sample these two distributions with corresponding errors. 

\begin{figure}[!t]
    \centering
    \includegraphics[width=0.45\textwidth]{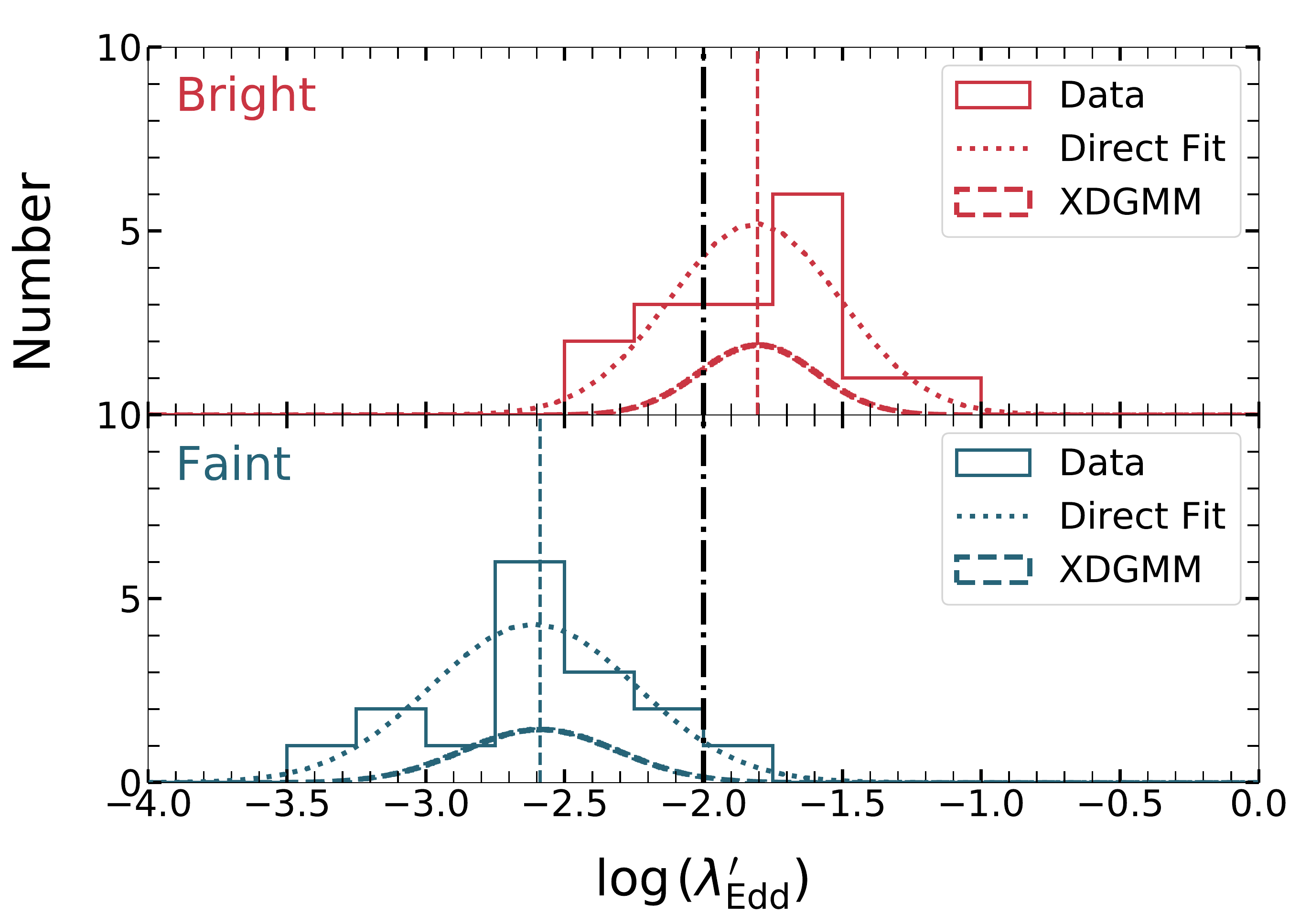}
    \caption{The distributions of Eddington ratios of bright and faint state changing-look quasars in this work and in \citet{Ruan2019arXiv} are shown in the upper and lower panels, respectively. The original bright and faint state distributions are shown in the solid histograms. The means and widths of the dotted Gaussian functions are derived from average Eddington ratios and standard deviations of Eddington ratios, without any uncertainties.
    The normalized re-sampled distributions from extreme deconvolution method (XDGMM) are shown in the thick dashed lines. Best-fit mean values from XDGMM are denoted by vertical thin dashed lines of corresponding color. The black dash-dot line denotes the critical Eddington ratio 1\%, predicted by the disk-wind model.}
\end{figure}  \label{fig:xdgmm_ledd}

\begin{deluxetable}{cccC}\centering 
\tabletypesize{\scriptsize}
\tablecolumns{6}
\tablecaption{Measured $\lambda_{\rm Edd}$ of Bright and Faint State Changing-look Quasars}
\tablewidth{0pt}
\tablehead{
\colhead{Target Name (SDSS)} & \colhead{State} & \colhead{Optical MJD} & \colhead{${\rm log}\;\!(\lambda_{\rm Edd})$}}
\colnumbers
\startdata
J000904.54-103428.6 & Bright & 52141 & -1.24^{+0.05}_{-0.05} \\
& Faint & 57726 & -1.92^{+0.05}_{-0.05}\\
 J002311.06+003517.5 
 & Bright & 55480 & -2.07^{+0.05}_{-0.05}\\
 & Faint & 57597 & -2.71^{+0.05}_{-0.05}\\
 J022556.08+003026.7 
 & Bright & 52944 & -1.95^{+0.37}_{-0.31} \\
 & Faint & 55208 & -2.73^{+0.37}_{-0.31} \\
 J132457.29+480241.2 & Bright & 52759 & -1.52^{+0.04}_{-0.04} \\
 & Faint & 58127 & -2.12^{+0.04}_{-0.04}\\
 J160111.25+474509.6 & Bright & 52354 & -1.95^{+0.04}_{-0.04} \\
 & Faint & 57895 & -2.71^{+0.04}_{-0.04}\\
 J164920.79+630431.3 & Bright & 51699 & -2.12^{+0.09}_{-0.06}\\
 & Faint & 58276 & -2.88^{+0.09}_{-0.06}\\
J214613.30+000930.8 
& Bright & 55478 & -2.29^{+0.07}_{-0.06}\\
& Faint & 57663 & -2.75^{+0.08}_{-0.07} \\
J220537.71-071114.5 & Bright & 52468 & -2.32^{+0.06}_{-0.06}\\
& Faint & 57989 & -3.26^{+0.06}_{-0.06}\\
J225240.37+010958.7 
& Bright & 55500 & -1.60^{+0.05}_{-0.05} \\
& Faint & 58814 &  -2.39^{+0.06}_{-0.06}\\
J233317.38-002303.5 
& Bright & 55447 &  -1.65^{+0.07}_{-0.08}\\
& Faint & 58429 & -2.38^{+0.08}_{-0.09}\\
\enddata
\tablecomments{(1) Name of CLQs in SDSS; (2) ``Bright" denotes the brightest optical spectrum among all the optical spectra of the same object, while ``Faint" denotes the faintest optical spectrum. (3) MJD of the optical data, see Table \ref{tab:Optical} for details; 
(4) Logarithmic Eddington ratio, using a bolometric luminosity calculated with only optical data (i.\,e.,\,Equation \ref{eq:lbolo}, $\lambda\;\!L{\rm {\scriptstyle5100\AA}}$ and ${\rm M_{BH}}$ can be found in Table \ref{tab:fitteddata}); 
All error bars denote $1\sigma$ errors.}
\vspace{-8mm}
\end{deluxetable}\label{tag:ledd}

Figure \ref{fig:xdgmm_ledd} displays the distributions of bright and faint state Eddington ratios. The original Eddington ratio distributions are in solid histograms. We also directly calculate the mean values and the standard deviations of bright and faint state Eddington ratios without uncertainties, and show them as dotted single Gaussian functions.
The best-fit mean values of the two distributions from the XDGMM lay above and below the 1\% Eddington ratio value, which is close to the critical Eddington ratios observed for many X-ray binaries undergoing accretion state transitions. Furthermore, from the bright state to the faint state, broad emission lines in the optical spectra significantly dim, and thus the best-fit mean Eddington ratios of the bright state to the faint state cross 1\% Eddington ratio. 
The disk-wind model predicts dramatic fading of the broad emission lines
below the 1\% Eddington ratio \citep{Elitzur2016MNRAS}. This is in good agreement with the 
measured Eddington ratios of CLQs from their bright state to their faint state, when their broad emission lines have dimmed significantly.

We further test whether this result will change when calculating Eddington ratios with a different bolometric correction. The linear bolometric correction (Equation \ref{eq:lbolo}) is derived from bright quasars \citep{Runnoe2012MNRAS}, and assumes the same SED shape for quasars at different Eddington ratios. This assumption should not be problematic for CLQs in the bright state, but might be biased for CLQs in the faint state, considering the SED shape changes at different Eddington ratios \citep{Ho1999ApJ,VF2007MNRAS,Ho2009ApJ,VF2009MNRAS}. We thus investigate nine faint state CLQs with sensitive X-ray data from \textit{Chandra}, and calculate their Eddington ratios with both optical and X-ray data (Equation \ref{eq:lbolxo}, see the column (5) in Table \ref{tag:aoxledd_1}). We find this bolometric correction results in higher Eddington ratios for the nine faint states by 0.5 dex on average. Applying this shift to the best-fit mean value of the faint state, we find the mean value of the faint state is still below the 1\% Eddington ratio. 
Thus, even when using a different bolometric correction, the mean of the Eddington ratio distributions in the bright and faint states still cross the 1\% value in Eddington ratio. 

Apart from the critical 1\% Eddington ratio, 
the disk-wind model also suggests that emission line profiles characteristic of a disk origin (flat-topped, red-asymmetric, and/or double-peaked) are expected for intermediate type AGN \citep{Chiang1996ApJ,MC1997ApJ,Flohic2012ApJ,Chajet2013MNRAS,Elitzur2014MNRAS}. Two of our ten CLQs (SDSS J022556.08+003026.7 and J225240.37+010958.7) in this study show a weak double-peaked feature in their broad H$\beta$ emission line (see Appendix \ref{Appendix:optical}), which may support the disk-wind model as the origin of broad emission lines. For the other eight CLQs, we do not detect a double peaked broad emission line. This might be attributed to the low signal-noise ratio spectra or lack of spectroscopy when those CLQs faded. Multi-epoch, high-quality spectra of CLQs are still needed to study the transformation in their emission line profiles in detail, and to further investigate the connection with the disk-wind model. 

\section{Conclusion}\label{sec:conclusion}
In this paper, we present optical and X-ray measurements of ten CLQs. By comparing their $\alpha_{\rm OX}$ and $\lambda_{\rm Edd}$ with simulated results from \citet{Sobolewska2011MNRASa}, based on \textit{RXTE} observations of a black hole X-ray binary, we find similar trends in spectral index changes between X-ray binaries in accretion state transitions in low/hard states and CLQs below 1\% Eddington ratios. This result bolsters the idea that quasars 
have analogous flows with those in X-ray binaries. 
This  $\alpha_{\rm OX}$ and $\lambda_{\rm Edd}$ anti-correlation below 1\% $\lambda_{\rm Edd}$ is also found by multi-epoch \textit{Swift} observations of the changing-look AGN NGC~2617 \citep{Ruan2019arXivb}. 
Furthermore, by measuring the Eddington ratios of CLQs before/after they show changes in broad emission lines, we find that CLQs appear to cross the 1\% Eddington ratio value when the strength of their broad emission lines changes drastically. 
Future multi-epoch observations and spectroscopy on individual CLQs may be able to investigate the disk-wind model in more detail, and shed more light on the origin of broad emission lines.

\acknowledgments
We thank the anonymous reviewer for their helpful and informative comments which improved and clarified this manuscript.
X.J., J.J.R., and D.H. acknowledge support from a NSERC Discovery Grant, a FRQNT Nouveaux Chercheurs Grant, and support from the Canadian Institute for Advanced Research (CIFAR).  J.J.R, S.F.A., A.D., and M.E. are supported by Chandra Award Number GO7-18033X and GO8-19090A, issued by the  {\it Chandra} X-ray Observatory center, which is operated by the Smithsonian Astrophysical Observatory for and on behalf of the National Aeronautics Space Administration (NASA) under contract NAS8-03060. C.L.M, P.J.G., S.F.A., and J.J.R. are supported by the National Science Foundation under Grants No. AST-1715763 and AST-1715121.  J.J.R.\ acknowledges funding from the McGill Trottier Chair in Astrophysics and Cosmology, the McGill Space Institute, and the Dan David Foundation.

The scientific results reported in this article are based to a significant degree on observations made by the Chandra X-ray Observatory, and the Chandra Data Archive.


This work uses the ROSAT Data Archive of the Max-Planck- Institut für extraterrestrische Physik (MPE) at Garching, Germany. This work also uses observations obtained with XMM-Newton, an ESA science mission with instruments and contributions directly funded by ESA member states and NASA.

This work uses observations obtained with the Apache Point Observatory 3.5 m telescope, which is owned and operated by the Astrophysical Research Consortium.

This work uses observations obtained at the MMT Observatory, a joint facility of the Smithsonian Institution and the University of Arizona.

This paper includes data gathered with the 6.5 meter Magellan Telescopes located at Las Campanas Observatory, Chile.

This work uses observations obtained with the Hobby-Eberly Telescope and The Low Resolution Spectrograph 2. The Hobby-Eberly Telescope (HET) is a joint project of the University of Texas at Austin, the Pennsylvania State University, Stanford University, Ludwig-Maximillians-Universit\"at Munchen, and Georg-August-Universit\"at G\"ottingen. The HET is named in honor of its principal benefactors, William P. Hobby and Robert E.
Eberly.

The Low Resolution Spectrograph 2 (LRS2) was developed and funded by
the University of Texas at Austin McDonald Observatory and Department
of Astronomy and by The Pennsylvania State University. We thank the
Leibniz-Institut f\"ur Astrophysik Potsdam (AIP) and the Institut
f\"ur Astrophysik G\"ottingen (IAG) for their contributions to the
construction of the integral field units.


Funding for the Sloan Digital Sky Survey IV has been provided by the Alfred P. Sloan Foundation, the U.S. Department of Energy Office of Science, and the Participating Institutions. The SDSS-IV acknowledges support and resources from the center for High-Performance Computing at the University of Utah. The SDSS website is \url{www.sdss.org}.

The SDSS-IV is managed by the Astrophysical Research Consortium for the Participating Institutions of the SDSS Collaboration, including the Brazilian Participation Group, the Carnegie Institution for Science, Carnegie Mellon University, the Chilean Participation Group, the French Participation Group, the Harvard-Smithsonian Center for Astrophysics, Instituto de Astrofísica de Canarias, The Johns Hopkins University, the Kavli Institute for the Physics and Mathematics of the Universe (IPMU) /University of Tokyo, the Korean Participation Group, Lawrence Berkeley National Laboratory, Leibniz Institut für Astrophysik Potsdam (AIP), Max-Planck-Institut für Astronomie (MPIA Heidelberg), Max-Planck-Institut für Astrophysik (MPA Garching), Max-Planck-Institut für Extraterrestrische Physik (MPE), National Astronomical Observatories of China, New Mexico State University, New York University, the University of Notre Dame, Observatário Nacional/MCTI, The Ohio State University, Pennsylvania State University, the Shanghai Astro- nomical Observatory, the United Kingdom Participation Group, Universidad Nacional Autónoma de México, the University of Arizona, the University of Colorado Boulder, the University of Oxford, the University of Portsmouth, the University of Utah, the University of Virginia, the University of Washington, the University of Wisconsin, Vanderbilt University, and Yale University.

IRAF is distributed by the National Optical Astronomy Observatory, which is operated by the Association of Universities for Research in Astronomy (AURA) under a cooperative agreement with the National Science Foundation.

%

\vspace{5mm}
\facilities{SDSS, MMT, Magellan Telescope, ARC 3.5m, HET, ROSAT, XMM (PN), CXO (ACIS-S)}
\software{CIAO \citep{Fruscione2006SPIE}, SAS (\url{https://www.cosmos.esa.int/web/xmm-newton/what-is-sas}), XIMAGE (\url{https://heasarc.gsfc.nasa.gov/docs/xanadu/ximage}), XSPEC \citep{Arnaud1996ASPC}, Chandra WebPIMMs (\url{https://cxc.harvard.edu/toolkit/pimms.jsp}), IRAF \citep{Tody1986SPIE,Tody1993ASPC}, emcee \citep{FM2013PASP}, XDGMM \citep{Holoien2017AJ}, Astropy \citep{Astropy2013A&A,Astropy2018AJ}, Matplotlib \citep{Hunter2007}, BCES \citep{AB1996ApJ,Nemmen2012Sci}}

\appendix

\section{Early X-ray Constraints from \textit{ROSAT}}\label{Appendix:ROSAT}

We obtain constraints in X-ray fluxes from \textit{ROSAT}, and connect them with the earliest SDSS observations to measure $\alpha_{\rm OX}$. Note that
the time lag between optical and \textit{ROSAT} X-ray observations is around 10 years, which is comparable to the timescale on which CLQs have been observed to change spectral types. 
Therefore, those $\alpha_{\rm OX}$ measurements should be viewed as crude constraints. In the following part of this section, we summarize the \textit{ROSAT} observations and present $\alpha_{\rm OX}$ from the \textit{ROSAT} observations and the earliest optical observations. 

We first search the coordinates of CLQs with a default 1{\arcmin} matching radius in the second \textit{ROSAT} all-sky survey source catalog \cite[2RXS;][]{Boller2016AA}, which is the newest public source catalog derived from \textit{ROSAT} position-sensitive proportional counter (PSPC) all-sky survey. The typical observation date of {\it ROSAT} PSPC observations is around 1990/1991, which is $\sim$10 years before the first SDSS spectra of our CLQs. 
Three sources (SDSS J000904.54-103428.6, SDSS J002311.06+003517.5 and SDSS J132457.29+480241.2) were detected by the 2RXS catalog. We adopt their 0.1--2.4~keV count rate reported in the 2RXS catalog in our analysis. 

For the remaining sources, we acquire 0.1--2.4~keV photon images and exposure maps from the {\it ROSAT} PSPC all-sky survey, and use the \texttt{SOSTA} tool in \texttt{XIMAGE}\footnote{\url{https://heasarc.gsfc.nasa.gov/docs/xanadu/ximage}} to extract the 0.1--2.4~keV count rate. We convert the 0.1--2.4~keV count rate into the rest-frame 2~keV flux density using the \textit{Chandra} WebPIMMs tool with an assumed photon index of 1.8.
For the sources which were not detected by {\it ROSAT} (i.\,e.,\,signal-to-noise ratio $<3$), we instead calculate their $3\sigma$ upper limit on the 0.1--2.4~keV count rate, and set a $3\sigma$ upper limit on the rest-frame 2~keV luminosity $\nu L{\rm \scriptstyle 2\;\!keV}$. We present our measurements in Table \ref{tab:xray_rosat}. 

\begin{deluxetable}{cccccc}\centering 
\tabletypesize{\scriptsize}
\tablecolumns{7}
\tablecaption{X-ray Properties of Changing-look Quasars from \textit{ROSAT}}
\tablewidth{0pt}
\tablehead{
\colhead{Target Name} & \colhead{Observation Date} & \colhead{Exp.} & \colhead{Count Rate} & \colhead{Ref.} & \colhead{$\nu L{\rm \scriptstyle 2\;\!keV}$}\\
\colhead{(SDSS)} & \colhead{(MJD)} & \colhead{(ks)} &   \colhead{$(10^{-2}\;{\rm cts\;s^{-1}})$} & \colhead{} & \colhead{$(10^{42}\;{\rm erg\;s^{-1}})$}
}
\colnumbers
\startdata
J000904.54-103428.6 & 48225 & 0.4 & $9\pm2$ & 2RXS &$140\pm30$\\
J002311.06+003517.5 & 48083 & 0.4 & $4\pm1$ & 2RXS &$260\pm70$\\ 
J022556.08+003026.7 & 48266 & 0.2 &  $<10$ & & $<1000$\\
J132457.29+480241.2 & 48214 & 0.6 & $14\pm2$ & 2RXS & $200\pm30$ \\
J160111.25+474509.6 & 48083 & 0.5 & $41\pm14$ & & $800\pm300$\\
J164920.79+630431.3 & 48083 & 2.7 & $<13$ &  &  $<400$\\
J214613.30+000930.8 & 48193 & 0.3 & $<28$ & &$<6600$\\
J220537.71-071114.5 & 48197 & 0.1 & $<64$ & & $<1900$\\
J225240.37+010958.7 & 48292 & 0.3 & $<32$ &  & $<4900$\\
J233317.38-002303.5 & 48224 & 0.3 & $<18$ &  & $<2300$\\
\enddata
\tablenotetext{\it a}{The photon index is derived from an absorbed power-law ({\tt wabs*powerlaw}) fitting.}
\tablecomments{(1) Name of CLQs in SDSS; 
(2) Date of X-ray observations in MJD; (3) Exposure time in kilo-seconds (ks). \textit{ROSAT} observations are from the PSPC all-sky survey, therefore they have relatively shallow exposure; (4) 0.1--2.4~keV count rate in units of $10^{-2}\;{\rm cts\;s^{-1}}$. 
(5) Reference for count rates in (4). 2RXS represents the ``Second \textit{ROSAT} all-sky Survey Source Catalogue" \citep{Boller2016AA}. If it is not specified, the count rate is measured from our work; (6) Unabsorbed rest-frame 2~keV luminosity, in units of $10^{42}\;{\rm erg\;s^{-1}}$, calculated by \textit{Chandra} WebPIMMs by assuming a fixed galactic absorption and a photon index of 1.8. All the errors are 1$\sigma$, and all the upper limits are 3$\sigma$.}
\end{deluxetable}\label{tab:xray_rosat}

\begin{deluxetable}{ccCCCc}\centering 
\tabletypesize{\scriptsize}
\tablecolumns{6}
\tablecaption{Measured $\alpha_{\rm OX}$ and $\lambda_{\rm Edd}$ of Changing-look Quasars from \textit{ROSAT} and SDSS}
\tablewidth{0pt}
\tablehead{
\colhead{Target Name (SDSS)} & \colhead{Optical MJD} & \colhead{X-ray MJD} & \colhead{$\alpha_{\rm OX}$} & \colhead{${\rm log}\;\!(\lambda_{\rm Edd})$} & \colhead{Label}}
\colnumbers
\startdata
J000904.54-103428.6 & 52141 & 48225 & 0.95^{+0.05}_{-0.04} & -1.24^{+0.05}_{-0.05} & A \\
 J002311.06+003517.5 & 51900 & 48083 & 0.96^{+0.05}_{-0.05} & -2.28^{+0.05}_{-0.05} & B \\
 J022556.08+003026.7 & 52200 & 48266 & >0.55 & -2.20^{+0.37}_{-0.31} & C \\
 J132457.29+480241.2 & 52759 & 48214 & 0.93^{+0.02}_{-0.02}  & -1.52^{+0.04}_{-0.04} & D \\
 J160111.25+474509.6 & 52354 & 48083 & 0.62^{+0.06}_{-0.06} & -1.95^{+0.04}_{-0.04} & E \\
 J164920.79+630431.3 & 51699 & 48083 & >0.70 & -2.12^{+0.09}_{-0.06} & F \\
J214613.30+000930.8 & 52968 & 48193 & >0.21 & -2.62^{+0.07}_{-0.06} & G \\
J220537.71-071114.5 & 52468 & 48197 & >0.56 & -2.32^{+0.06}_{-0.06} & H  \\
J225240.37+010958.7 & 52178 & 48292 & >0.29 &  -2.02^{+0.05}_{-0.05} & I \\
J233317.38-002303.5 & 52199 & 48224 & >0.34 & -2.11^{+0.07}_{-0.08}  & J  \\ 
\enddata
\tablecomments{(1) Name of CLQs in SDSS; 
(2) MJD of the optical data, see Table \ref{tab:Optical} for details; (3) MJD of the X-ray data, see Table \ref{tab:xray} for details; (4) Optical/UV-X-ray spectral indices $\alpha_{\rm OX}$ ($\lambda L{\rm \scriptstyle 2500\AA}$ can be found in Table \ref{tab:fitteddata}, and $\nu L{\rm \scriptstyle 2\;\!keV}$ can be found in Table \ref{tab:xray_rosat}); (5) Logarithmic Eddington ratio, using a bolometric luminosity calculated with only optical data (i.\,e.,\,Equation \ref{eq:lbolo}, $\lambda\;\!L{\rm {\scriptstyle5100\AA}}$ and ${\rm M_{BH}}$ can be found in Table \ref{tab:fitteddata}); (6) Label of orange square data points presented in Figure \ref{fig:aoxledd_rosat}. All error bars denote $1\sigma$ errors, and all the lower limits are $3\sigma$.}
\end{deluxetable}\label{tag:aoxledd_rosat}

\begin{figure}[!ht]\centering
\includegraphics[width=0.45\textwidth]{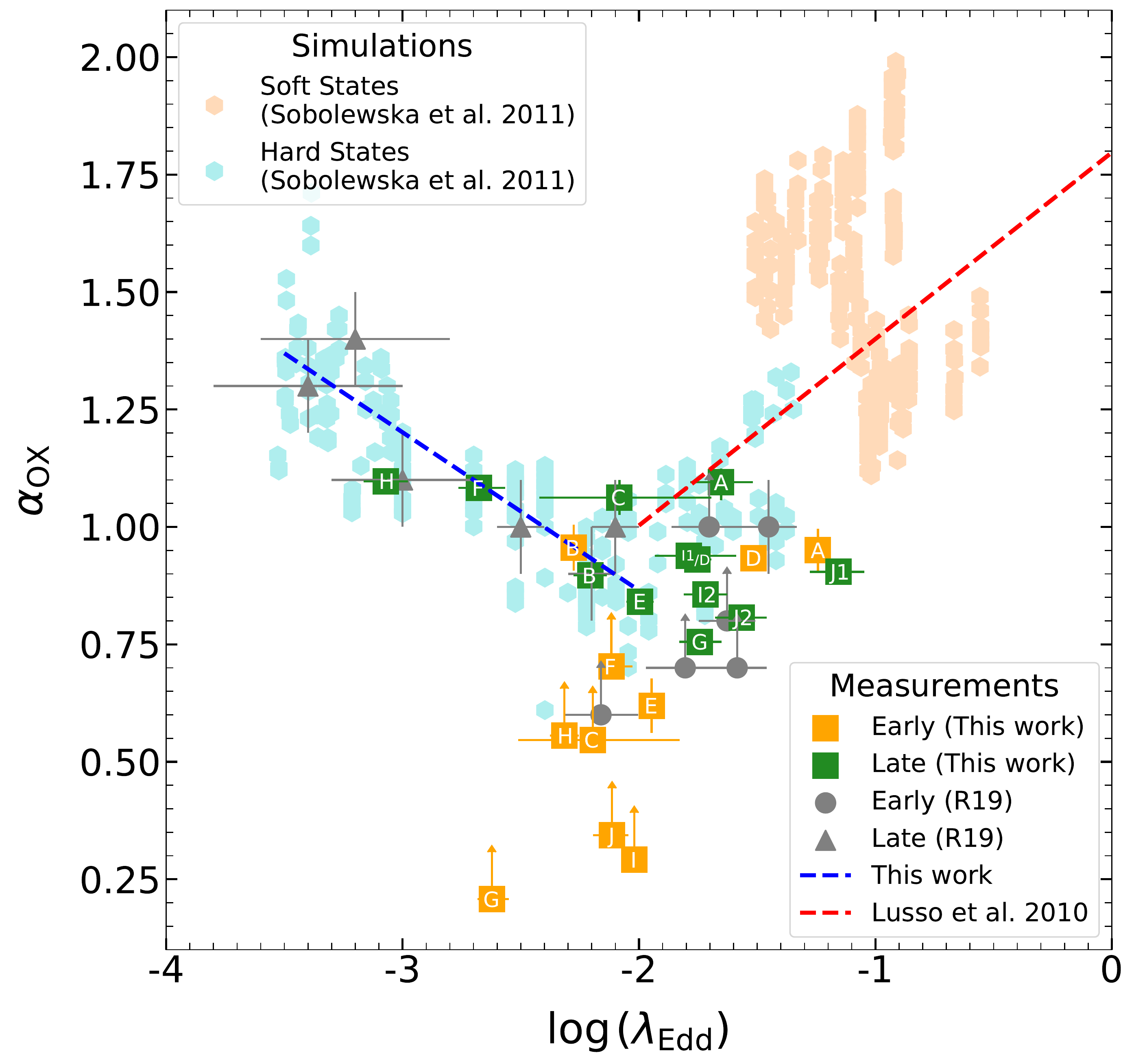}
\caption{Measured $\alpha_{\rm OX}$ and $\lambda_{\rm Edd}$ of changing-look quasars, including X-ray constraints from \textit{ROSAT} observations. Orange squares are measurements from \textit{ROSAT} observations and earliest optical observations (see Table \ref{tag:aoxledd_rosat}). Labels of individual changing-look quasars can be found in the column (6) of Table \ref{tag:aoxledd_1} and Table \ref{tag:aoxledd_rosat}. Early measurements of $\alpha_{\rm OX}$ and $\lambda_{\rm Edd}$ from \citet{Ruan2019arXiv} are denoted by grey circles. Other symbols and lines have the same definitions with Figure \ref{fig:aoxledd}.}
\end{figure}\label{fig:aoxledd_rosat}


We connect the \textit{ROSAT} observations with the earliest optical observations from SDSS. 
We then calculate $\alpha_{\rm OX}$ based on Equation \ref{eq:aox}. 
For those measurements, the \textit{ROSAT} observations 
predate the optical spectra by $\gtrsim10$~years.
Thus, we can only measure the bolometric luminosity from the optical data, by adopting a linear bolometric correction in Equation \ref{eq:lbolo}. We present our measured $\alpha_{\rm OX}$ and $\lambda_{\rm Edd}$ in Table \ref{tag:aoxledd_rosat} and in Figure \ref{fig:aoxledd_rosat}. The measurements using the earliest optical spectra and the \textit{ROSAT} observations are referred to as the ``early" measurements, denoted by orange squares in Figure \ref{fig:aoxledd_rosat}, because they are prior in time than the measurements shown in \S\ref{subsec:aox_ledd}. The measurements presented in \S\ref{subsec:aox_ledd}, using late optical spectra and \textit{Chandra} or \textit{XMM-Newton} observations, are referred to as the ``late" measurements. Note that ``early/late" measurements are different than ``bright/faint" states mentioned in \S\ref{subsec:ledd_changes}. 

We also include $\alpha_{\rm OX}$ and $\lambda_{\rm Edd}$ from \citet{Ruan2019arXiv}. 
That work adopts a different bolometric correction in the ``early" measurements than our work. To ensure consistency, 
we re-calculate the Eddington ratios of these six ``early" measurements, based on their reported $\lambda\;\!L{\rm {\scriptstyle5100\AA}}$. These early measurements should be viewed as crude constraints on $\alpha_{\rm OX}$, because the \textit{ROSAT} and SDSS observations may be separated by up to 13 years. 


\section{Optical spectra} \label{Appendix:optical}

We present spectral decomposition and quasar spectral fitting of other nine CLQs in Figure \ref{fig:optical_data_misc_3}, as described in \S\ref{sec:decompose} and \S\ref{sec:qsofitting}.

\begin{figure*}[!htb]
    \centering
    \includegraphics[width=0.8\textwidth]{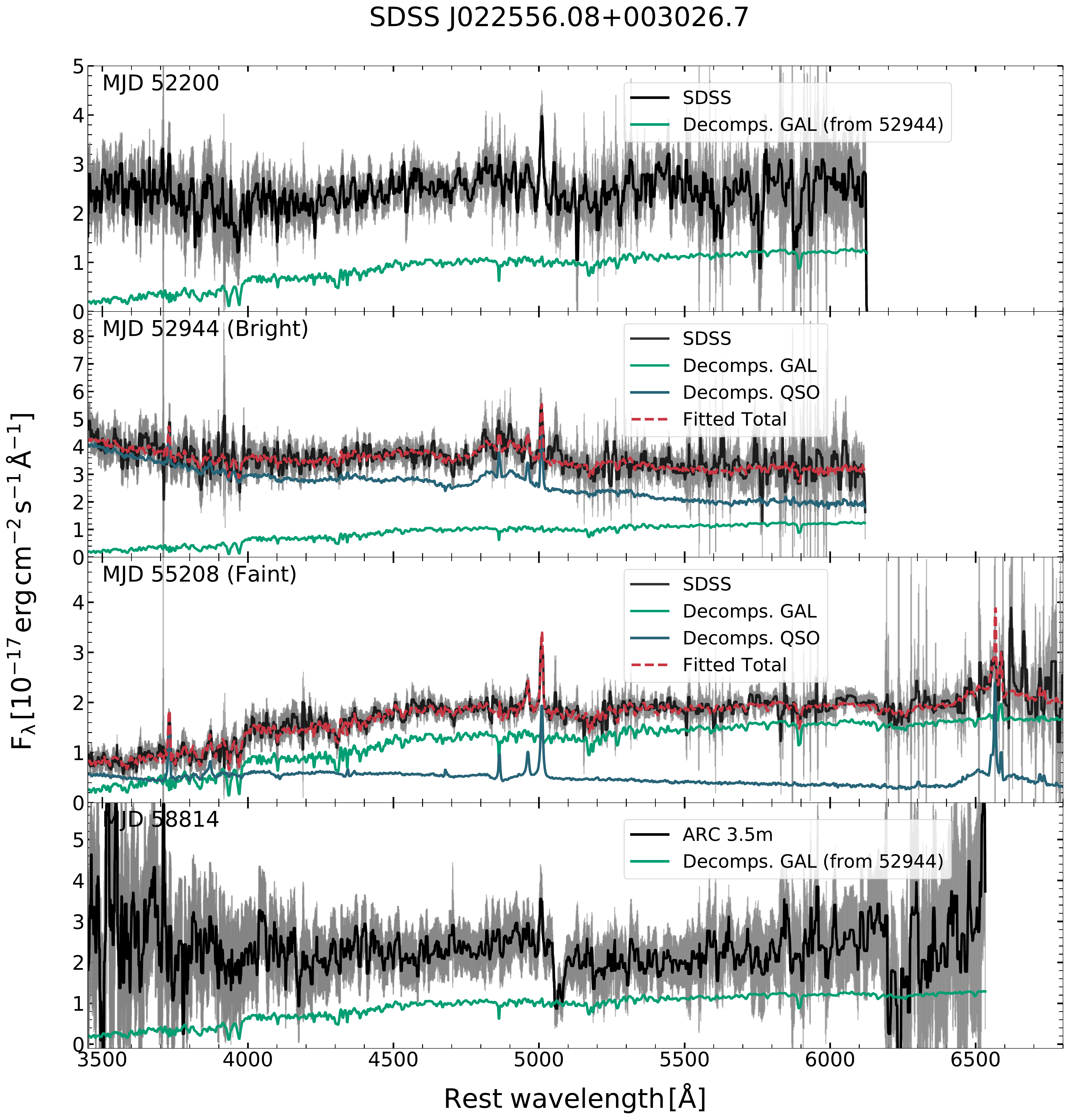}
    \caption{Spectral decomposition for SDSS J022556.08+003026.7.}
    \label{fig:optical_data_misc_3}
\end{figure*}

\renewcommand{\thefigure}{\arabic{figure} (Cont.)}
\addtocounter{figure}{-1}

\begin{figure*}[!htb]
    \centering
    \includegraphics[width=0.8\textwidth]{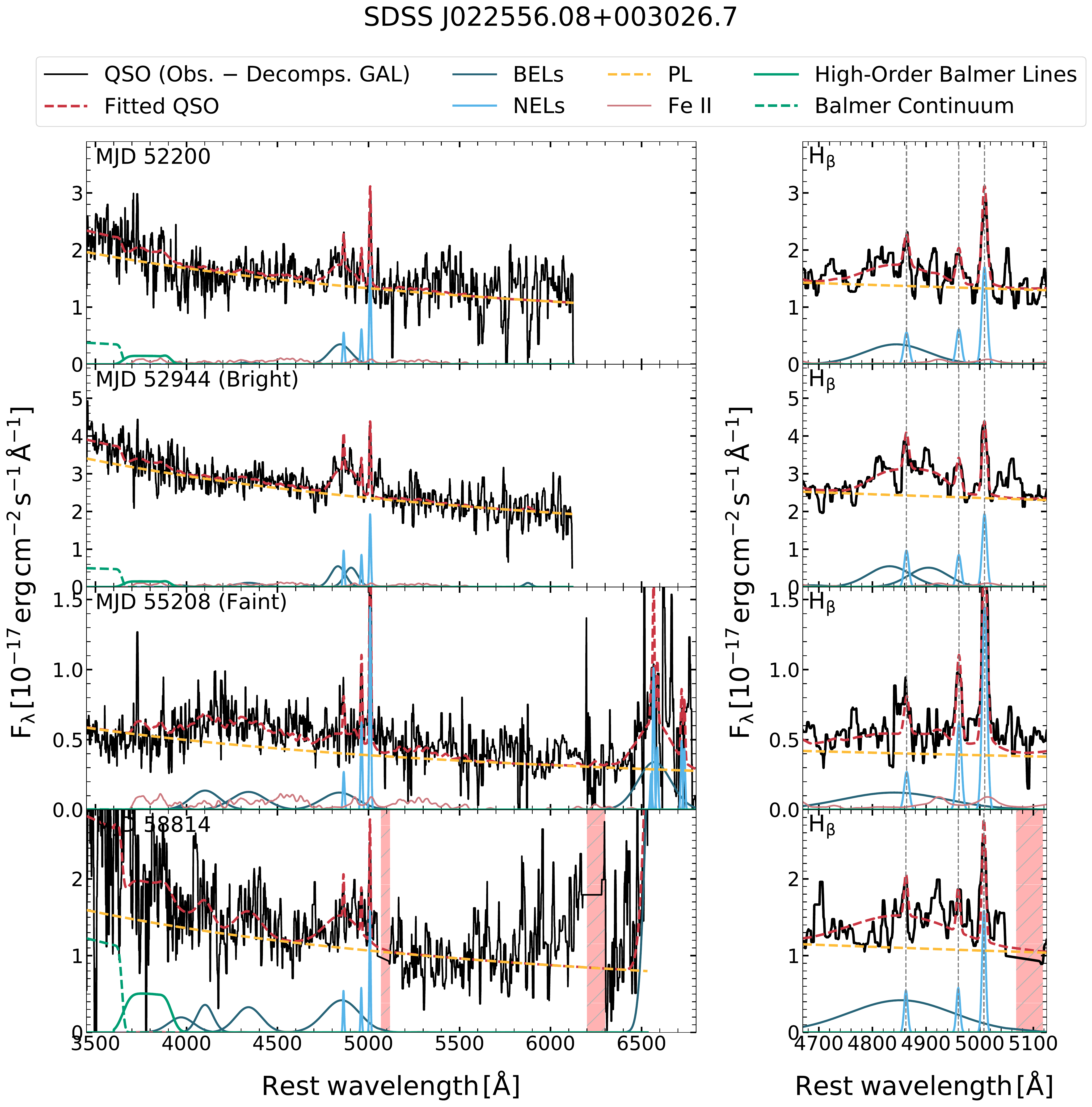}
    \caption{Quasar spectral fitting for SDSS J022556.08+003026.7. Pink shaded regions are masked when we perform the quasar spectrum fitting.}
    \label{fig:optical_data_misc_3}
\end{figure*}

\renewcommand{\thefigure}{\arabic{figure} (Cont.)}
\addtocounter{figure}{-1}

\begin{figure*}[!htb]
    \centering
    \includegraphics[width=0.8\textwidth]{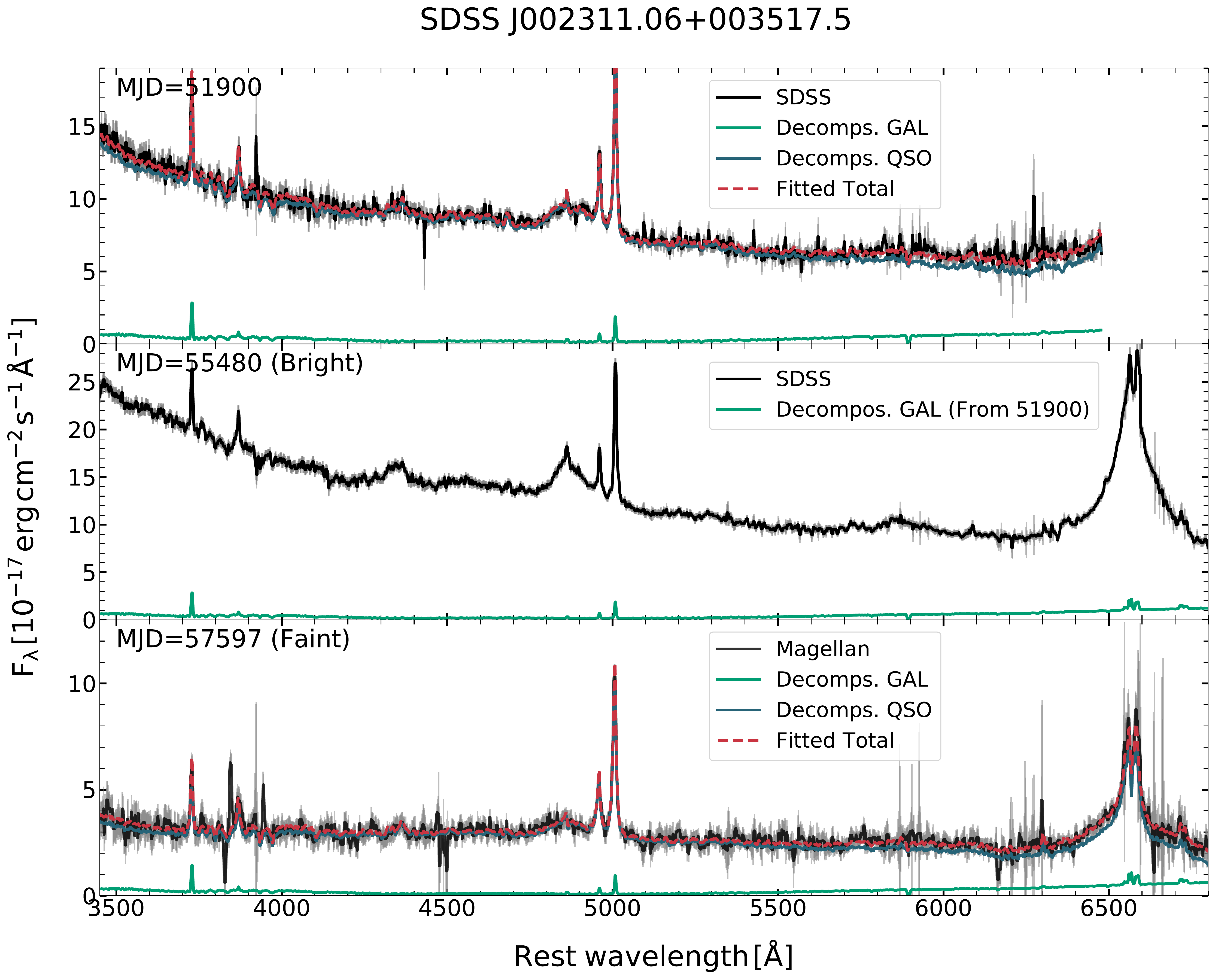}
    \includegraphics[width=0.8\textwidth]{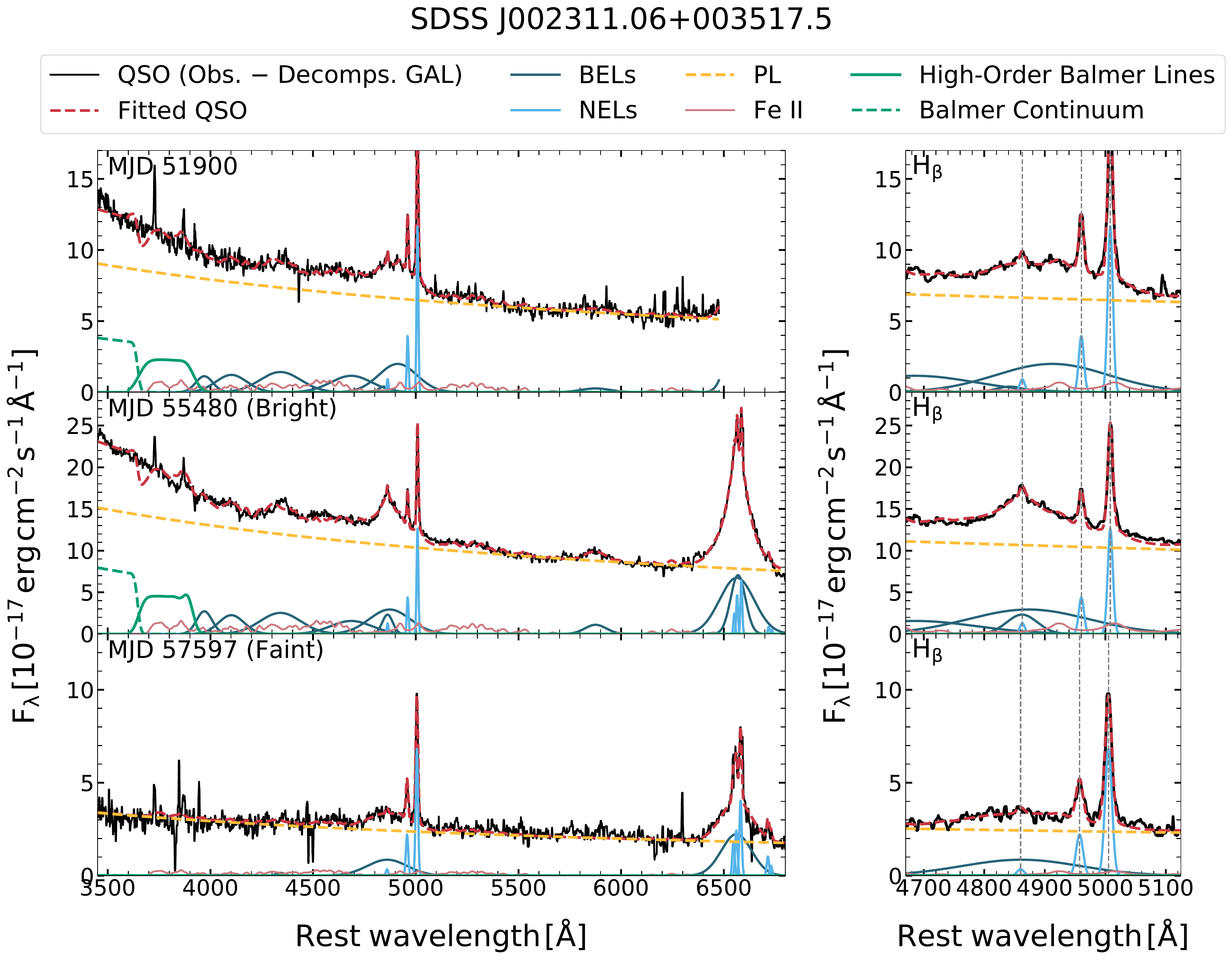}
    \caption{Spectral decomposition and the quasar spectral fitting at ${\rm H\beta}$ for SDSS J002311.06+003517.5.}
    \label{fig:optical_data_misc_3}
\end{figure*}

\renewcommand{\thefigure}{\arabic{figure} (Cont.)}
\addtocounter{figure}{-1}

\begin{figure*}[!htb]
    \centering
    \includegraphics[width=0.8\textwidth]{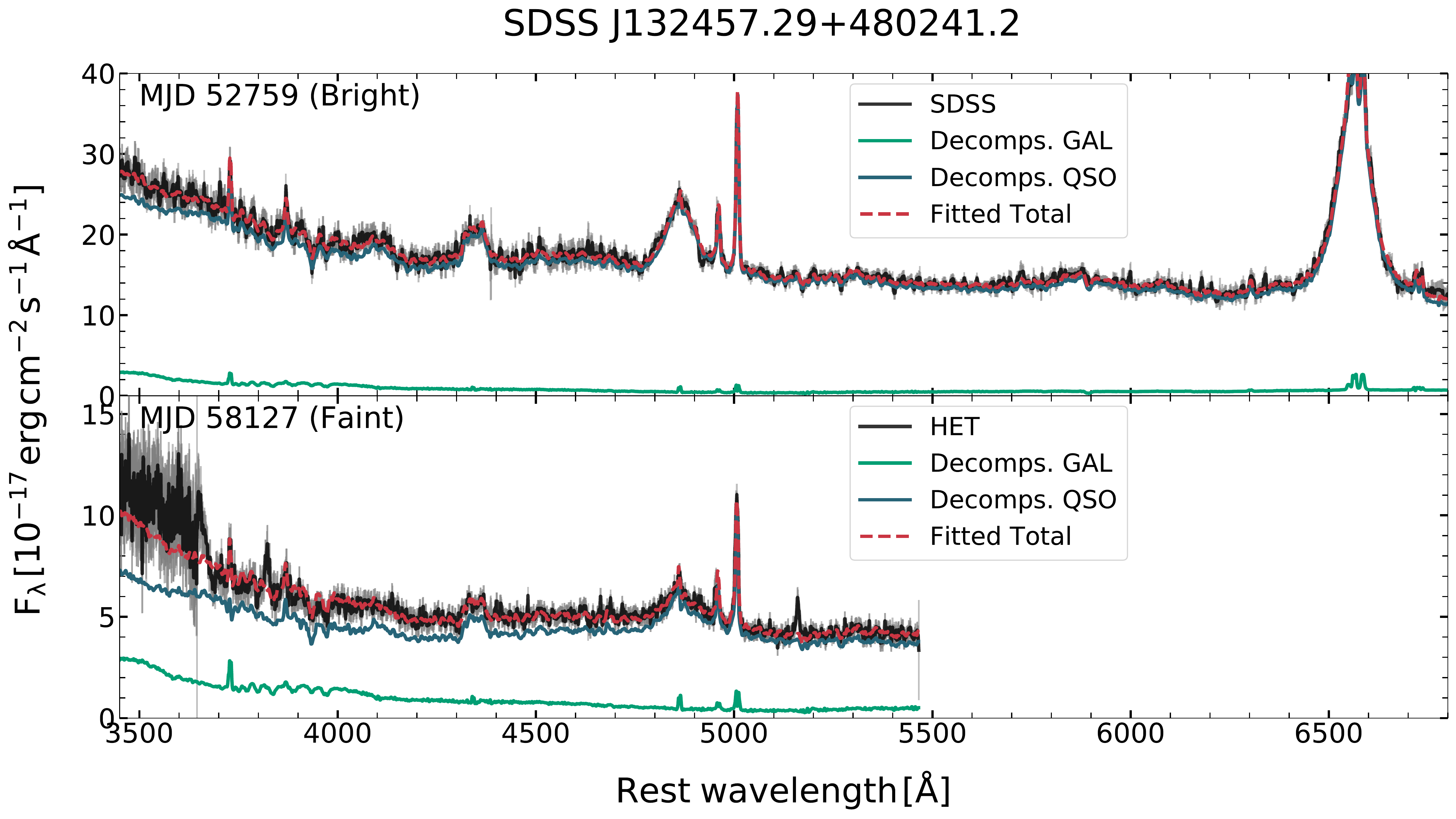}
    \includegraphics[width=0.8\textwidth]{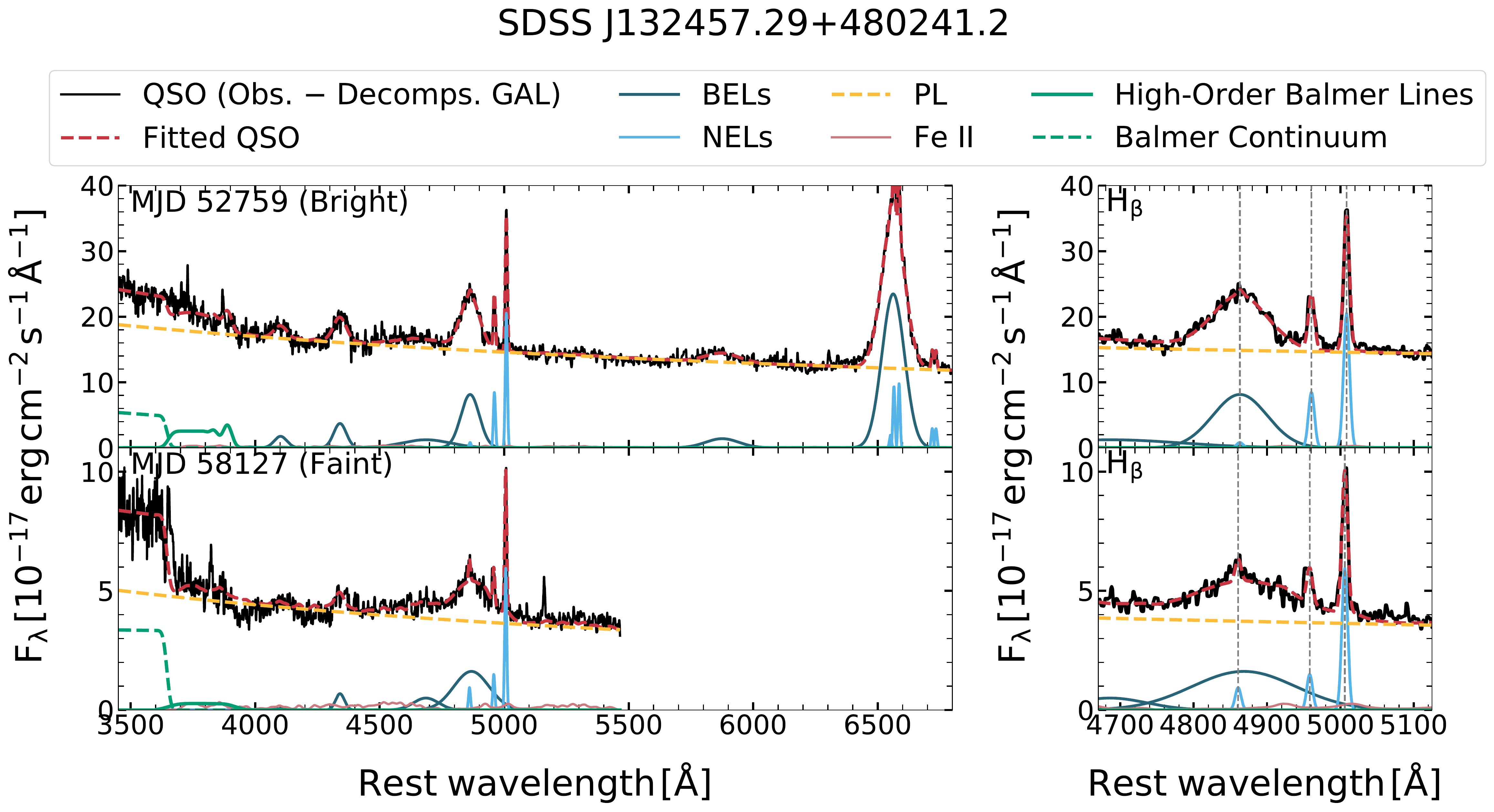}
    \caption{Spectral decomposition and the quasar spectral fitting for SDSS J132457.29+480241.2. Pink shaded regions are masked when we perform the quasar spectrum fitting.}
    \label{fig:optical_data_misc_3}
\end{figure*}

\renewcommand{\thefigure}{\arabic{figure} (Cont.)}
\addtocounter{figure}{-1}

\begin{figure*}[!htb]
    \centering
    \includegraphics[width=0.8\textwidth]{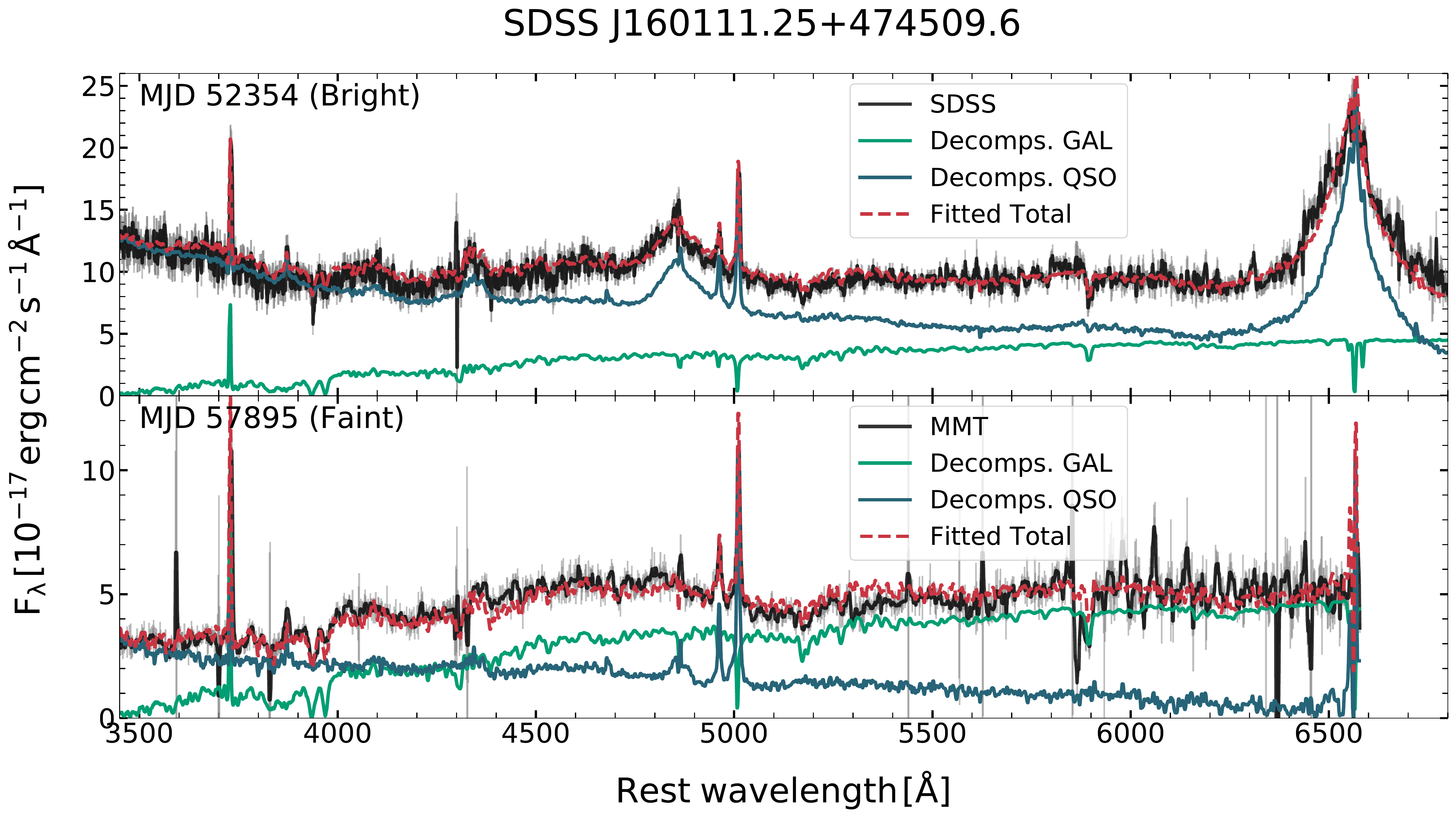}
    \includegraphics[width=0.8\textwidth]{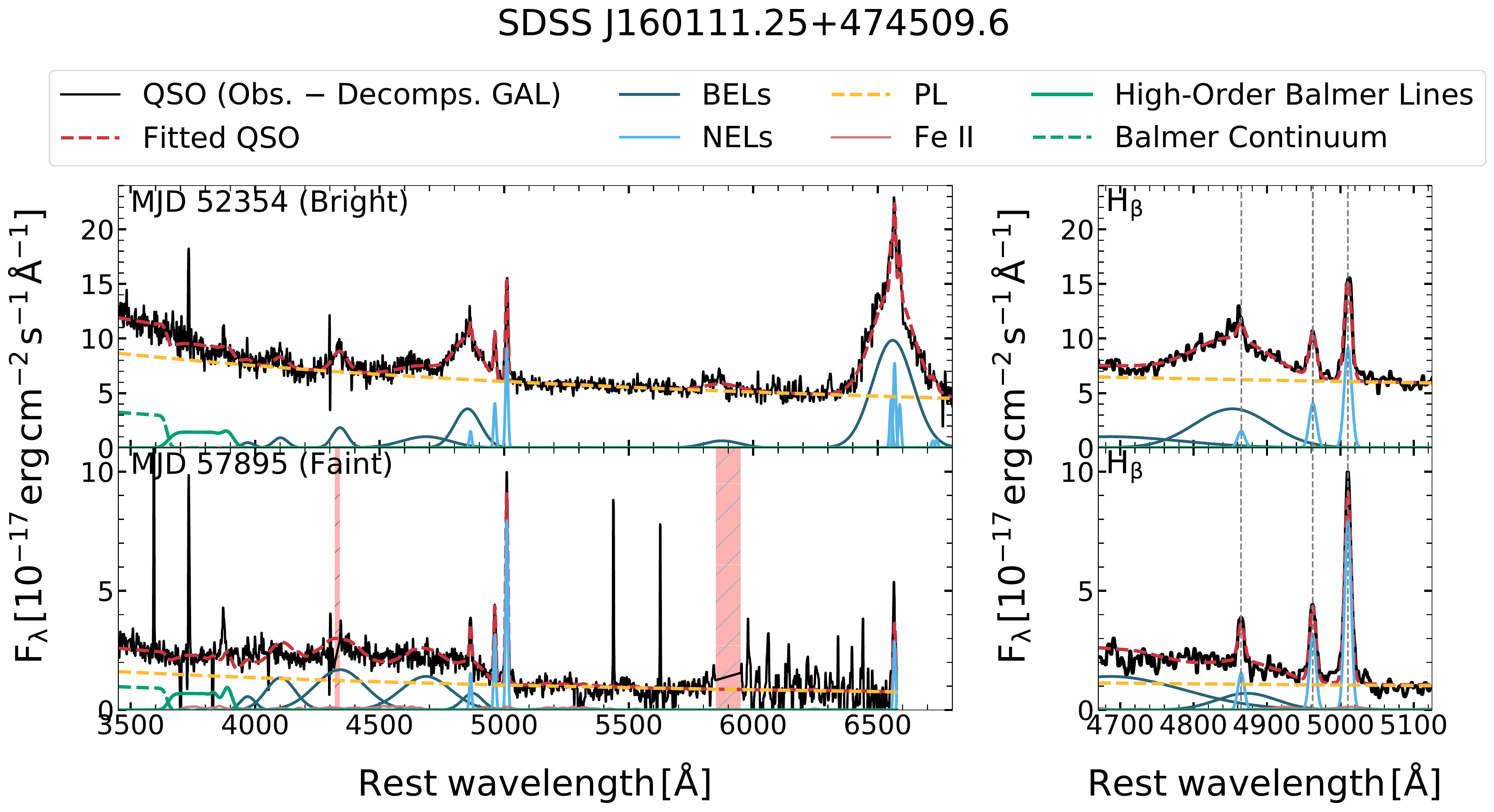}
    \caption{Spectral decomposition and the quasar spectral fitting for SDSS J160111.25+474509.6. Pink shaded regions are masked when we perform the quasar spectrum fitting.}
    \label{fig:optical_data_misc_3}
\end{figure*}

\renewcommand{\thefigure}{\arabic{figure} (Cont.)}
\addtocounter{figure}{-1}

\begin{figure*}[!htb]
    \centering
    \includegraphics[width=0.8\textwidth]{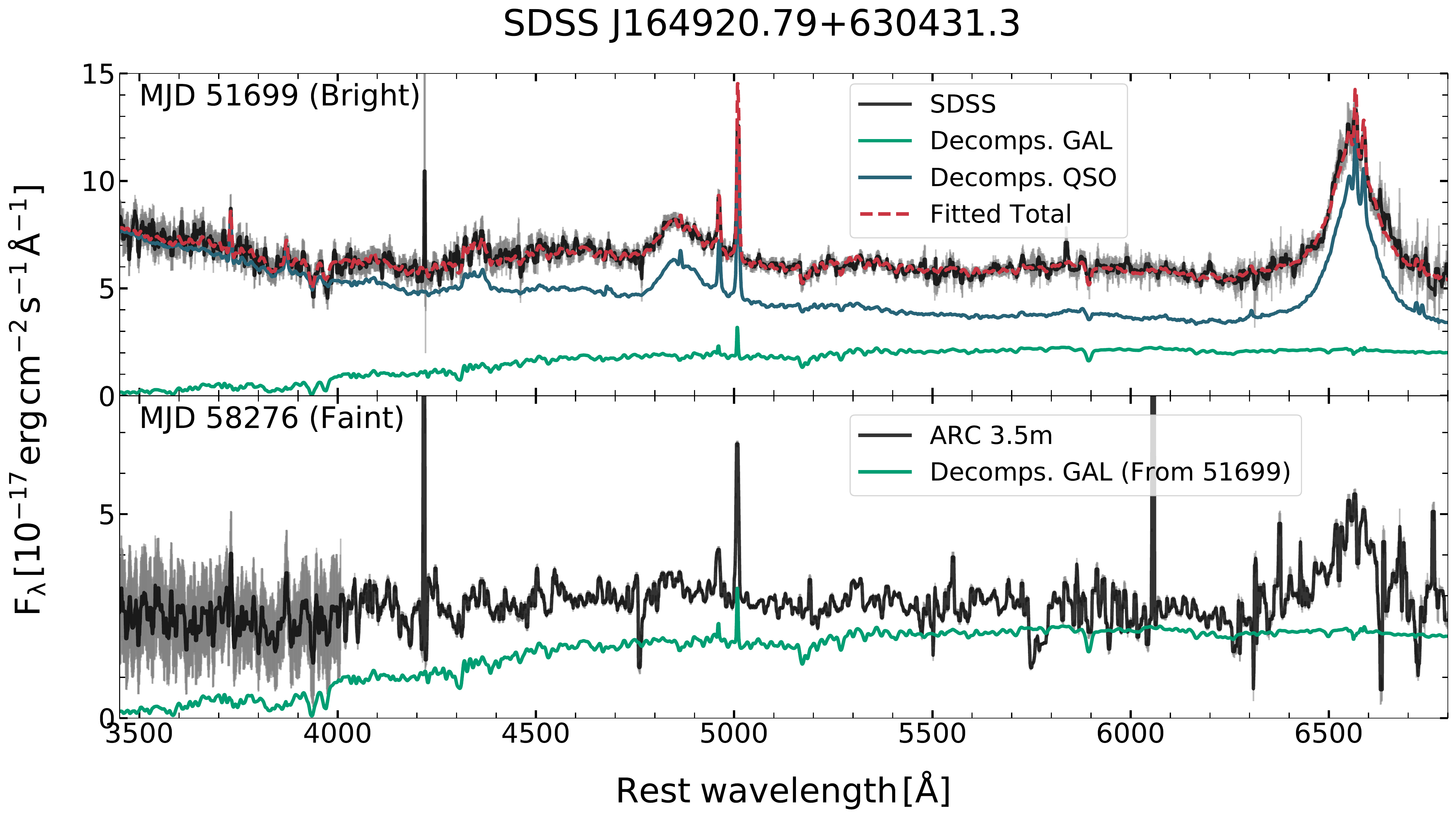}
    \includegraphics[width=0.8\textwidth]{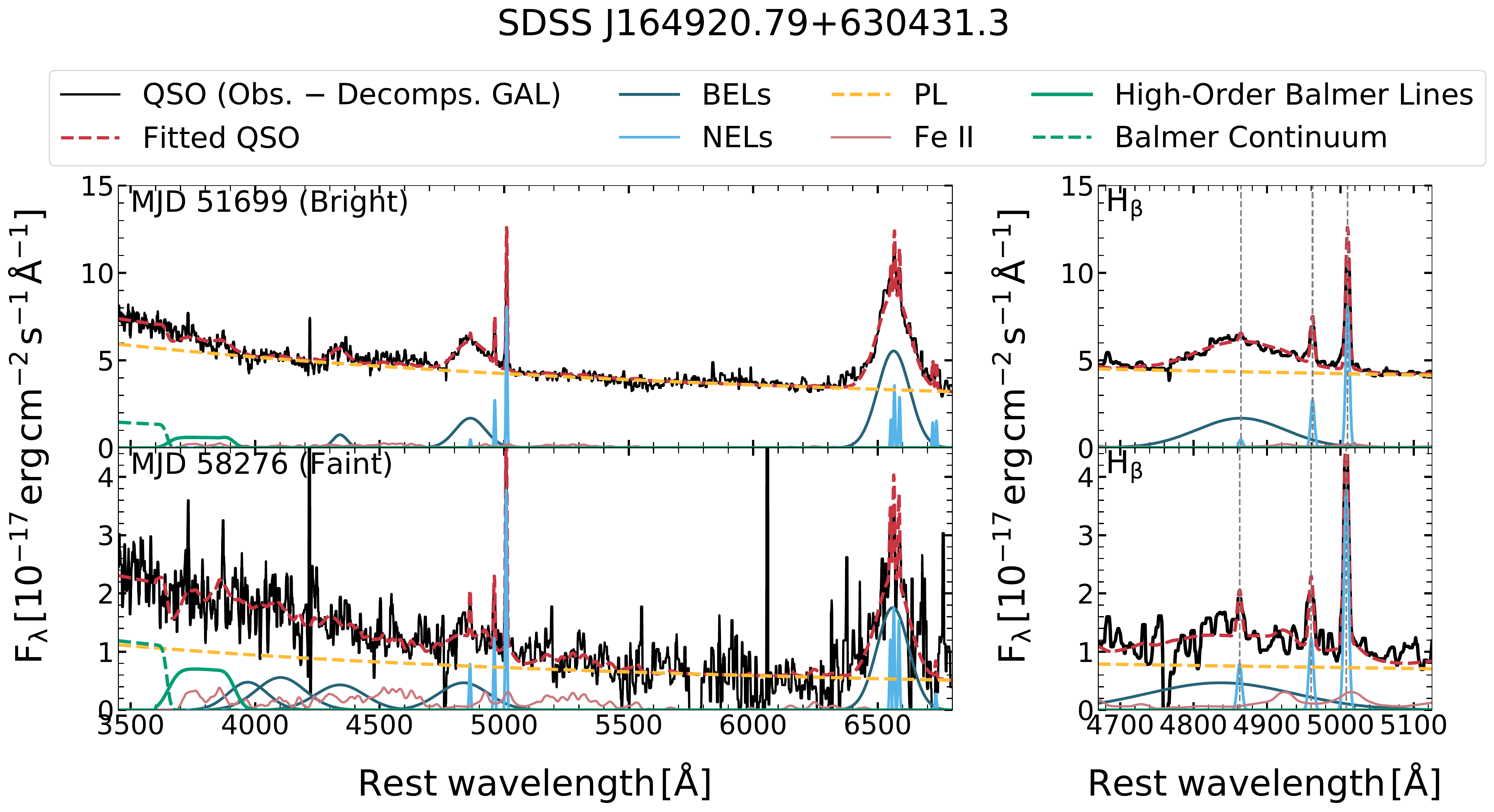}
    \caption{Spectral decomposition and the quasar spectral fitting for SDSS J164920.79+630431.3.}
    \label{fig:optical_data_misc_3}
\end{figure*}

\renewcommand{\thefigure}{\arabic{figure} (Cont.)}
\addtocounter{figure}{-1}

\begin{figure*}[!htb]
    \centering
    \includegraphics[width=0.8\textwidth]{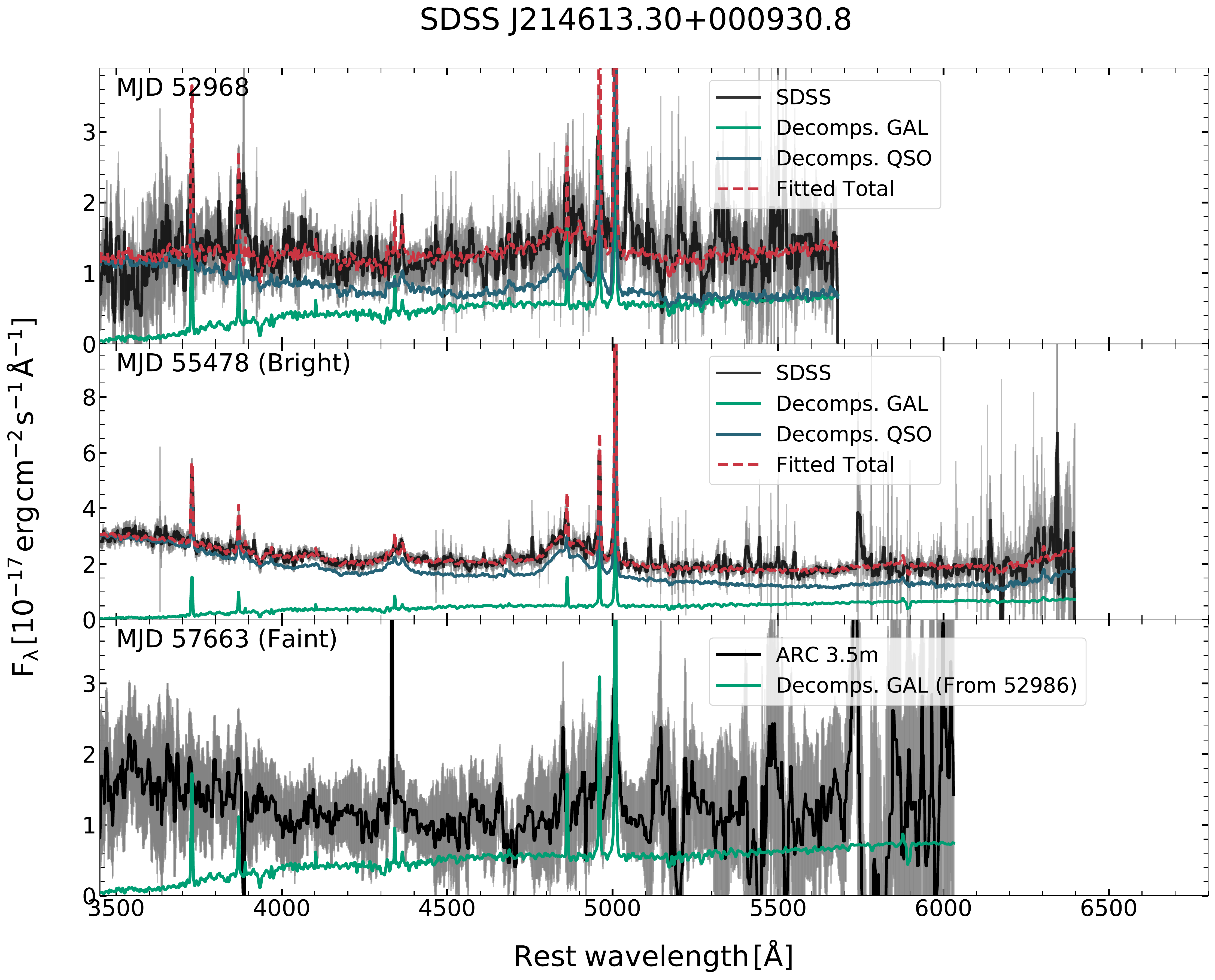}
    \includegraphics[width=0.8\textwidth]{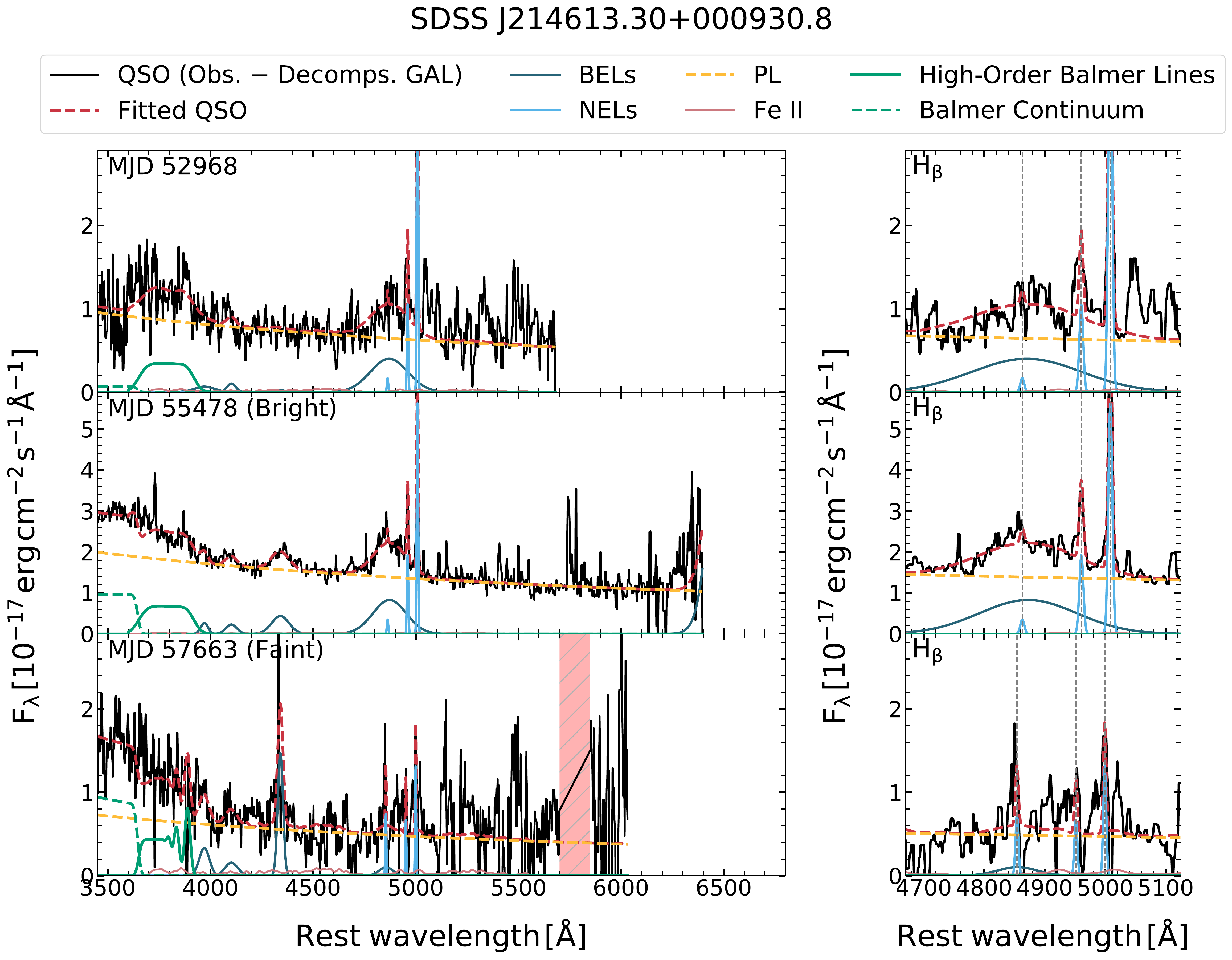}
    \caption{Spectral decomposition and the quasar spectral fitting for SDSS J214613.30+000930.8.}
    \label{fig:optical_data_misc_3}
\end{figure*}

\renewcommand{\thefigure}{\arabic{figure} (Cont.)}
\addtocounter{figure}{-1}

\begin{figure*}[!htb]
    \centering
    \includegraphics[width=0.8\textwidth]{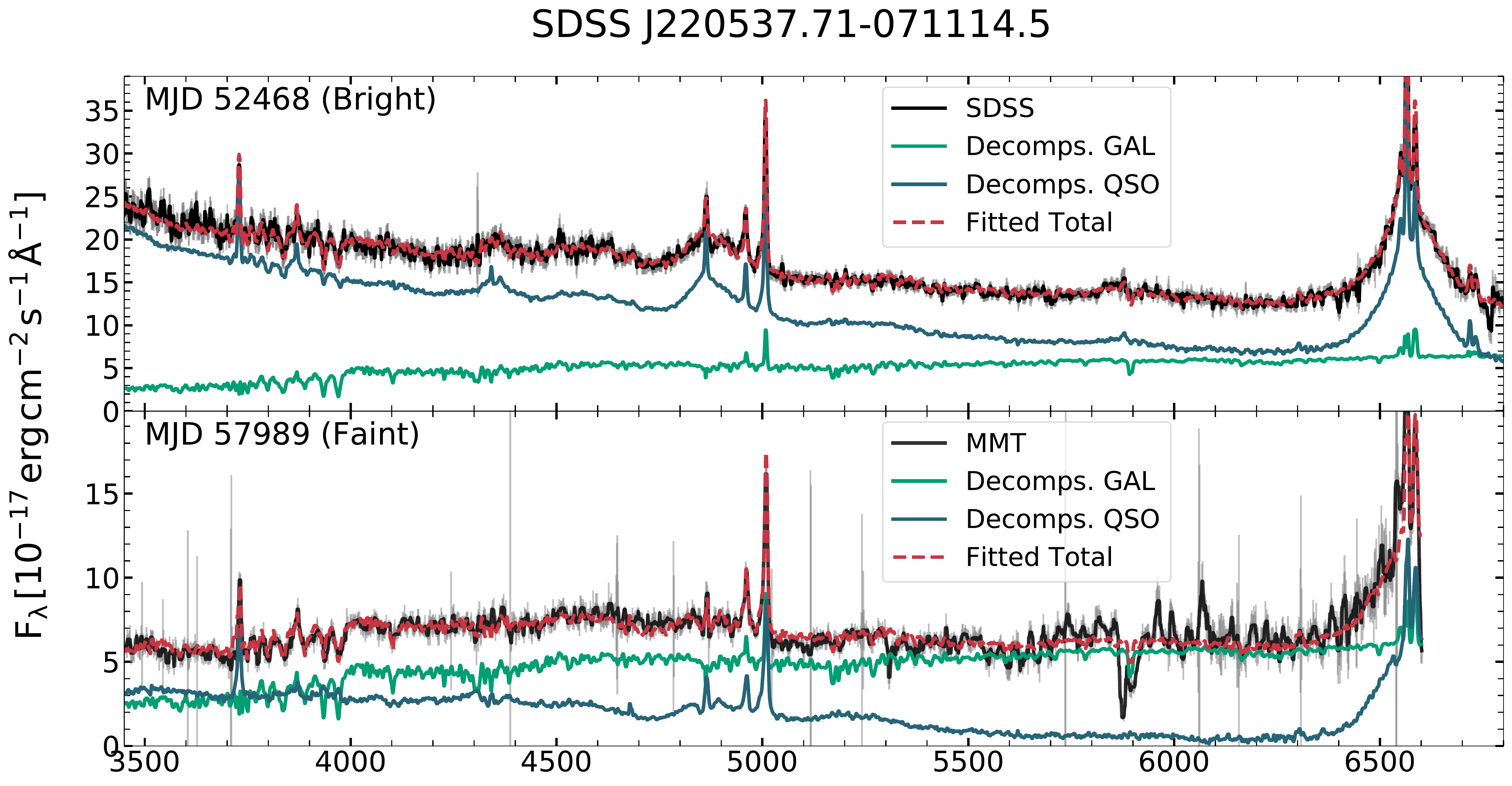}
    \includegraphics[width=0.8\textwidth]{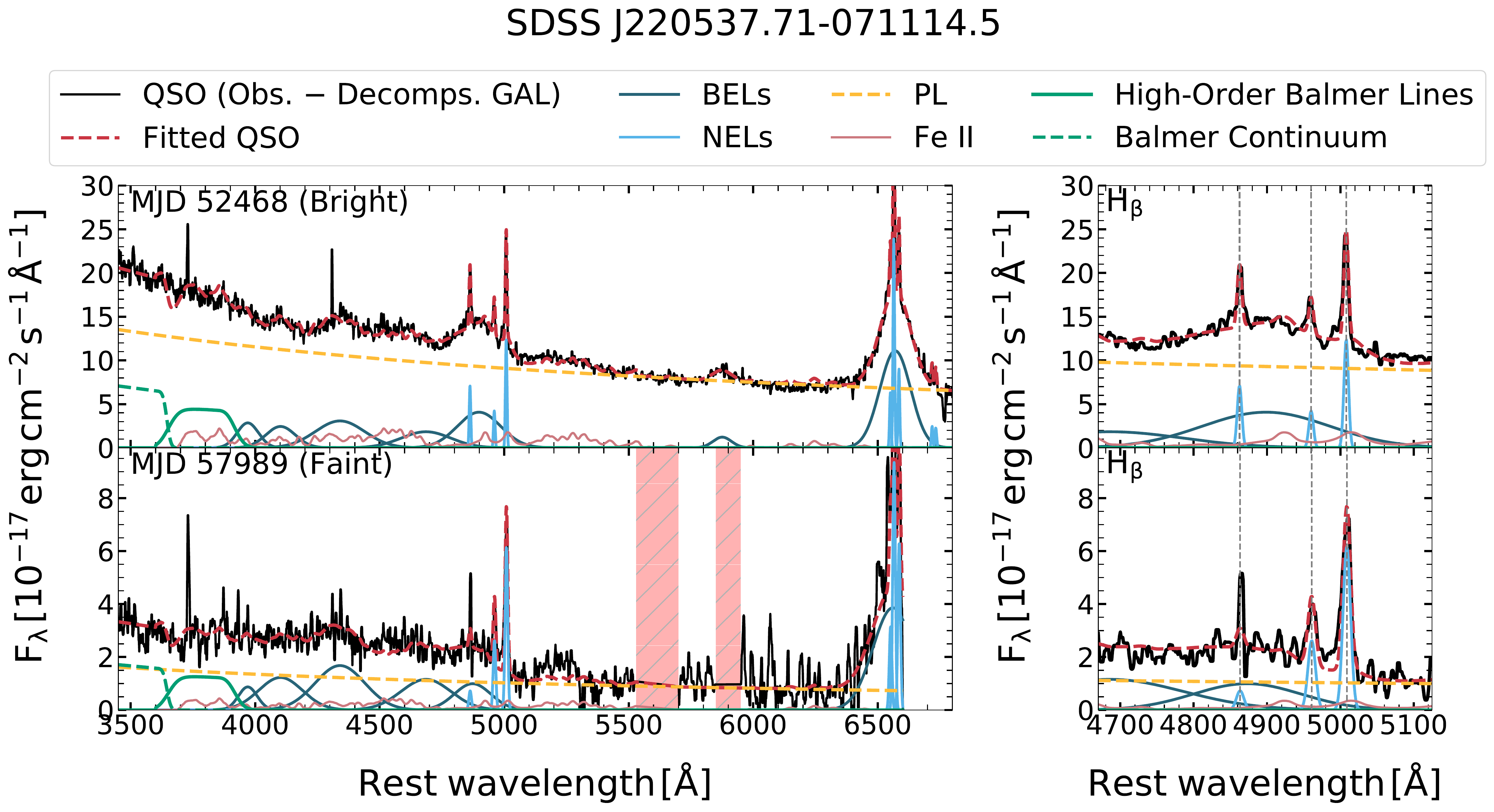}
    \caption{Spectral decomposition and the quasar spectral fitting for SDSS J220537.71-071114.5. Pink shaded regions are masked when we perform the quasar spectrum fitting.}
    \label{fig:optical_data_misc_3}
\end{figure*}

\renewcommand{\thefigure}{\arabic{figure} (Cont.)}
\addtocounter{figure}{-1}

\begin{figure*}[!htb]
    \centering
    \includegraphics[width=0.8\textwidth]{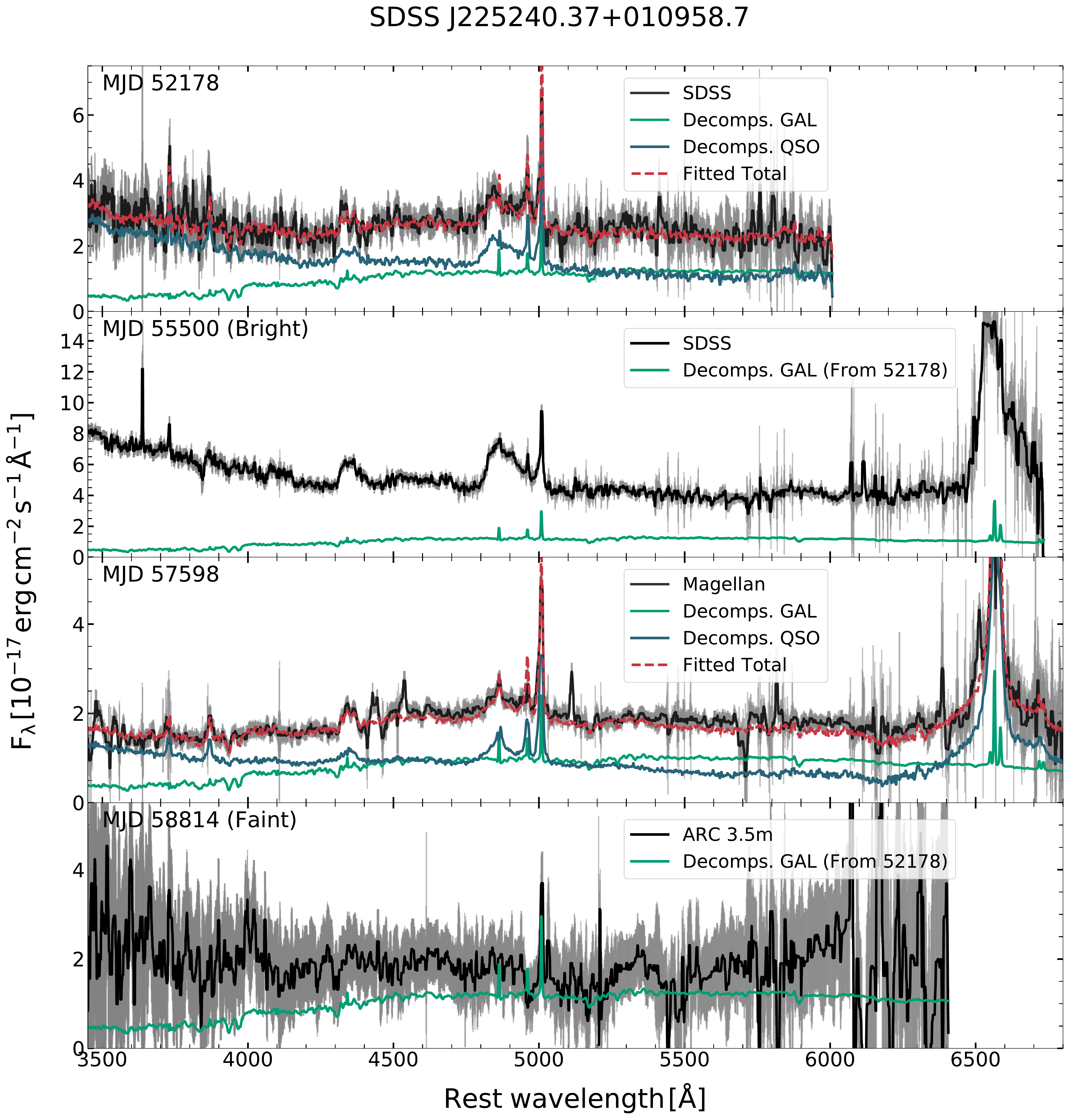}
    \caption{Spectral decomposition for SDSS J225240.37+010958.7. Pink shaded regions are masked when we perform the quasar spectrum fitting.}
    \label{fig:optical_data_misc_3}
\end{figure*}

\renewcommand{\thefigure}{\arabic{figure} (Cont.)}
\addtocounter{figure}{-1}

\begin{figure*}[!htb]
    \centering
    \includegraphics[width=0.8\textwidth]{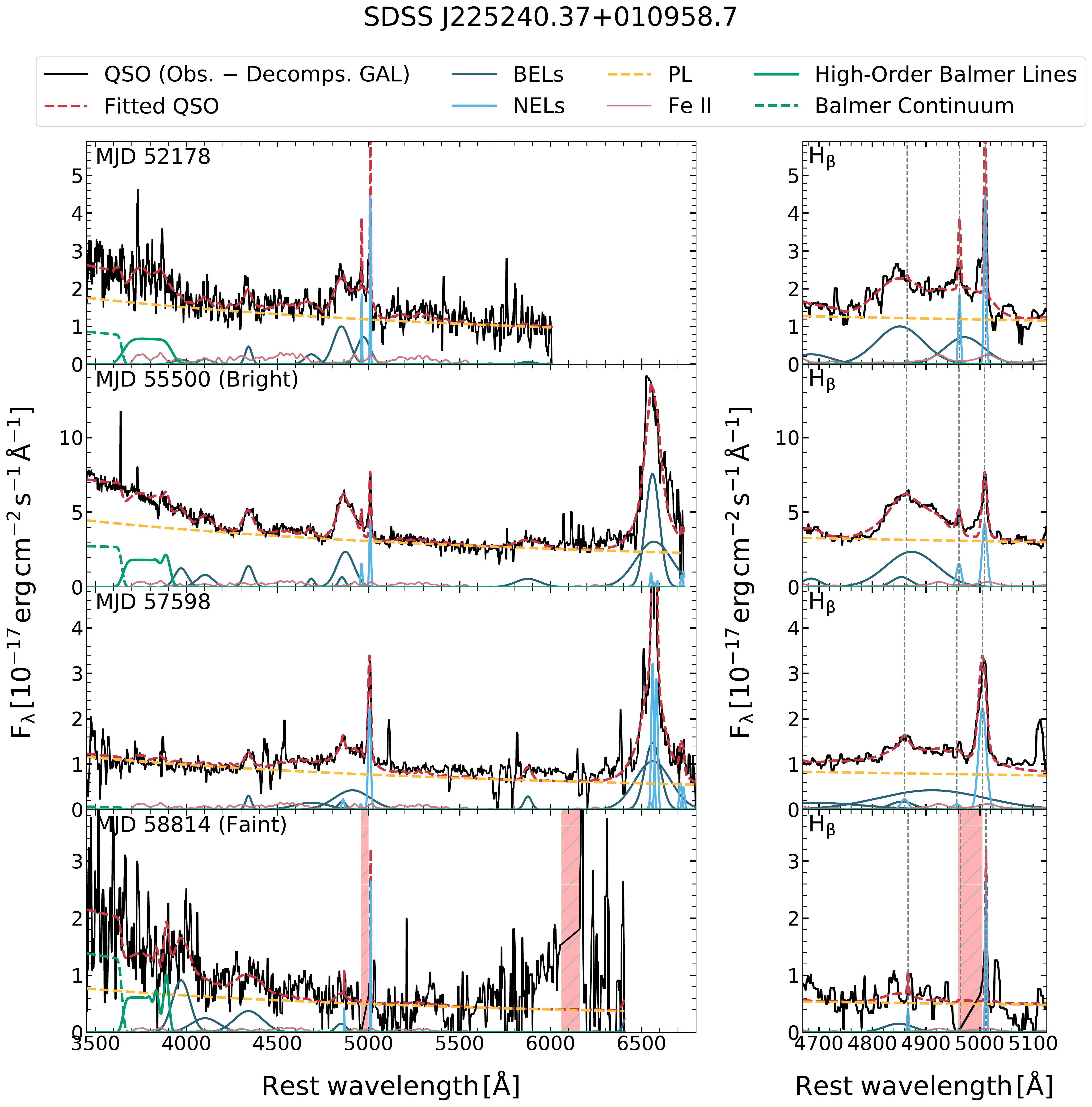}
    \caption{Quasar spectral fitting for SDSS J225240.37+010958.7. Pink shaded regions are masked when we perform the quasar spectrum fitting.}
    \label{fig:optical_data_misc_3}
\end{figure*}

\renewcommand{\thefigure}{\arabic{figure} (Cont.)}
\addtocounter{figure}{-1}

\begin{figure*}[!htb]
    \centering
    \includegraphics[width=0.8\textwidth]{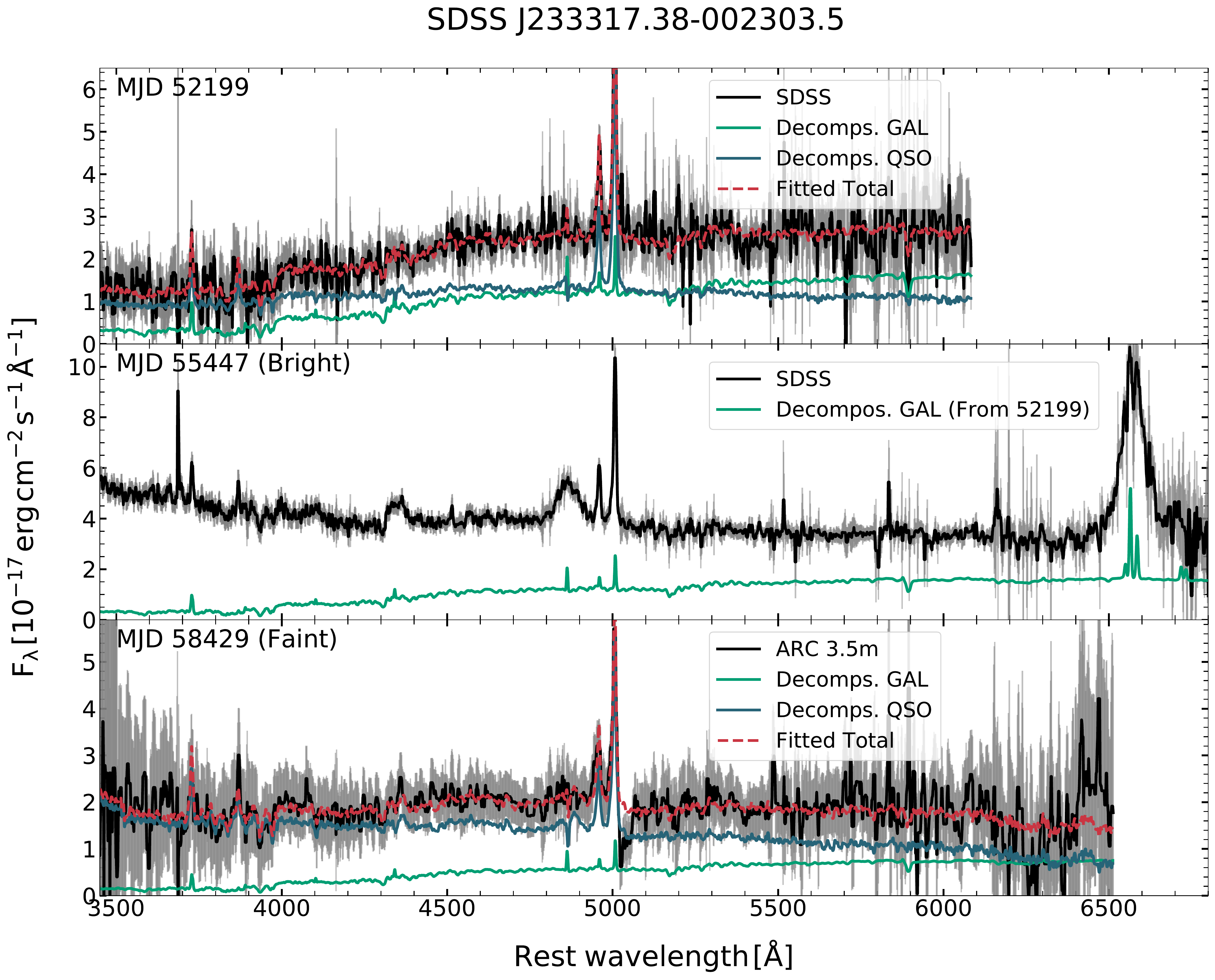}
    \includegraphics[width=0.8\textwidth]{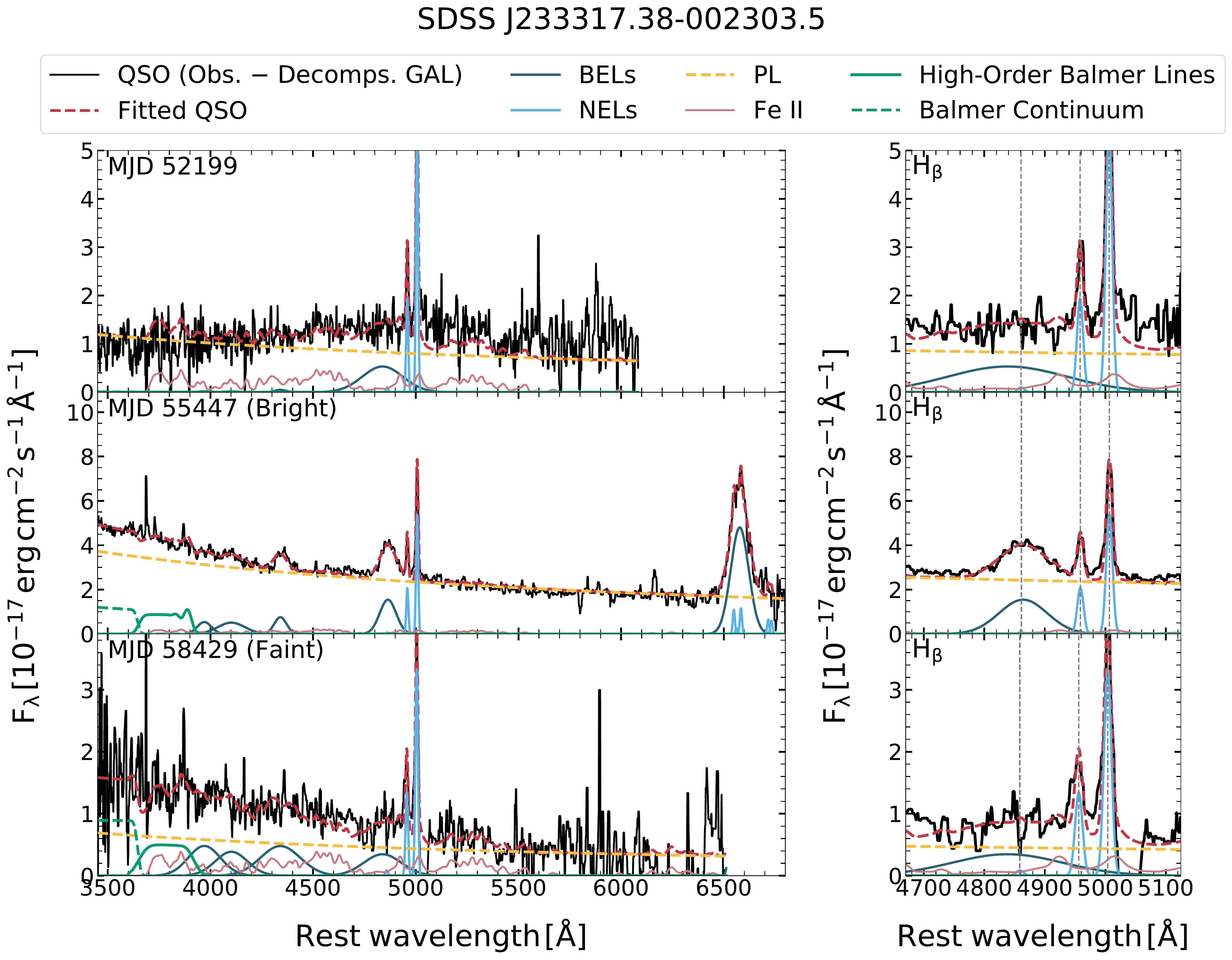}
    \caption{Spectral decomposition and the quasar spectral fitting for SDSS J233317.38-002303.5. Pink shaded regions are masked when we perform the quasar spectrum fitting.}
    \label{fig:optical_data_misc_3}
\end{figure*}

\renewcommand{\thefigure}{\arabic{figure} (Cont.)}
\addtocounter{figure}{-1}

\renewcommand{\thefigure}{\arabic{figure}}



\end{document}